\documentclass[reprint,twocolumn,prm,aps]{revtex4-2}

\usepackage{xcolor}
\usepackage[english]{babel}
\usepackage{graphicx}
\usepackage[colorlinks,allcolors=blue]{hyperref}
\usepackage{epstopdf}
\usepackage{amssymb}
\usepackage{amsmath}
\usepackage{bm}
\usepackage{natbib}
\usepackage{fixmath}
\usepackage{soul}
\usepackage[version=4]{mhchem}

\DeclareMathAlphabet\mathbfcal{OMS}{cmsy}{b}{n}

\newcommand{\etal}{\textit{et al.}}

\usepackage{pdfpages} 
\usepackage{pgffor} 

\makeatletter
\AtBeginDocument{\let\LS@rot\@undefined}
\makeatother

\def\supplementfilename{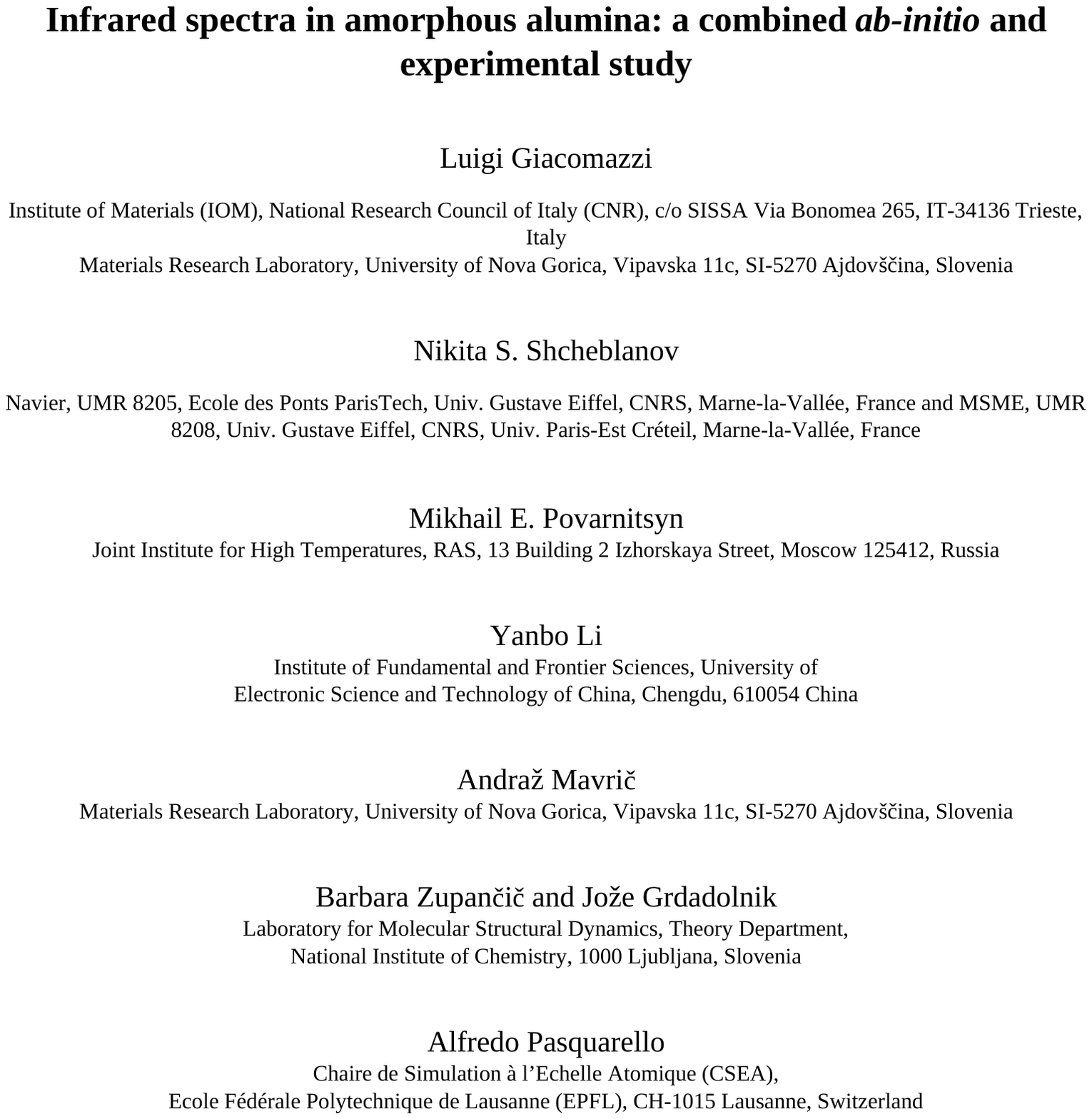}

\pdfximage{\supplementfilename}
\def\numbersupplementpages{\the\pdflastximagepages}

\newif\ifarXiv
\arXivtrue 

\begin{document} 

\title{Infrared spectra in amorphous alumina: a combined \emph{ab initio} and experimental study}

\author{Luigi Giacomazzi}
\email{giacomazzi@iom.cnr.it}
\affiliation{Institute of Materials (IOM), National Research Council of Italy (CNR), c/o SISSA Via Bonomea 265, IT-34136 Trieste, Italy}
\affiliation{Materials Research Laboratory, University of Nova Gorica, Vipavska 11c, SI-5270 Ajdov\v{s}\v{c}ina, Slovenia}

\author{Nikita S. Shcheblanov}
\email{n.s.shcheblanov@gmail.com}
\affiliation{Navier, UMR 8205, Ecole des Ponts ParisTech, Univ. Gustave Eiffel, CNRS, Marne-la-Vall{\'e}e, France}
\affiliation{MSME, UMR 8208, Univ. Gustave Eiffel, CNRS, Univ. Paris-Est Cr{\'e}teil, Marne-la-Vall{\'e}e, France}

\author{Mikhail E. Povarnitsyn}
\affiliation{Joint Institute for High Temperatures, RAS, 13 Building~2 Izhorskaya Street, Moscow 125412, Russia}

\author{Yanbo Li}
\affiliation{Institute of Fundamental and Frontier Sciences, University of Electronic Science and Technology of China, Chengdu, 610054 China}

\author{Andra\v{z} Mavri\v{c}}
\affiliation{Materials Research Laboratory, University of Nova Gorica, Vipavska 11c, SI-5270 Ajdov\v{s}\v{c}ina, Slovenia}

\author{Barbara Zupan\v{c}i\v{c} }
\author{Jo\v{z}e Grdadolnik}
\affiliation{Laboratory for Molecular Structural Dynamics, Theory Department, National Institute of Chemistry,
1000 Ljubljana, Slovenia}

\author{Alfredo Pasquarello}
\affiliation{Chaire de Simulation \`a l'Echelle Atomique (CSEA), Ecole F\'ed\'erale Polytechnique de Lausanne (EPFL), CH-1015 Lausanne, Switzerland}

\begin{abstract}
We present a combined study based on experimental measurements of infrared (IR) dielectric function and first-principles calculations of IR spectra and vibrational density of states (VDOS) of amorphous alumina (am-\ce{Al2O3}). In particular, we show that the main features of the imaginary part of the dielectric function $\epsilon_2(\omega)$ at $\sim$380 and 630~cm$^{-1}$ are related to the motions of threefold coordinated oxygen atoms, which are the vast majority of oxygen atoms in am-\ce{Al2O3}. Our analysis provides an alternative point of view with respect to an earlier suggested assignment of the vibrational modes, which relates them to the stretching and bending vibrational modes of \ce{AlO}$_{n}$ ($n=$~4, 5, and 6) polyhedra. Our assignment is based on the additive decomposition of the VDOS and $\epsilon_2(\omega)$ spectra, which shows that: (i) the band at $\sim$380~cm$^{-1}$ features oxygen motions occurring in a direction normal to the plane defined by the three nearest-neighbor aluminum atoms, i.e. \textit{out-of-plane} motions of oxygen atoms; (ii) Al--O stretching vibrations (i.e. \textit{in-plane} motions of oxygen atoms) appear at frequencies above $\sim$500~cm$^{-1}$, which characterize the vibrational modes underlying the band at $\sim$630~cm$^{-1}$. Aluminum and fourfold coordinated oxygen atoms contribute uniformly to the VDOS and $\epsilon_2(\omega)$ spectra in the frequency region $\sim$350--650~cm$^{-1}$ without causing specific features. Our numerical results are in good agreement with the previous and presently obtained experimental data on the IR dielectric function of am-\ce{Al2O3} films. Finally, we show that the IR spectrum can be modeled successfully by assuming isotropic Born charges for aluminum atoms and fourfold coordinated oxygen atoms, while requiring the use of three parameters, defined in a local reference frame, for the anisotropic Born charges of threefold coordinated oxygen atoms.
\end{abstract}

\maketitle
\section{Introduction}
%
%
Aluminium oxide (\ce{Al2O3}) is a technologically relevant material used in a wide variety of applications~\cite{Hung2022,Scott2018,Reuna2021,Garcia2016,Robertson2006,Zhao2022}. In particular, polymorphs of \ce{Al2O3} possess chemical inertness, thermal stability, and high dielectric properties~\cite{Mavric19}. In recent decades, the electronic properties of amorphous alumina (am-\ce{Al2O3}) have made it an attractive dielectric-insulator material for the use as a high-$\kappa$ material~\cite{Robertson2006} and for the development of supercapacitors~\cite{Fuku2018,Mavric19}. Also, the nanocomposites based on the am-\ce{Al2O3} matrix have been explored for applications in harsh environments, for instance, to produce fuel cladding for Generation~IV systems~\cite{Garcia2016,Zabo2021}, and as scratch-resistant coatings~\cite{valant2016}. Finally, the heat transport properties of am-\ce{Al2O3} are also an active field of research~\cite{Scott2018,harper2023}, the understanding of which is ultimately associated with the correct description of the vibrational modes.

%
%
\ce{Al2O3} presents a variety of crystalline phases (e.g. $\alpha$,  $\gamma$, $\delta$, $\eta$, $\theta$, $\kappa$). This structural polymorphism reflects the predominant ionic nature of the Al-O bond, which entails a large variability in the O-Al-O bond angle and in local structures, which can be composed of both tetrahedral and octahedral Al units.
%
%
Amorphous \ce{Al2O3} can be produced in different forms, e.g. anodic or evaporated films. The atomic structure of am-\ce{Al2O3} thin films has been the object of many experimental investigations in the last two decades. For instance, high-resolution solid-state NMR studies have shown that the fractions of fourfold and fivefold coordinated Al atoms (i.e. \ce{AlO4} and \ce{AlO5} units) are dominant in these films reaching $\sim$92--95\%~\cite{Lee2010,Lee2009,Kim2014}. A subsequent first-principles study~\cite{Liz2011} confirmed this fact for the densities of model structures of 2.9, 3.1, and 3.3~g/cm$^3$. 
The fivefold coordination of Al atoms can be regarded as a fingerprint of the amorphous phase~\cite{Nakamura2013}, which eventually might be relevant for catalysis applications~\cite{Zhao2022}. Crystallization phenomena leads to formation of edge-sharing sixfold coordinated Al polyhedra (i.e. \ce{AlO6} units) together with threefold and fourfold coordinated O atoms (i.e. \ce{OAl3} and \ce{OAl4} units) while the fivefold coordinated Al atoms undergo annihilation~\cite{Lee2018,Nakamura2013}. 
Recent structural investigations (X-ray, neutron, and NMR) of amorphous and deeply supercooled liquid alumina were carried out by Shi~\etal~\cite{Shi2019} and Hashimoto~\etal~\cite{Hash2022}, which also support the presence of high populations of \ce{AlO4} and \ce{AlO5} units predominantly linked by triply shared O atoms.

%
%
%
Unlike various glass-forming oxides (e.g. \ce{SiO2}, \ce{GeO2}, \ce{B2O3}, \ce{P2O5}, etc.), for am-\ce{Al2O3}, as far as we know, there are still no experimental data on inelastic scattering of neutrons or X-rays that provide a vibrational density of states (VDOS), with the exception for nanoparticle shells in the Boson peak frequency range~\cite{Cortie2020}. 
Infrared (IR) spectroscopy is the most widely used experimental technique to investigate vibrational properties in am-\ce{Al2O3} thin films~\cite{Erik1981,Ohw1999,Arai1991,Chu1988,Orosco2018}, but despite this fact, the vibrational bands in the frequency range from $\sim$300 to 700~cm$^{-1}$ are still assigned in the literature to not so well specified \ce{AlO4}, \ce{AlO5}, and \ce{AlO6} stretching vibrational modes~\cite{Li2020ACS}. Moreover, previous theoretical approaches ranging from \emph{classical} and \emph{ab initio} molecular dynamics (MD)~\cite{Momi2006,Vash2008,Gut2010} to neural network schemes~\cite{Li2020} do not lead to a reliable analysis for the assignment of the vibrational modes. 
 
To the best of our knowledge, experimentally derived complex dielectric functions, $\epsilon(\omega)=\epsilon_1(\omega)+i \epsilon_2(\omega)$, for high-quality am-\ce{Al2O3} films have been provided only in a few cases~\cite{Erik1981,Bege1997}. 
In particular, in Ref.~\onlinecite{Erik1981} the dielectric function was derived from the experimental transmittance and reflectance data by using Fresnel's equations. 
However, there is a problem with these experimental data, in particular that these data extrapolate to a rather low static dielectric limit, $\epsilon_1(\omega\to 0)\sim7$, compared to typical values obtained for beam-evaporated alumina films ($\sim$8--9)~\cite{Kubler1991,Eis1975,Sham2004,Mikhael1998}.
It should be noted that the technique used in Ref.~\onlinecite{Erik1981} requires a non-reflecting substrate, high film homogeneity, low surface roughness, and sufficiently large thickness to ensure that no light returns from the inner sample to the sample surface. In addition, the quality of the experimental data depends on the proper choice and treatment (subtraction) of the substrate as well as from the evaluation of the refraction index $n=\sqrt{\epsilon_\infty}$. All of these factors can be the cause of the aforementioned inconsistencies. As a result, all these contradictions, both experimental and theoretical, prompt us to undertake a new combined numerical-experimental study.

%
%
In this work, we primarily focus on the origin of the vibrational modes of am-\ce{Al2O3} giving rise to the experimental IR features peaking at $\sim$380~cm$^{-1}$ and $\sim$600--630~cm$^{-1}$ in the $\epsilon_2(\omega)$ dielectric function.
In order to elucidate the physics behind these features, we perform a complementary analysis based on high-quality experimental IR measurements (of am-\ce{Al2O3} thin films prepared by using an electron-beam evaporation), and first-principles calculations based on density functional theory (DFT). 
We provide detailed numerical modeling of the IR and VDOS spectra, and reveal the contributions of the chemical species (O and Al) and various atomic units through additive decomposition approach. In particular, we point out the primary role of the motions of threefold coordinated O atoms in directions normal and parallel to the plane of the neighboring Al atoms. Such a comprehensive analysis provides a solid ground for the assignment of the vibrational modes and specific IR features of am-\ce{Al2O3}.
Based on the analysis of the calculated Born effective charges, we outline a parametric model that, despite its simplicity, reflects the underlying physics and allows us to calculate the dielectric function at a level comparable to \emph{ab initio} quality. 
%

%

This paper is organized as follows. 
In Sec.~\ref{Sec_Methods} we provide computational details of the DFT calculations (Sec.~\ref{Sec_CompDet}), and the physical properties of the am-\ce{Al2O3} models used for this study are discussed in Sec.~\ref{Sec_Models}. Experimental details concerning the synthesis, structural analysis, and IR spectra measurements are given in Sec.~\ref{Sec_ExptMeth}. 
In Sec.~\ref{Sec_StructAnal}, we address structural properties such as coordination numbers and composition of the \ce{OAl3} units for threefold coordinated oxygen atoms, and we also analyse short- and medium-range orders using the Al-O bond-length and void distributions. 
Section~\ref{Sec_VDOS} is devoted to vibrational modes, and, in particular, we present an analysis of the VDOS spectra. 
In Sec.~\ref{Sec_Infra}, we present results for the calculated and measured IR dielectric functions, the static and high-frequency dielectric constants, while in Sec.~\ref{Sec_BornC} we provide an analysis of the dynamical Born charges needed to define a parametric model for calculating IR spectra.
Section~\ref{Sec_Disc} is devoted to the discussion of the results. The conclusions of this work are drawn in Sec.~\ref{Sec_Concl}.
\section{Methods}\label{Sec_Methods} 
%
%
\subsection{\emph{ab initio} computational details}\label{Sec_CompDet}
All the calculations for am-\ce{Al2O3} carried out in this work are based on DFT, using the generalized gradient approximation (GGA) [i.e. the Perdew-Burke-Ernzerhof (PBE)~\cite{PBE_funct} functional] and norm-conserving pseudopotentials for O and Al atoms~\cite{PseudiNota,EXC_Nota}.
The Kohn-Sham wave functions are expanded in a basis of plane waves up to a kinetic energy cutoff of 70~Ry at the sole $\mathrm{\Gamma}$-point of the Brillouin zone, as justified by the large size and the large band gap of the considered system~\cite{Coll2015}.
The structural models of am-\ce{Al2O3} (Sec.~\ref{Sec_Models}) are relaxed by adopting a force threshold of 0.0025~eV/\AA, which allows for a proper harmonic treatment of the vibrational modes. We calculate the vibrational frequencies and eigenmodes as well as the high-frequency dielectric constant and the dynamical Born charges within the linear response approach~\cite{QE}. In addition, the vibrational modes are obtained under the constraint that the force constants satisfy the \emph{acoustic sum rule}. The codes used here for calculating structural and vibrational properties are freely available with the Quantum ESPRESSO package~\cite{QE}. 
\subsection{am-\ce{Al2O3} model structures}\label{Sec_Models}
%
%
In our study, we use several structural models of am-\ce{Al2O3}. The first structural model (hereafter named model~I) was generated in Ref.~\onlinecite{Coll2015} via Born-Oppenheimer MD simulations following a quench-from-the-melt approach. 
The initial configuration was obtained by adapting a 160-atom model of $\kappa$-\ce{Al2O3} to an orthorhombic supercell. The alumina melt was equilibrated at 3000~K for 1~ps and subsequently quenched to 300~K in 14~ps.
The atomic structure was then further relaxed by constraining the shape of the supercell and by allowing the lattice constants to relax freely in order to remove the residual stress. 
The final model contains 160 atoms in a (periodic) orthorombic simulation cell with sides: 11.475, 11.244, and 12.778~\AA, corresponding to a density of 3.3~g/cm$^{3}$. The \emph{neutron} radial distribution function (RDF) calculated for this model~\cite{Coll2015} is found in fair agreement with experimental data from Ref.~\onlinecite{Lamp97}. 
For the present work, we have just further relaxed the atomic configuration (generated in Ref.~\cite{Coll2015} and available in Ref.~\cite{Guo2019}) as explained in Sec.~\ref{Sec_CompDet}. 

%
%
Two other models, hereafter named model~II and model~III, were generated in Ref.~\onlinecite{Momi2006} (and named there as model G and model H, respectively). Models II and III both contain 120 atoms in an orthorhombic periodic supercell with a density of 3.27 and 3.22~g/cm$^{3}$~\cite{Momi2006}, respectively.
The models were generated by means of classical MD through a quench-from-the-melt procedure using the empirical pair-wise potential developed by Matsui in Ref.~\onlinecite{Matsui96}, and then employed in DFT calculations within the local density approximation (LDA) to obtain the dielectric response~\cite{Momi2006,Momi2007b}. 
Structure factors and RDFs of models~II and III have been discussed in Ref.~\cite{Momi2006}, and found to be consistent with other classical MD results and experiments. 
In the present work, the atomic configurations of these models have been only further relaxed by first-principles as explained in Sec.~\ref{Sec_CompDet}.

%
To analyze the medium-range-order structure of our am-\ce{Al2O3} models, we perform a void size distribution (VSD) analysis by using the Zeo++ code~\cite{zeopp}, an open source software that carries out a geometry-based analysis of porous materials and their voids (pores). In this analysis, every atom is considered as a sphere, and the radius of Al and O atoms are 1.23 and 0.73~\AA, respectively~\cite{hung2006}. A probe radius of 0.1~\AA~is used to scan empty space and find voids, i.e. spheres that can be inserted in contact with surrounding atoms without intersecting with them. Thus, the resulting VSD shows us how much void space corresponds to certain void sizes.

\subsection{Experimental methods}\label{Sec_ExptMeth}

Amorphous \ce{Al2O3} films are deposited on the surface of a 500 $\mu$m-thick undoped Si single-crystal wafer by using an electron-beam evaporation (Angstrom Engineering AMOD) with the substrate at room temperature and base pressure of $5\times10^{-5}$~Torr. The evaporation source is high-purity \ce{Al2O3} and the deposition rate is 1~\AA/s and controlled by a quartz crystal microbalance. The nominal thickness is 300~nm for FTIR spectroscopy measurements and 100~nm for X-ray reflectivity (XRR) measurements.
%
The high resolution transmission electron microscopy (HR-TEM) imaging and X-ray diffraction are used to reveal the amorphous nature of the \ce{Al2O3} film. Applying the XRR technique gives a density of 2.74~g/cm$^3$ and shows a low surface roughness of 1.1~nm. The reader is referred to Fig.~S1 and related text in the Supplementary Material 
\cite{SupplInfo} for more information on film characterization.

%
%
IR spectra are recorded using Bruker Vertex 70V IR spectrometer in transmission mode under vacuum. 64 scans are averaged with a resolution of 2~cm$^{-1}$. A blank Si wafer is used for background recording. The traces of atmospheric water and \ce{CO2} were removed using atmospheric compensation in Opus software (Bruker Optik GmbH).
The spectra are smoothed by using the Savytzky-Golay algorithm and leveled to zero absorption at 4000~cm$^{-1}$. 
In addition to the broad absorption between $\sim$100 and $\sim$1100~cm$^{-1}$, there is a weak but broad absorption between $\sim$3000 and $\sim$3600~cm$^{-1}$. This band pattern can be assigned to the stretching vibrations of surface hydroxyl groups~\cite{Abadleh2003}. 

For calculating optical constants $n(\omega)$ and $k(\omega)$, related to the complex refraction index $\underline{n}(\omega)=n(\omega)+ik(\omega)$, we follow the procedure developed by Gerakines and Hudson~\cite{Gera2020}. This iterative procedure requires as input a transmission spectrum with a reference refractive index of the film, its thickness, and optical constants for the substrate over the same wavenumber range as the transmission spectrum. At each iteration, the Kramers-Kronig relation is used to obtain $n(\omega)$ from $k(\omega)$ for each wavenumber.
The method is preferred in systems where obtaining a pure substance is challenging (e.g. in thin films), because it is hard to avoid multiple reflections and inferences when using reflective substrates.

\section{Results}\label{Sec_Results}
%
%
\subsection{Structural analysis}\label{Sec_StructAnal}
Before addressing the vibrational spectra of am-\ce{Al2O3} presented in the next subsections, we here provide some basic information about the structure of the am-\ce{Al2O3} models used in this study. In particular, in Table~\ref{Tab_Coord} we provide the coordination numbers of Al and O atoms, which is the most essential information for describing the local atomic arrangement (short-range order) in am-\ce{Al2O3}. 
Model~I features a large fraction ($\sim$70\%) of fivefold and sixfold coordinated Al atoms with a smaller content of fourfold coordinated Al atoms ($\sim$30\%), in contrast to models~II and III where the fraction of fourfold coordinated Al atoms ($\sim$50\%) is comparable to the total amount of fivefold and sixfold coordinated Al atoms. 
In model~I, the vast majority (75\%) of the O atoms are threefold coordinated (to Al atoms), while the remaining 25\% show fourfold coordination. The fractions of O atoms that are threefold coordinated are even larger in models~II and III, reaching up to 91.7\%, although some of the O atoms in these models are even found to be twofold coordinated (1.4 and 8.3\% in models~II and III, respectively).
%
%
%
\begin{table}
\caption{\label{Tab_Coord} Frequency (\%) of coordination numbers of Al and O atoms in the am-\ce{Al2O3} models used in this study, obtained after \emph{ab initio} relaxation in PBE~\cite{RelaxNota}. Experimental coordination frequencies, obtained from the $^{27}$Al MAS NMR spectra of Ref.~\onlinecite{Hash2022} (anodized alumina) and Ref.~\onlinecite{Lee2018} (PVD film), are also given. 
The cutoff radius of $R_{\rm cut}\simeq2.25$~\AA\ is used.}
\begin{center}
\begin{ruledtabular} 
\begin{tabular}{lcccccc} 
& \multicolumn{3}{c}{Al} & \multicolumn{3}{c}{O} \\
 Coordination	 & 4 & 5 & 6 & 2 & 3 & 4 \\
 \hline
Model~I & 29.7 & 53.1 & 17.2 & $-$ & 75.0 & 25.0 \\ 
Model~II & 50.0 & 39.6 & 10.4 & 1.4 & 91.7 & 6.9 \\
Model~III & 52.1 & 43.7 & 4.2 & 8.3 & 82.0 & 9.7 \\ 
Expt. \cite{Hash2022} & 37.5 & 52.1 & 10.3 & & &   \\
Expt. \cite{Lee2018}  & 48$\pm$4 & 45$\pm$4 & 7$\pm$2   &   &  & 
\end{tabular}
\end{ruledtabular}
\end{center}
\end{table}
%
%
\begin{table*}
\caption{Composition (\%) of \ce{OAl3} units by the coordination number of neighboring Al atoms of the central threefold coordinated O atom. Minor fractions related to \ce{OAl3} units with only highly coordinated Al atoms (i.e. O$^{[6]}\!$Al$_3$) are omitted for clarity. For each type of \ce{OAl3} units, the average degree of planarity $\pi_d$ is indicated. The cutoff radius of $R_{\rm cut}\simeq2.25$~\AA\ is used.}
\label{Tab_Coord_O3}
\begin{center}
\begin{ruledtabular} 
\begin{tabular}{lccccccc} 
 	 & $^{[4]}\!$Al$_3$ & $^{[4]}\!$Al$_2^{[5]}\!$Al & $^{[4]}\!$Al$^{[5]}\!$Al$_2$ & $^{[4]}\!$Al$_2^{[6]}\!$Al & $^{[4]}\!$Al$^{[6]}\!$Al$_2$ & $^{[4]}\!$Al$^{[5]}\!$Al$^{[6]}\!$Al & $^{[5]}\!$Al$_3$ \\
 \hline
Model~I & 2.8 & 11.1 & 36.1 & 4.2 & 4.2 & 15.3 &  11.3 \\ 
Model~II & 9.1 & 18.2 & 22.7 & 15.2 & 3.0 & 22.7 & 3.0 \\
Model~III & 8.5 & 30.5 & 25.4 & 8.5 & 1.7 & 11.9 & 11.9 \\
$\pi_d$ & 0.987 & 0.978 & 0.961 & 0.972 &  0.957 & 0.969 & 0.972  
\end{tabular}
\end{ruledtabular}
\end{center}
\end{table*}
Models~II and III show Al coordination numbers in reasonable agreement with the experimental estimate of Refs.~\onlinecite{Lee2009,Lee2018,Hash2022}, with minor differences with respect to the LDA-DFT results of Ref.~\onlinecite{Momi2006}, of which the most noticeable one concerns the low number of fivefold coordinated Al atoms, which has been found to increase by 10$\%$ upon the generation of model~II~\cite{Momi2006}. 
On the other hand, model~I shows a larger average Al coordination due to the larger number of sixfold and fivefold coordinated Al atoms. 
Incidentally, we note that, in spin-coated am-\ce{Al2O3} films, large fractions ($\sim$20\%) of sixfold coordinated Al atoms have been estimated by using NMR spectroscopy~\cite{Cui2018}. Indeed, the structure of am-\ce{Al2O3} films depends on the method of synthesis and on the deposition conditions~\cite{Lee2018,Lee2014,Cui2018}, which can lead to structural changes, typically corresponding to fluctuations up to $\sim$10\% of the coordination frequencies of Al (see Table~\ref{Tab_Coord}). 
 In particular, am-\ce{Al2O3} resulting from solution processing such as sol-gel and spin-coating can exhibit large fractions of sixfold coordinated Al atoms due to the presence of hydroxyl groups, residual hydrogen or hydration~\cite{Lee2018}.

%
%
The NMR investigation of Ref.~\onlinecite{Lee2018} provided evidence that threefold coordinated O atoms in am-\ce{Al2O3} can be distinguished from threefold coordinated O atoms in transition alumina phases, like $\theta$- and $\gamma$-alumina. In fact, the threefold coordinated O atoms in the crystal structure are found to form a particular structural motif with one fourfold and two sixfold coordinated Al atoms (i.e. O$^{[4]}\!$Al$^{[6]}\!$Al$_2$). By contrast, Ref.~\onlinecite{Lee2018} suggests that in am-\ce{Al2O3} the dominant configurations are O$^{[4]}\!$Al$_2^{[5]}\!$Al and O$^{[4]}\!$Al$^{[5]}\!$Al$_2$, while the occurrence of configurations featuring only sixfold coordinated Al atoms, such as O$^{[6]}\!$Al$_3$, should be negligible. 
%
%
\begin{figure}[b]
 \includegraphics[width= 0.95\columnwidth]{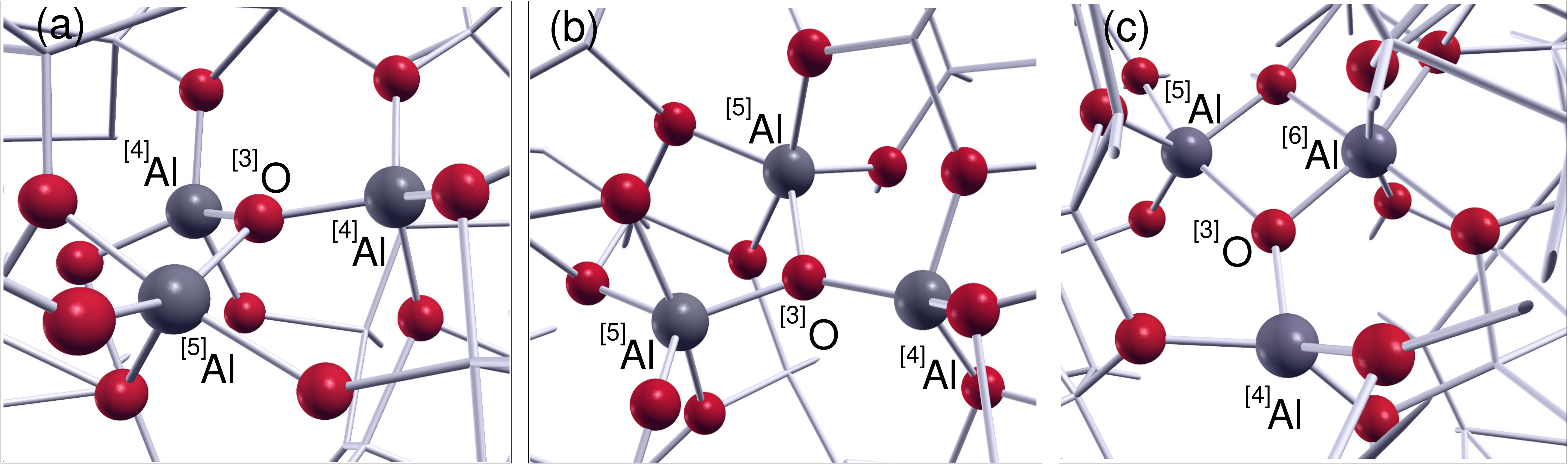}
 \caption{Ball and stick models~\cite{Kokal99} of the dominant \ce{OAl3} units as found in the investigated am-\ce{Al2O3} models:
(a) an O$^{[4]}\!$Al$_2\,^{[5]}\!$Al unit, (b) an O$^{[4]}\!$Al$^{[5]}\!$Al$_2$ unit, and (c) an O$^{[4]}\!$Al$^{[5]}\!$Al$^{[6]}\!$Al unit. Oxygen atoms (red balls) and aluminum atoms (grey balls) are shown. For clarity only atoms belonging to the \ce{OAl3} unit are displayed.}
 \label{Fig_O3_struct}
\end{figure}

In Table~\ref{Tab_Coord_O3} we give the results of the analysis of \ce{OAl3} units in our models, and in particular we confirm that the dominant configurations (Fig.~\ref{Fig_O3_struct}) in am-\ce{Al2O3} models are O$^{[4]}\!$Al$_2^{[5]}\!$Al and O$^{[4]}\!$Al$^{[5]}\!$Al$_2$ as they sum up to $\sim$50\% of the \ce{OAl3} units ($\sim$56\% in model~III). 
Yet, we also find an important fraction of O$^{[4]}\!$Al$^{[5]}\!$Al$^{[6]}\!$Al units, which have to be considered as typical of the amorphous phase because of the occurrence of a fivefold coordinated Al atom, and which could be relevant in the context of crystallization~\cite{Lee2018}. The low concentration of O$^{[4]}\!$Al$^{[6]}\!$Al$_2$ in our models indicates a marked structural difference with respect to $\theta$- or $\gamma$-alumina crystalline phases, and it is consistent with the amorphous nature of our model structures~\cite{Lee2018}. 

%
%
The threefold coordinated O atoms are found to approximately lie on the plane of their three Al neighbors. To evaluate the planarity of these units, we define the ratio $\pi_d$ as obtained by dividing the sum over the three Al-O-Al angles by 360$^{\circ}$, which at most equals 1 for an ideal planar unit. About 80\% of \ce{OAl3} units in our models show a quite high degree of planarity, with a $\pi_d$ larger than 0.95 (i.e. with a sum of angles of 342$^{\circ}$), comparable to the degree of planarity of \ce{OAl3} units in $\theta$-alumina. In particular, we find that O$^{[4]}\!$Al$_3$ and O$^{[4]}\!$Al$_2^{[5]}\!$Al are highly planar with an average $\pi_d$ of 0.987 and 0.978, respectively. 
%
%
\begin{figure}
 \includegraphics[width= 0.95\columnwidth]{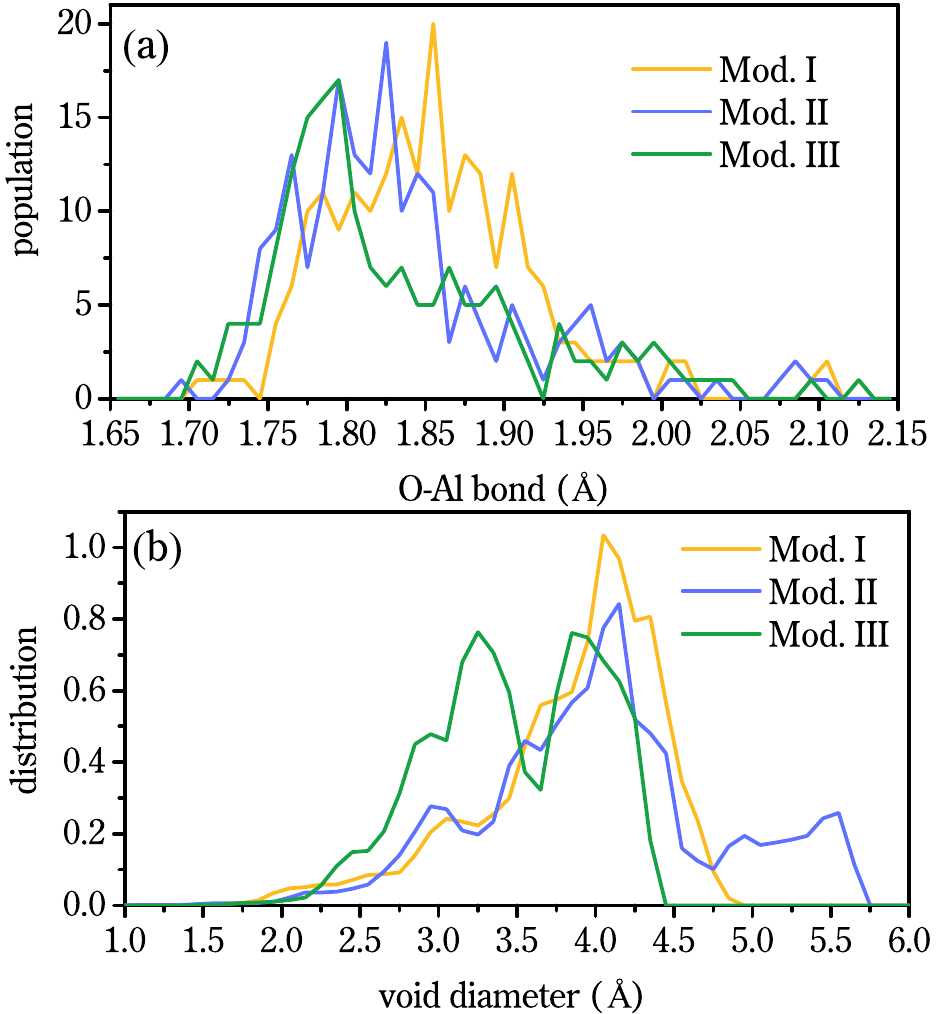}
 \caption{(a) Distribution of O-Al bond lengths for threefold coordinated oxygen atoms for models~I, II and III. (b) Void size distribution for the same models.}
 \label{Fig_bond_voids}
\end{figure}

The mean Al-O bond lengths for models~I, II, and III are 1.88, 1.85, and 1.84~\AA, respectively. The average Al-O bond lengths for models~II and III are in rather good agreement with the peak of the experimental neutron RDF of Ref.~\onlinecite{Hash2022} 
located at 1.81~\AA. Compared to the LDA-DFT simulation~\cite{Momi2006}, we register a slight increase (by $\sim$0.10~\AA) of the mean Al-O bond length for models~II and III. Moreover, the calculated average Al-O bond-lengths of the \ce{AlO4}, \ce{AlO5} and \ce{AlO6} polyhedra in our models are 1.79, 1.89, and 1.96~\AA, respectively. 
Hence, the slightly longer mean Al-O bond length of model~I reflects the higher number of \ce{AlO6} polyhedra and the smaller number of \ce{AlO4} tetrahedra in this model as compared to models~II and III (see Table~\ref{Tab_Coord}). 
On the other hand, the calculated average O-Al bond lengths in our models for twofold, threefold and fourfold coordinated O atoms are 1.72, 1.85, and 1.95~\AA, respectively, with some minor variations among models ($\sim$0.01~\AA) and with a rather large spread of about 0.04~\AA\, in the bond distributions.

In Figs.~\ref{Fig_bond_voids}(a) and (b) we provide the distribution of O-Al bond lengths for threefold coordinated O atoms and the void distribution analysis for the three alumina models. The Al-O bond length distributions of models~II and III show peaks at $\sim$1.82~\AA\, and $\sim$1.78~\AA, respectively. The Al-O bond distribution of model~I shows a shift towards longer bonds, as indicated by its peak located at $\sim$1.86~\AA\, with respect to the distributions of models~II and III. This value overestimates by 0.05~\AA\, the experimental bond length~\cite{Hash2022}. This is also consistent with the differences between the mean Al-O bond lengths in our models, indicating that threefold coordinated O are still rather well representing the O-Al bond distribution. 
 
The density of alumina films may vary due to its porosity and this can affect the dielectric function~\cite{Bege1997}. Although a discussion of porosity in alumina is beyond the scope of the present paper, it is of some interest to evaluate the distribution of voids in our models. This will also give a measure of the ``openness'' of the alumina network structure. Despite the fact that all the models have similar material densities, although slightly smaller in model~III than in model~I, we note that the VSD shown in Fig.~\ref{Fig_bond_voids}(b) is substantially different in the three models. 
Moreover, from the positions of the main peaks of the VSDs in Fig.~\ref{Fig_bond_voids}(b), one is brought to assume that a greater porosity corresponds to longer bonds, i.e. that the population of bond lengths correlates with the VSD.

\subsection{Vibrational density of states }\label{Sec_VDOS}
The vibrational density of states $\rho(\omega)$ is calculated as
\begin{equation} \label{eq:vdos}
 \rho(\omega)=\frac{1}{3N}\sum_n \delta(\omega-\omega_n),
\end{equation} 
where $N$ is the total number of atoms in the model and $\omega_n$ is the vibrational frequency of the $n$-th vibrational eigenmode $\mathbold{\xi}_I^{n}$. In addition, we consider the atomic contributions of the $\ce{Al}$ and differently coordinated O subpopulations:
\begin{equation}
 \rho_{\alpha}(\omega)=\frac{1}{3N}\,\sum_n\sum_{I\in\alpha} \left|\mathbold{\xi}_I^{n}\right|^2\delta(\omega-\omega_n),
 \label{eq:rho}
\end{equation}
where $\alpha=\ce{Al}$ or O and the index $I$ enumerates the atoms.

\begin{figure}
\begin{center}
 \includegraphics[width= 0.90\columnwidth, angle=0]{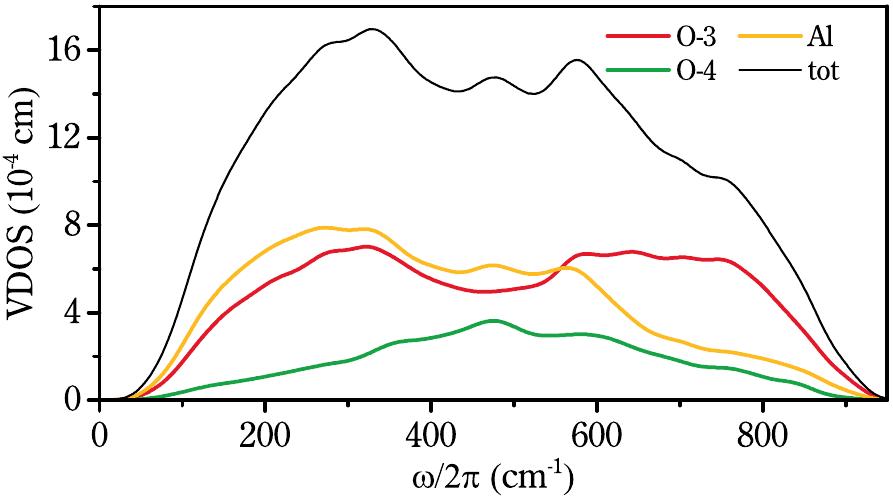} 
 \caption{Total VDOS (solid black) with its decomposition, as defined by Eq.~\eqref{eq:rho}, in terms of the contribution of aluminum (solid yellow), threefold (solid red), and fourfold (solid green) coordinated oxygen atoms. Analysis made for the model~I of Ref.~\onlinecite{Coll2015}. A Gaussian broadening of 20~cm$^{-1}$ is used.}
 \label{Fig_vDOS}
 \end{center}
\end{figure}

In Fig.~\ref{Fig_vDOS}, we show the total VDOS for model~I of the am-\ce{Al2O3} along with its decomposition in terms of oxygen and aluminium contributions as defined in Eq.~\eqref{eq:rho}. The calculated VDOS spectrum extends up to $\sim$950~cm$^{-1}$. Several bands can be discerned and interpreted thanks to the atomic decomposition. The two bands peaked at $\sim$320 and $\sim$590~cm$^{-1}$ reflect underlying features related to threefold coordinated O atoms. The central band peaking at $\sim$470~cm$^{-1}$ corresponds to the contribution of fourfold coordinated O atoms to the total VDOS. 
The localization of vibrational modes is studied by using the mode participation ratio~\cite{bell72,Nik2019}, which is shown in Fig.~S2 in the Supplementary Material \cite{SupplInfo}. It can be noted that the modes between 200 and 400~cm$^{-1}$ are not spatially localized on individual \ce{OAl3} units, while the stretching modes above 500~cm$^{-1}$ show a tendency to become much more spatially localized.

%
%

%
%
In Fig.~\ref{Fig_AlO_Stretch}, we consider threefold coordinated O atoms, and demonstrate that there is a correlation between the O-Al bond-lengths associated with these O atoms and the O-Al stretching frequencies $\omega_{\mathrm{max}}$. These frequencies are calculated by first finding the projections ${\mathcal P}_{\rm OAl}(\omega)$ of the oxygen displacements, $\mathbold{\xi}_I^{n}$, onto the O-Al directions for each one of the three neighboring Al atoms, and then by looking at the frequency position of the peak of ${\mathcal P}_{\rm OAl}(\omega)$. 
The linear interpolation obtained by fitting the stretching frequencies $\omega_{\mathrm{max}}$ for all threefold coordinated O atoms of the three models under study shows that the O-Al stretching frequency $\omega_{\mathrm{max}}$ decreases with increasing bond lengths with a slope of $\approx$1250~cm$^{-1}$/\AA.
Since Al atoms with increasing oxygen coordination show longer Al-O bond-lengths (Sec.~\ref{Sec_StructAnal}), this correlation allows us to explain the redshifts of the stretching bands in the spectra of our models as discussed in Sec.~\ref{Sec_Infra}.

\subsection{Infrared dielectric function}\label{Sec_Infra}
\begin{figure}
 \includegraphics[width= 0.95\columnwidth]{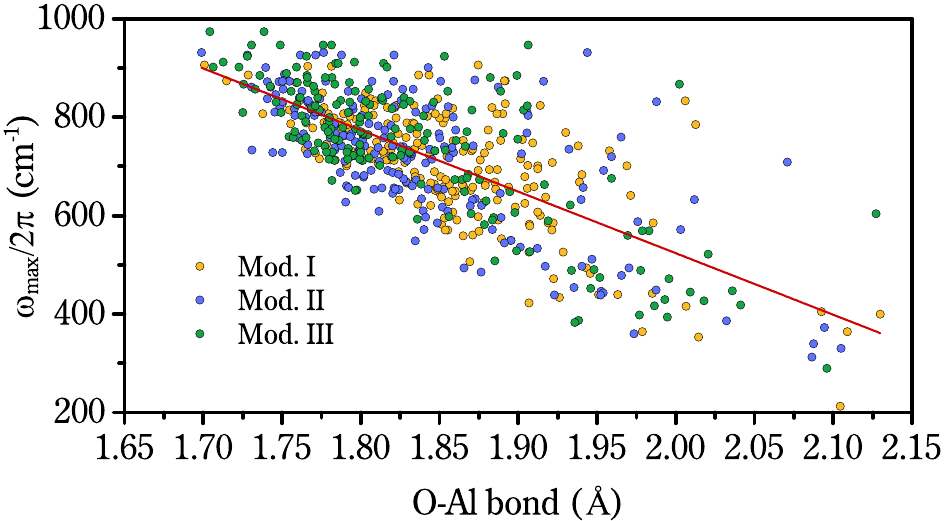}
 \caption{Stretching frequency vs O-Al bond-length for models I (yellow discs), II (blue discs), and III (green discs) [see text for more details]. The solid (red) line corresponds to a linear fit of all the data.}
 \label{Fig_AlO_Stretch}
\end{figure}
The real and imaginary parts of the dielectric response function, $\epsilon_1(\omega)$ and $\epsilon_2(\omega)$, are here calculated with the following expressions~\cite{ThorpeLeeuw,PC97}: 
\begin{eqnarray}
\label{EqEps1}
 \epsilon_1(\omega)&=&\epsilon_{\infty}-\frac{4\pi}{3 \mathrm{\Omega}}\sum_n
 \frac{|\boldsymbol{\mathcal F}^{n}|^2}{\omega^2-\omega^2_n}, \\ 
 \label{EqEps2}
 \epsilon_2(\omega)&=&\frac{4\pi^2}{3 \mathrm{\Omega}}\sum_n \frac{|\boldsymbol{\mathcal F}^{n}|^2}{2\omega_n}
 \delta(\omega-\omega_n),
\end{eqnarray}
where $\mathrm{\Omega}$ is the volume of the periodic simulation cell, and $\epsilon_\infty=\frac{1}{3}{\rm Tr}(\boldsymbol{\epsilon}_\infty)$ is the high-frequency (electronic) dielectric constant. 
The oscillator strengths $\boldsymbol{\mathcal F}^{n}$ can be calculated from the Born charges ${Z}_I^{*}$ and eigenmodes $\mathbold{\xi}_I^{n}$ as:
\begin{equation}
 {\mathcal F}^{n}_j = \sum_{I}{\mathcal F}^{n}_{Ij}=\sum_{I}\sum_{k}Z^{*}_{I,jk}
 \frac{\xi^{n}_{I k}}{\sqrt{m_I}},
\end{equation}
where $m_I$ is the atomic mass of the atom $I$. We also introduce a coupling function as follows: 
\begin{equation}
 C_{\textrm{IR}}(\omega) = \epsilon_2(\omega)/\rho(\omega),
 \label{eq:cir}
\end{equation}
where $\rho(\omega)$ is the VDOS given by Eq.~\eqref{eq:vdos}. 

Similar to the Raman additive decomposition (see Refs.~\onlinecite{Nik2019,misha2020}), the IR spectrum can be recast as:
\begin{equation}
\epsilon_2(\omega)=\frac{4\pi^2}{3 \mathrm{\Omega}}\sum_n\sum_{\alpha} \frac{\{\boldsymbol{{\mathcal F}}^{n}\}^2_{\alpha}}{2\omega_n}
 \delta(\omega-\omega_n),
 \label{eq:eps2:decomp}
\end{equation}
where
\begin{equation}
 |\boldsymbol{\mathcal F}^{n}|^2=\sum_{\alpha}\{\boldsymbol{{\mathcal F}}^{n}\}^2_{\alpha}=\sum_{\alpha} \left \{ \sum_{\alpha'}\left(\boldsymbol{\mathcal F}^{n}_{\alpha}\cdot\boldsymbol{\mathcal F}^{n}_{\alpha'}\right) \right\},
 \label{eq:eps2:int}
\end{equation}
with the $\alpha$-th oscillator strengths given by:
\begin{equation}
 \boldsymbol{\mathcal F}^{n}_{\alpha}={\mathcal F}^{n}_{\alpha j}=\sum_{I\in\alpha}{\mathcal F}^{n}_{Ij}.
\end{equation}
We evaluate the static dielectric constant $\epsilon_0$, which can be written in terms of the electronic $\epsilon_\infty$ and lattice $\epsilon_{\rm lat}$ contributions~\cite{PC97} as:
\begin{eqnarray}\label{StatDielConst}
 \epsilon_0 & = &\epsilon_\infty + \epsilon_{\rm lat}, \\ 
 \epsilon_{\rm lat}&=& \frac{4\pi}{3 \mathrm{\Omega}} \sum_n \frac{| \boldsymbol{\mathcal{F}}^{n}|^2}{\omega_n^2}.
\end{eqnarray}
The dielectric function described above gives access to all the dielectric properties. In particular, to access longitudinal optical modes, we hereafter address the so-called energy loss function, which can be written in terms of the complex dielectric function $\epsilon(\omega)$ by using directly Eqs.~\eqref{EqEps1} and \eqref{EqEps2}:
\begin{equation}
 -{\mathrm{Im}}\left[\frac{1}{\epsilon(\omega)}\right]=\frac{\epsilon_2(\omega)}{\epsilon^2_1(\omega)+\epsilon^2_2(\omega)}.
 \label{Eq:EnLoss}
\end{equation}
However, the loss function can also be obtained through a direct calculation~\cite{ThorpeLeeuw,PumaDRM} by using the longitudinal modes as follows:
\begin{equation}
 -{\mathrm{Im}}\left[\frac{1}{\epsilon(\omega)}\right]=\frac{4\pi^2}{ \mathrm{\Omega} (\epsilon_{\infty})^2}\sum_n 
 \frac{(\hat{\mathbf q}\cdot \boldsymbol{\mathcal F}^{n})^2}{2\omega_n}\, \delta(\omega-\omega_n),
 \label{EnLoss}
\end{equation}
where $\boldsymbol{\mathcal F}^{n}$ is calculated by using $\omega_n$ and $\mathbold{\xi}_I^{n}$ obtained by diagonalizing the full dynamical matrix ${\mathcal D}^{\mathbf {q}\to0}_{Ii,Jj}$~\cite{PumaDRM}. For isotropic systems, the energy loss function in Eq.~\eqref{EnLoss} can be averaged over all directions $\hat{\mathbf q}$. Here, we use the three Cartesian directions for this average. 

\begin{table}
\caption{\label{Tab_DielConst} Density, $\varrho$ (g/cm$^3$), static dielectric constant, $\epsilon_0$, high-frequency dielectric constant, $\epsilon_\infty$, and lattice contribution, $\epsilon_{\rm lat}$, as calculated for models~I, II, and III of am-\ce{Al2O3} in this work. The experimental data are presented for am-\ce{Al2O3} films obtained by various methods. Data for $\varrho$ are taken from Refs.~\cite{Bege1997,Lee1995,Segda2001}, those for $\epsilon_0$ from Refs.~\cite{Gus2001,Kubler1991,Eis1975}, and those for $\epsilon_\infty$ from Refs.~\cite{Sham2004,Mikhael1998,Gama2015}.}
\begin{center}
\begin{ruledtabular}
\begin{tabular}{llllr} 
 & $\varrho$ & $\epsilon_0$ & $\epsilon_{\rm lat}$ & $\epsilon_\infty$ \\ 
	 \hline 
Model~I & 3.30 & 11.44 & 8.26 & 3.18 \\
Model~II & 3.27 & 10.33 & 7.21 & 3.12 \\
Model~III & 3.22 & 11.11 & 8.01 & 3.10 \\
Expt. & 2.5$-$3.4 & 8$-$11 & & 2.5$-$2.9 \\
Expt. (present) & 2.74 & $\sim$9 & & $\sim$2.72 
\end{tabular} 
\end{ruledtabular}
\end{center} 
\end{table}
%
%
In Table~\ref{Tab_DielConst}, we give the static dielectric constants calculated for models~I, II, and III, together with the high-frequency dielectric constants and the lattice contributions, in comparison with experimental data. The values of the static dielectric constant given in the literature depend on the method of preparation, e.g. $\epsilon_0$ is 8.3 for electron-beam evaporated films~\cite{Sham2004,Eis1975,Kubler1991}, while Ref.~\onlinecite{Evan2017} gives values ranging from 9.8 to 13.3 for barrier-type aluminium oxides. For films obtained by atomic layer deposition (ALD), one finds in Ref.~\onlinecite{Gus2001} a value of 11, while K\"ubler ~\cite{Kubler1991} provides $\approx$9.0 below 10~Hz at room temperature for films prepared by ion-assisted evaporation. The densities of am-\ce{Al2O3} films also varies significantly depending on the production method, e.g. densities of $\sim$3.1--3.4~g/cm$^{3}$ have been reported for magnetron-sputtered alumina films~\cite{Segda2001,Lee1995}, while a value of 2.45~g/cm$^{3}$ has been measured for sol-gel prepared alumina films~\cite{Bege1997}. Beam-evaporated alumina films, as those investigated in this study, typically show a density range of $\sim$2.6--3.1~g/cm$^{3}$~\cite{Kubler1991,Parf1995}. 
%
%
The experimental values of the high-frequency electronic dielectric constant $\epsilon_\infty$ of am-\ce{Al2O3} films range from 2.5 to 2.9, which corresponds to a high-frequency refractive index $n$ between $\sim$1.6 and $\sim$1.7 (at wavelengths around 1 $\mu$m)~\cite{Sham2004,Mikhael1998,Gama2015,Segda2001}, with slight variations depending on the preparation method.

%
%
The calculated density of the studied am-\ce{Al2O3} models is about 3.2--3.3~g/cm$^{3}$ and lies in the range of experimentally recorded densities (Table~\ref{Tab_DielConst}). 
The electronic dielectric constant $\epsilon_\infty$ shows only minor variations between models ($\sim$3\%), which have slightly different densities (see Table~\ref{Tab_DielConst}). Also, the PBE-DFT results for models~II and III show only minor differences ($\lesssim$1.5\%) compared to the LDA-DFT results~\cite{Momi2007b}.
In contrast, the lattice contribution $\epsilon_{\rm lat}$ is quite sensitive to the details of the VDOS, as indicated by the rather large differences between the models in Table~\ref{Tab_DielConst} (see also Ref.~\onlinecite{Momi2006}). Compared to the LDA-DFT results of Ref.~\cite{Momi2006}, we have an increase of about 20\% for the $\epsilon_{\rm lat}$ in model~III, meanwhile for model~II we register a decrease of $\sim$2\%, likely due to the fact that our VDOS does not show the strong peaks in the frequency range 50--100~cm$^{-1}$ found for the corresponding model G~\cite{Momi2006,Momi2007b}. 
In Ref.~\onlinecite{Evan2017}, by using finite electric fields, a value of 3.03 was found for the high-frequency permittivity $\epsilon_\infty$ of model I, which only slightly underestimates the linear-response result in Table~\ref{Tab_DielConst}. This is consistent with the results obtained with different methods for models containing about a hundred atoms~\cite{PUMA2003}. 

%
%
\begin{figure}[!ht]
\begin{center}
 \includegraphics[width= 0.95\columnwidth]{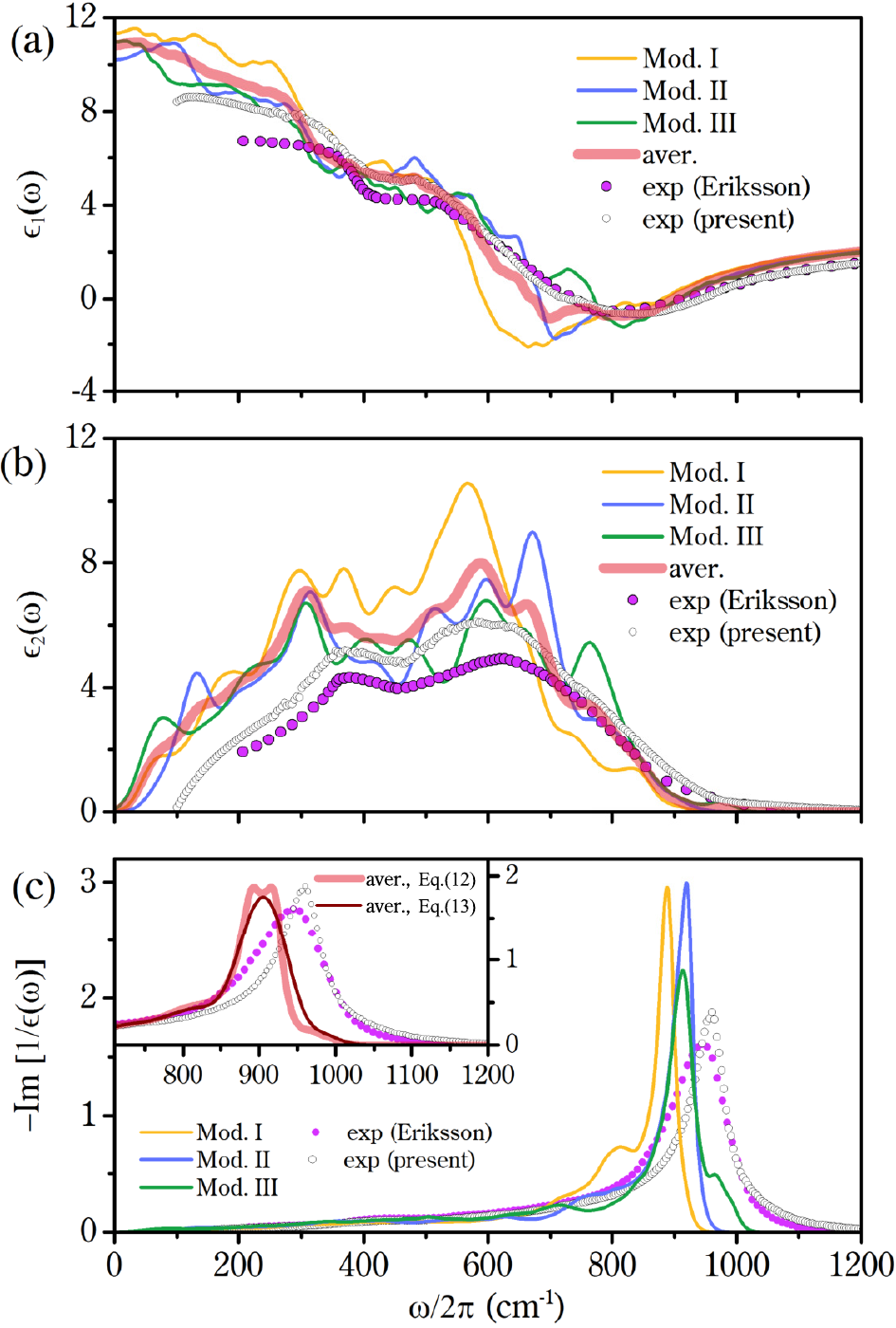}
 \caption{(a) The real part of the dielectric function, $\epsilon_1(\omega)$, for model~I (yellow curve), model~II (blue curve), model~III (green curve), and averaged over the three models (red). Experimental data from~\cite{Erik1981,Erik1982,NotaDigit} (violet circles) and this work (black empty circles). (b) The same notations as in (a) but for the imaginary part, $\epsilon_2(\omega)$. (c) The same notations as in (a) but for the energy loss function, $-\mathrm{Im}[1/\epsilon(\omega)]$, calculated by using Eq.~\eqref{Eq:EnLoss}. The inset in panel (c) shows a zoom of the loss function averaged over the three models calculated by using Eq.~\eqref{Eq:EnLoss} (red curve) and Eq.~\eqref{EnLoss} (brown curve). A Gaussian broadening of 20~cm$^{-1}$ is used in all calculations.}
 \label{Fig_ImEps_Models}
\end{center}
\end{figure}

In Figs.~\ref{Fig_ImEps_Models}(a) and (b), the dielectric functions $\epsilon_1(\omega)$ and $\epsilon_2(\omega)$ calculated for the three am-\ce{Al2O3} models are compared with the experimental ones obtained from the data on the transmittance and reflectance by using the Fresnel equations~\cite{Erik1981}, and those measured in this work. While the experimental $\epsilon_1(\omega)$ of Refs.~\onlinecite{Erik1981,Erik1982} gives a static limit of $\sim$7, the data measured in this work entails a larger value $\sim$9. 
The shapes of the $\epsilon_1(\omega)$ spectra shown in Fig.~\ref{Fig_ImEps_Models}(a) are similar for both sets of experimental data, with a rather smooth step at $\sim$360--380~cm$^{-1}$ and a very broad minimum at $\sim$800~cm$^{-1}$. The dielectric function $\epsilon_1(\omega)$ calculated for all three models is rather noisy due to the finite-size effect, which limits the number of vibrational modes at low frequencies. For all the models the static limit ($\sim$10--11) is considerably larger than the experimental ones, because of (i) the slightly larger density, which may account for an increase of $\epsilon_\infty$ by $\sim$0.4 (Table~\ref{Tab_DielConst}), and of (ii) the finite size of our models, which leads to vibrational modes being less accessible at low frequencies ($\lesssim 50$~cm$^{-1}$, i.e. Boson peak region). Despite these limitations, the shape of the $\epsilon_1(\omega)$ spectrum obtained by averaging over the three models agrees reasonably well with the experimental data.

%
%
The experimental $\epsilon_2(\omega)$ spectrum reported in Refs.~\onlinecite{Erik1981,Erik1982} shows a very broad peak with a maximum at $\sim$630~cm$^{-1}$ together with an additional feature at $\sim$380~cm$^{-1}$. The experimental data obtained in this work present similar broad bands: the main band forming a \textit{plateau} from $\sim$570 to 620~cm$^{-1}$, peaking at $\sim$585~cm$^{-1}$), and the additional band showing a peak at $\sim$370~cm$^{-1}$. The intensity of both bands turns out to be significantly increased compared to previous measurements~\cite{Erik1981}.
%
%
The main features of the $\epsilon_2(\omega)$ spectrum calculated for model~I comprise a double peak at 305 and 370~cm$^{-1}$ and a principal band peaked at $\sim$570~cm$^{-1}$, which exhibits a redshift with respect to the experimental data displayed in Fig.~\ref{Fig_ImEps_Models}(b). 
%
%
The average Al-O bond length in model~I is larger than in models~II and III
by about 0.03~\AA. Because of the trend shown in Fig.~\ref{Fig_AlO_Stretch}, this produces a redshift in the stretching region (500--900~cm$^{-1}$) of the $\epsilon_2(\omega)$ spectrum of model~I, compared to the spectra of the other two models.
%
The spectra of models~II and III show the first band at $\sim$320~cm$^{-1}$ and the second band extending from $\sim$500 to $\sim$700~cm$^{-1}$, the main peaks of which are located at $\sim$600 and 670~cm$^{-1}$ in models~III and II, respectively.
%
The dielectric response for models~II and III should be compared with the LDA-DFT results of Ref.~\onlinecite{Momi2007b}, since our GGA-PBE models~II and III are obtained by relaxation of the LDA-DFT models (see details in Sec.~\ref{Sec_Models}).
For model~III, the results are very close to those reported in Ref.~\onlinecite{Momi2007b}. By contrast, for model~II, our results do not show neither any strong peak below $\sim$100~cm$^{-1}$, as was found in Ref.~\onlinecite{Momi2007b}, nor any important feature at $\sim$480~cm$^{-1}$. We attribute these differences to minor structural changes that occur upon the structural relaxation described in Sec.~\ref{Sec_StructAnal} and to the different DFT setups.
In addition, we note that porous am-\ce{Al2O3} can exhibit features in the $\epsilon_2(\omega)$ spectrum at low frequency, i.e. $\sim$100~cm$^{-1}$, but these do not occur in high-quality films of am-\ce{Al2O3}~\cite{Bege1997}.
Overall, averaging over the three models gives an $\epsilon_2(\omega)$ spectrum in good agreement with the experimental one, except for a slight underestimation of the frequencies of the main peaks (by $\sim$10\%) and an overestimation of the band intensities, which is typical for amorphous models of such size~\cite{LG2009b}.

%
%
In Fig.~\ref{Fig_ImEps_Models}(c), we show the loss function, $-\mathrm{Im}[1/\epsilon(\omega)]$, calculated through Eq.~\eqref{Eq:EnLoss}.
The transverse and longitudinal modes are revealed by the frequency position of the maxima in $\epsilon_2(\omega)$ and $-\mathrm{Im}[1/\epsilon(\omega)]$ spectra, respectively~\cite{Gal76}.
A dominant high-frequency peak of $-\mathrm{Im}[1/\epsilon(\omega)]$ at $\sim$900~cm$^{-1}$ is clearly visible for all models.
In models~II and III, the peaks occur at very close frequencies (920 and 915~cm$^{-1}$), while model~I exhibits a peak at a slightly redshifted frequency 890~cm$^{-1}$, as a likely consequence of the trend explained hereabove for the $\epsilon_2(\omega)$ spectrum. 
The peaks obtained by averaging the spectra [calculated by Eqs.~\eqref{Eq:EnLoss} and \eqref{EnLoss}] over all models [see inset on the panel (c)], are located at $\sim$900~cm$^{-1}$, which is in good agreement (within $\sim$7\%) with our experimental position of the high-frequency peak at $\sim$970~cm$^{-1}$ obtained by using Eq.~\eqref{Eq:EnLoss}. Moreover both the experimental data obtained in this work and in Ref.~\onlinecite{Erik1981} produce peaks at close frequencies at about 950-970~cm$^{-1}$. 
%
%
\begin{figure}
 \includegraphics[width= 0.95\columnwidth]{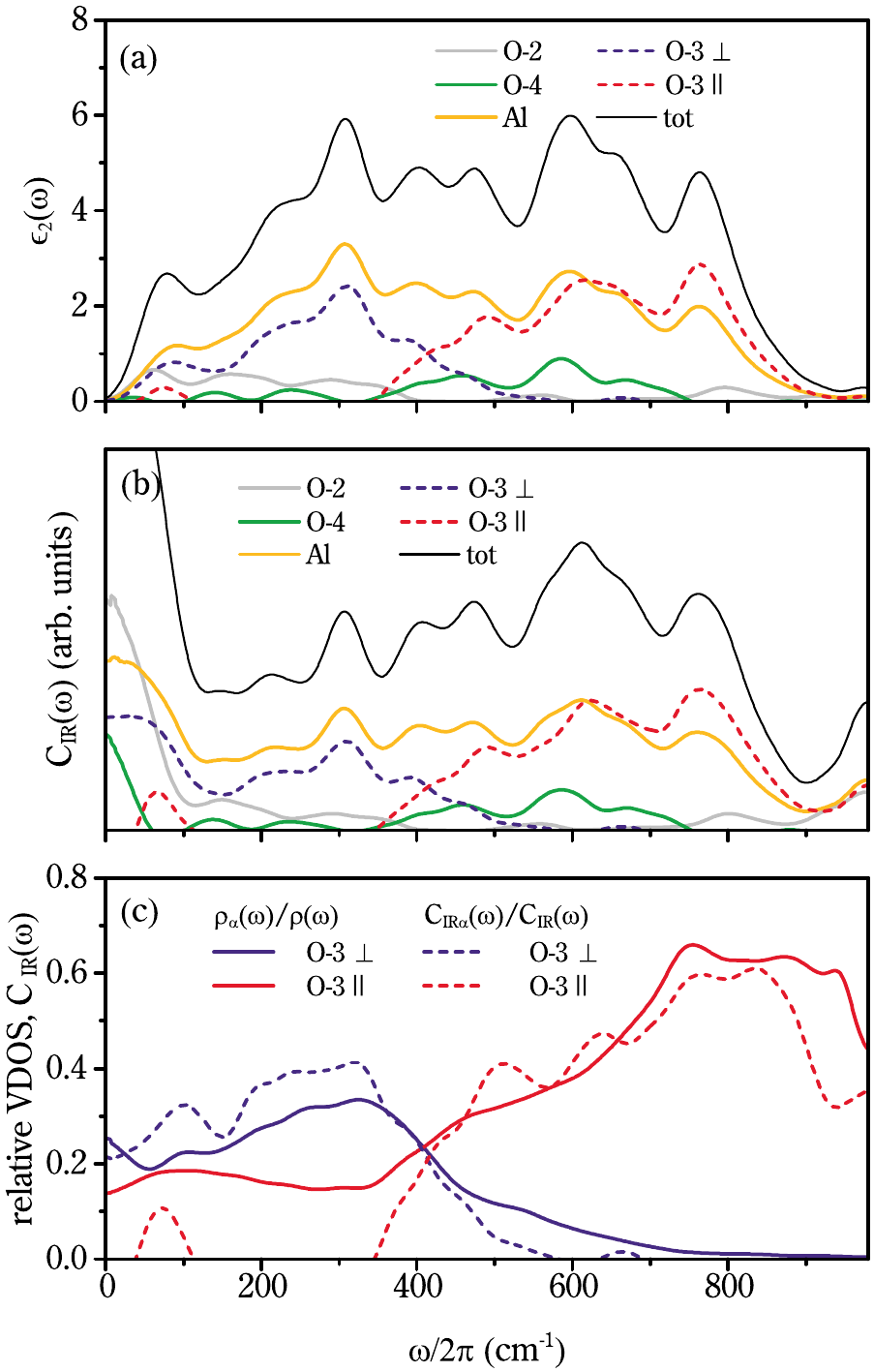}
 \caption{Results for model~III. (a) Decomposition of $\epsilon_2(\omega)$ into contributions from Al atoms (yellow) and from twofold (solid light-gray), fourfold (solid green), and threefold coordinated O atoms. Contribution of the latter is split into \textit{out-of-plane} i.e. perpendicular (dashed, blue) and \textit{in-plane} i.e. parallel (dashed, red) motions. (b) Decomposition of the IR coupling function $C_{\text{IR}}(\omega)$ with the same notations as in panel (a). (c) Comparison between the contributions of \textit{out-of-plane} (blue) and \textit{in-plane} (red) motions of threefold coordinated O atoms in terms of the relative VDOS (solid) and relative IR coupling function (dashed) spectra.
 }
 \label{Fig_e1_e2_rel}
\end{figure}

%
%
In the remainder of this subsection, in order to discuss the origin of the IR vibrational bands, we provide an atomic decomposition of the IR spectrum by using Eqs.~\eqref{eq:cir} and \eqref{eq:eps2:decomp}. Next, we analyse the two main components underlying the IR spectrum: (i) the IR coupling functions given by Eq.~\eqref{eq:cir}; (ii) the squared displacement spectra in terms of relative weights of partial VDOSs given by Eq.~\eqref{eq:rho}.
In Fig.~\ref{Fig_e1_e2_rel}(a), we show such a decomposition focusing on the imaginary part of the dielectric function, $\epsilon_2(\omega)$, of model III, the results for the other two models being similar.
It can be seen that the main features come from contributions of \textit{in-plane} and \textit{out-of-plane} (perpendicular) motions of threefold coordinated O atoms, the plane being defined by the three nearest neighbor Al atoms. The vibrations related to the \textit{out-of-plane} motion of these O atoms occur in the range $\lesssim 500$~cm$^{-1}$ (dash blue), and give a maximal contribution to $\epsilon_2(\omega)$ around $\sim$320~cm$^{-1}$. Vibrational modes featuring \textit{in-plane} motions of threefold coordinated oxygens cover the range $\gtrsim 400$~cm$^{-1}$ (dash red) and become dominant above $\sim$500~cm$^{-1}$.
The distinct separation of these contributions is due to different force constants associated with motions of threefold coordinated O atoms in parallel (\textit{stiff}) and perpendicular (\textit{soft}) directions. 
Al atoms, being highly coordinated, do not possess prominent features and contribute to the $\epsilon_2(\omega)$ spectrum rather uniformly over the entire frequency domain ($\sim$50\%).
Twofold coordinated O atoms provide a small contribution for frequencies below 400~cm$^{-1}$, and also at about 800~cm$^{-1}$ (O-Al stretching motions).
Fourfold coordinated O atoms also give a rather minor contribution to the $\epsilon_2(\omega)$ spectrum in the frequency range going from 400 to 700~cm$^{-1}$.
Thus, twofold and fourfold coordinated O atoms do not produce specific features in the $\epsilon_2(\omega)$ spectrum, because of their rare occurrence in all our am-\ce{Al2O3} models (as compared to threefold coordinated O atoms, see Table~\ref{Tab_Coord}) and also because of the wide O-Al bond distribution associated with fourfold coordinated O atoms, as indicated by its large spread of 0.04~\AA. 
The atomic decomposition of the coupling function,
$C_{\text{IR}}(\omega)$, 
presented in Fig.~\ref{Fig_e1_e2_rel}(b), shows that the amplitude relations for the contributions of twofold, threefold, and fourfold coordinated O atoms, are quite similar to the ones displayed by the $\epsilon_{2 \alpha} (\omega)$ in Fig.~\ref{Fig_e1_e2_rel}(a), consistently with the fact that the VDOS is rather constant, without showing prominent peaks in the range from $\sim300$ to 800~cm$^{-1}$.
In Fig.~\ref{Fig_e1_e2_rel}(c) we show the relative partial VDOSs (or squared displacement spectra) $\left|\mathbold{\xi}_{\alpha}(\omega)\right|^2=\rho_\alpha(\omega)/\rho(\omega)$ that highlights the frequency ranges of \textit{in-plane} and \textit{out-of-plane} motions of threefold coordinated O atoms.
We also display the relative coupling functions [i.e. $C_{\text{IR}\alpha}(\omega)/C_{\rm IR}(\omega)=\epsilon_{2\alpha}(\omega)/\epsilon_2(\omega)$], which present a substantial similarity to the relative partial VDOSs. 
The good agreement between the curves highlights (i) the dominant role of the \emph{pure} \textit{in-plane} and the \emph{pure} \textit{out-of-plane} vibrations, i.e. the interference effect, given by Eq.~\eqref{eq:eps2:int}, is marginal, and (ii) the minor influence of the anisotropy and the orientation of the Born effective charge tensors. 
Thus, the functional dependence of the IR spectrum $\epsilon_2(\omega)$ on frequency is mainly determined by the weighted squared displacement spectra $\overline{Z_{\alpha}}^2\left|\mathbold{\xi}_{\alpha}(\omega)\right|^2$ and VDOS $\rho(\omega)$, i.e. $\epsilon_2(\omega)=\rho(\omega)\sum_{\alpha}{C_{\text{IR}\alpha}(\omega)} \sim \rho(\omega)\sum_{\alpha} \overline{Z_{\alpha}}^2 \left|\mathbold{\xi}_{\alpha}(\omega)\right|^2$. 
Equivalently, we can recast $\epsilon_2(\omega)$ as a superposition of weighted partial VDOS spectra, i.e. $\epsilon_2(\omega) \sim \sum_{\alpha} \overline{Z_{\alpha}}^2\rho_{\alpha}(\omega)$, where $\overline{Z_{\alpha}}^2$ are average Born effective charges corresponding to $\alpha$ species, and where the O contribution can be further decomposed into \textit{in-plane} and \textit{out-of-plane} contributions. The next subsection is devoted to a detailed discussion of the Born effective charges.

%
%
\subsection{Born charge tensors and parametric model}\label{Sec_BornC}

We here analyse the Born charges and discuss a parametric model that allows for the straightforward calculation of IR spectra provided that the atomic configuration, the vibrational modes and their frequencies are available.

The overwhelming majority of O atoms in am-\ce{Al2O3} are threefold coordinated with Al atoms (see Table~\ref{Tab_Coord}), while the O atom lies in the plane of its Al neighbors (see Sec.~\ref{Sec_StructAnal}).
Then, to analyze their Born charge tensors $Z^{*}$, it is convenient to introduce the following local reference frame: the first basis vector is chosen along the bisector of an Al-Al-Al angle ($x$ direction) of the triangle with the Al neighbors at the vertices; the second basis vector is taken along the direction perpendicular to the plane defined by the three nearest neighbors Al atoms ($y$ direction); the last one can be found by the cross product of the first two basis vectors ($z$ direction). 
Thus, \ce{OAl3} units have an approximate \emph{trigonal symmetry} (C$_{3v}$), with the symmetry axis lying in the $y$ direction, according to our definition hereabove.
In this local reference system, we calculate $\overline{Z^*}$ for the threefold coordinated O atom (hereafter $^{[3]}{\rm O}$) as an average over the three possible choices of the basis and over the three models:
\begin{equation}\label{Eq_ZO3} 
\overline{Z^*_{^{[3]}\!{\rm O}}}=\begin{pmatrix}
 -2.11 & 0.00 & 0.01 \\
 0.00 & -1.46 & 0.00 \\
 0.00 & -0.01 & -2.07 
\end{pmatrix}.
\end{equation}

The average isotropic charge, i.e. $Z_{\textrm{iso}}^{*}=\frac{1}{3}{\rm Tr}(\overline{Z^*})$, is found to be $\approx-1.89$. For each $^{[3]}{\rm O}$ atom, we perform a decomposition of its Born charge tensor in terms of the representations of the spatial rotations~\cite{LG2009}:
\begin{equation}\label{Eq_Zdec}
  Z^*_{^{[3]}\!{\rm O}} = Z_{\ell=0}^{*} + Z_{\ell=1}^{*} + Z_{\ell=2}^{*} \,,
\end{equation}
where the first term $Z_{\ell=0}^{*}$ is the isotropic term,
\begin{equation}
 Z_{\ell=0}^{*} = Z_{\textrm{iso}}^{*}\delta_{ij} \,,
\end{equation}
and the two traceless terms $Z_{\ell=1}^{*}$ and $Z_{\ell=2}^{*}$ are given by:
%
\begin{align}
Z_{\ell=1}^{*} &= \frac{1}{2}\left(Z^*_{^{[3]}\!{\rm O}} - {Z^{*}_{^{[3]}\!{\rm O}}}^T\right)\,, \\
%
%
Z_{\ell=2}^{*} &= \frac{1}{2}\left(Z^*_{^{[3]}\!{\rm O}} + {Z^{*}_{^{[3]}\!{\rm O}}}^T\right) - Z_{\ell=0}^{*} \,,
\end{align}
%
with the superscript $T$ indicating the transpose matrix. We adopt a standard matricial norm (or matrix norm)~\cite{PC97} to quantify the respective weights of each $\ell$ component in Eq.~\eqref{Eq_Zdec}. On average, we find that $Z_{\ell=1}^{*}$ is negligible (weight $\sim$0.03\%), while the traceless symmetric term $Z_{\ell=2}^{*}$ (weight $\sim$3\%) is well described by a diagonal matrix 
\begin{equation}\label{Eq_Zell2}
Z_{\ell=2}^{*}=
\begin{pmatrix}
\lambda & 0 & 0\\
0 & \mu & 0 \\
0 & 0 & -(\lambda+\mu)
\end{pmatrix}.
\end{equation}
By averaging over all $^{[3]}{\rm O}$ atoms in our models and also keeping into account the three possible local bases for each O atom, we obtain the values $\lambda = -0.23$ and $\mu=0.42$ with standard deviations of $\approx$0.1. 
Due to the approximate trigonal symmetry $-(\lambda+\mu) \simeq \lambda$, which leads to the fact that the tensor $Z^*$ can be regarded as isotropic in the $xz$ plane. Hence, we adopt the following Born charge tensor for $^{[3]}{\rm O}$ atoms, defined in their local reference frame:
\begin{equation}\label{Eq_ZO3_model} 
Z^*_{^{[3]}\!{\rm O}}= \begin{pmatrix} 
	-2.1 & 0 & 0 \\
	 0 & -1.5 & 0 \\
	 0 & 0 & -2.1 
\end{pmatrix}.
\end{equation}

For fourfold, fivefold, and sixfold coordinated Al atoms, the isotropic component ($\ell=0$) accounts on average for 98.9\%, 99.2\% and 99.6\% of the norm of the Born charge tensors $Z^*_{\rm Al}$, respectively. For fourfold coordinated O atoms, the average isotropic component accounts for 98.5\% of the Born charge, which is noticeably higher than for $^{[3]}{\rm O}$ atoms (96.6\%). It was found that the average isotropic dynamical charge of fourfold coordinated O atoms $Z_{\rm iso}^{*} = -1.95$, which is close to typical values given in the literature for O atoms in $\alpha$-\ce{Al2O3}~\cite{Heid2000}.
%
%
%
\begin{table}[b]
\caption{\label{Tab_ModelInfra} 
Average isotropic Born charges $\frac{1}{3}{\rm Tr}(\overline{Z^*})$ for Al atoms $(\overline{Z^*_{\rm Al}})$ with four-, five-, and sixfold coordination and for O atoms $(\overline{Z^*_{\rm O}})$ with two-, three- and fourfold coordination. 
The given values correspond to the average values over atoms of all three models. The Born charges of $^{[3]}{\rm O}$ atoms are described by the matrix given by Eq.~\eqref{Eq_ZO3_model}. Standard deviations are given in parentheses.}
\begin{center}
\begin{ruledtabular}
\begin{tabular}{lccc} 
 	Al & 4 & 5 & 6 \\
  \hline
 $\overline{Z^*_{\rm Al}}$ & 2.73 (0.1) & 2.84 (0.08) & 2.93 (0.07) \\  
 O & 2 & 3 & 4 \\
 \hline
 $\overline{Z^*_{\rm O}}$ & $-$1.82 (0.03) & $-$1.89 (0.05)& $-$1.95 (0.03)  \\
\end{tabular}
\end{ruledtabular}
\end{center}
\end{table}
Since the twofold coordinated O atoms are rare~\cite{Dicks2019} in am-\ce{Al2O3} (see Table~\ref{Tab_Coord}), we here just retain, for sake of simplicity, their average isotropic Born charge ($Z_{\rm iso}^{*} = -1.82$).
All average isotropic charges of Al and O atoms
to be used as parameters in the evaluation of IR spectra are given in Table~\ref{Tab_ModelInfra}.

\begin{figure}
\begin{center}
  \includegraphics[width= 0.95\columnwidth]{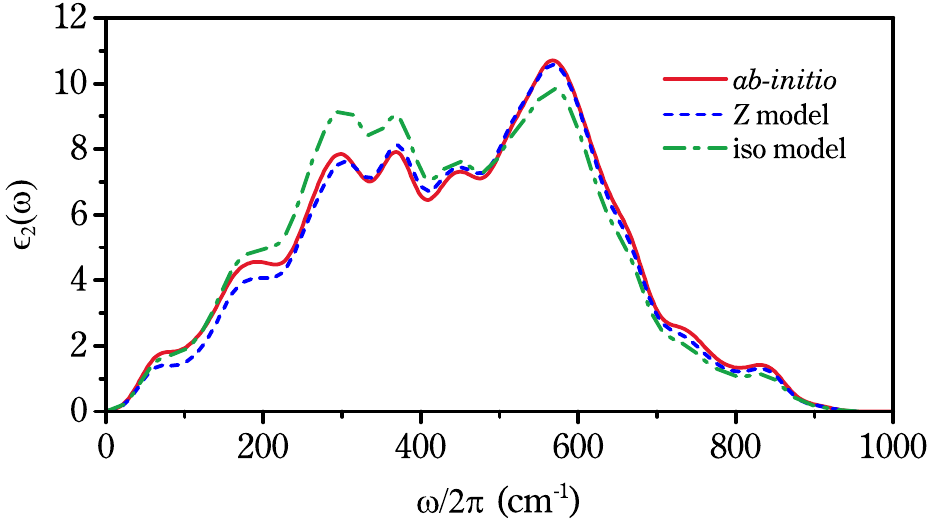} 
  \caption{The dielectric function $\epsilon_2(\omega)$ for model~I calculated by using three alternative sets of Born effective charges: \emph{ab initio} result (red, solid), parametric model (Z model) with values from Table~\ref{Tab_ModelInfra} (blue, dashed), and fully isotropic model (iso model) with 
$\overline{Z^*_{\rm O}}=-1.91$ and $\overline{Z^*_{\rm Al}}=2.85$ (green, dot-dashed).} 
 \label{Fig_eps2_zmodel}
\end{center}
\end{figure}

By taking advantage of Eqs.~\eqref{Eq_Zdec} and \eqref{Eq_Zell2}, and by assuming an isotropic charge for fourfold coordinated O atoms and for Al atoms, it is possible to define a basic \emph{parametric} model for the calculation of the Born charge tensors in am-\ce{Al2O3}. 
In Fig.~\ref{Fig_eps2_zmodel}, we show $\epsilon_2(\omega)$ spectra calculated with the parameters given in Table~\ref{Tab_ModelInfra} compared with the \emph{ab initio} result. For convenience, the spectrum achieved with the parametric coupling is also compared with the result from a calculation with sole average \emph{isotropic} charges, i.e. $\overline{Z^*_{\rm O}}=-1.91$ and $\overline{Z^*_{\rm Al}}=2.85$ for all O and Al atoms, respectively. As we can see, the \emph{isotropic} model is qualitatively correct, which is also implied by Sec.~\ref{Sec_Infra}, but it overestimates the band at $\sim$350~cm$^{-1}$ and underestimates the band at 570~cm $^{-1}$, showing quite large discrepancies from 8 to 15\% relative to the \emph{parametric} model, which accurately reproduces the intensity of all the main features of the \emph{ab initio} spectrum.
The higher accuracy of the \emph{parametric} model compared to the \emph{isotropic} one is due to the fact that the \emph{parametric} model accounts for the difference in coupling charges of $^{[3]}$O atoms for \textit{in-plane} and \textit{out-of-plane} motions (the plane defined by its Al neighbors).
As can be seen in Eq.~\eqref{Eq_ZO3_model}, this difference amounts to $\sim$30\%, with the coupling charge along the \textit{out-of-plane} (normal) direction being smaller than along \textit{in-plane} directions. As a result, in the $\epsilon_2(\omega)$ spectrum, the anisotropic Born charge of $^{[3]}$O atoms [see Eq.~\eqref{Eq_ZO3_model}] leads to a lowering of the band intensity at $\sim$300--370~cm$^{-1}$ with respect to that at 570~cm$^{-1}$ (see Fig.~\ref{Fig_eps2_zmodel}). 
%

\section{Discussion}\label{Sec_Disc}
%
%
IR spectroscopy is a common technique for analyzing am-\ce{Al2O3} films. In particular, absorption or transmission spectra are typically collected in the mid- and far-IR regions. These spectra are underlain by the dielectric function, the knowledge of which is essential for the correct description and interpretation of the features that appear in the experimental spectra.

%
%
Relying on the decomposition shown in Figs.~\ref{Fig_vDOS} and \ref{Fig_e1_e2_rel}, we associate the two main features of the $\epsilon_2(\omega)$ spectrum with the motions of $^{[3]}$O atoms. In particular, the band peaking at $\sim$380~cm$^{-1}$ reflects the occurrence of oxygen motions in the direction normal to the plane of its Al neighbors (\textit{out-of-plane} motions) in the \ce{OAl3} units, while the band peaking at $\sim$600--630~cm$^{-1}$ mainly results from \textit{in-plane} motions of the same O atoms (i.e. from O-Al stretching vibrations).
Hence, the $\epsilon_2(\omega)$ function (or equivalently, the absorption) of am-\ce{Al2O3} shows the presence of two main bands, and not of three bands, as e.g. in other oxides glasses, such as vitreous silica~\cite{Giacomazzi2007}. This is clearly a consequence of the dominant occurrence of \ce{OAl3} subunits in the structure of am-\ce{Al2O3} (see Table~\ref{Tab_Coord}).
We note that the presence of triply coordinated O atoms ($^{[3]}$O) in am-\ce{Al2O3} was recently discovered by using a special NMR-spectroscopy study~\cite{Lee2018}. This study shows that these $^{[3]}$O atoms carry detailed NMR characteristics that are peculiar to am-\ce{Al2O3} and differ substantially from those of $^{[3]}$O atoms in crystalline counterparts, such as $\gamma$-alumina. In accordance with the remarks made in Ref.~\onlinecite{Lee2018}, we here find that in our models the dominant \ce{OAl3} subunits contain one or two fivefold coordinated Al atoms (see Table~\ref{Tab_Coord_O3}), which do not occur in crystalline systems.
%
%
Subunits containing threefold coordinated atoms also critically affect the shape of the $\epsilon_2(\omega)$ spectrum in other materials, such as, e.g., amorphous \ce{Si3N4}, where nitrogen atoms are mostly threefold coordinated with silicon atoms~\cite{LG2009b}.
As in am-\ce{Al2O3}, the $\epsilon_2(\omega)$ spectrum in amorphous \ce{Si3N4} exhibits two bands, located at $\sim$480~cm$^{-1}$ and $\sim$820~cm$^{-1}$~\cite{LG2009b}. These bands are sharper than those of am-\ce{Al2O3}, which can be explained by the fact that the structure of amorphous \ce{Si3N4} is dominated by silicon-centered tetrahedra, while the structure of am-\ce{Al2O3} is more complex, with several competing types of Al-centered polyhedra. 

%
%
Even if the $\epsilon_2(\omega)$ spectrum in am-\ce{Al2O3} might at first sight resemble that of $\alpha$-\ce{Al2O3}, which consists of several sharp features located in the ranges 380--440~cm$^{-1}$ and 570--635~cm$^{-1}$~\cite{Schubert2000}, the origin of the vibrational modes at $\sim$380~cm$^{-1}$ in am-\ce{Al2O3} is essentially different. 
In $\alpha$-\ce{Al2O3} all the O atoms are fourfold coordinated, while the vast majority of O atoms in am-\ce{Al2O3} are threefold coordinated, which leads to the appearance of a broad band peaking at $\sim$380~cm$^{-1}$ in the $\epsilon_2(\omega)$ spectrum and featuring oxygen motion in the direction normal to the plane of three Al neighbors.
The modes at $\sim$380~cm$^{-1}$ in am-\ce{Al2O3} inevitably differ from the modes arising upon bending modes of single \ce{AlO6} octahedra in $\alpha$-\ce{Al2O3}~\cite{Boumaza2009}, where there are no $^{[3]}$O  atoms.
In fact, all bending modes of the \ce{AlO4}, \ce{AlO5}, and \ce{AlO6} polyhedra present in am-\ce{Al2O3} should be considered, but a detailed analysis based on all vibrations of polyhedra is beyond the scope of this work.
Nevertheless, the presence of structural motifs characteristic of the $\alpha$-\ce{Al2O3} phase, i.e. edge-sharing polyhedra and sixfold Al polyhedra, in particular, can accidentally lead to a spatially local analogy between the vibrational modes of crystalline phases 
and the vibrational modes observed in the amorphous phase.

Among the transition alumina phases~\cite{Liz2011}, $\theta$-alumina features the largest fraction (50\%) of \ce{AlO4} tetrahedra and \ce{AlO6} octahedra and, rather similarly to the amorphous phase, it also features a remarkable fraction (66\%) of threefold coordinated O atoms (\ce{OAl3} units in $\theta$-alumina are of the O$^{[4]}\!$Al$^{[6]}\!$Al$_2$ and O$^{[4]}\!$Al$_2^{[6]}\!$Al kinds showing a degree of planarity of 0.96 and 0.99, respectively) and a 
rather similar density ($\sim$3.6 g/cm$^3$ \cite{Kao2000}). 
IR-spectroscopy studies of transition alumina phases assign a doublet with features at $\sim$330 and 370~cm$^{-1}$ in the absorbance spectrum to the occurrence of the $\theta$-alumina phase~\cite{Boumaza2009,Dorsey1968,Tarte1967}. 
Considering that $\theta$-alumina and am-\ce{Al2O3} have some structural similarity, as noted above, one can conclude that vibrational modes around $\sim$330~cm$^{-1}$ in $\theta$-alumina~\cite{Boumaza2009} should have similar characteristics.
In fact, in the present work we find (Figs.~S3 and S4 in the Supplementary Material \cite{SupplInfo}) that in the frequency range 280--340~cm$^{-1}$ the vibrational modes of the $\theta$-alumina show significant \textit{out-of-plane} motions of threefold coordinated O atoms. In our IR calculations (PBE-DFT), the highest frequency peak of the peculiar doublet is underestimated by about 40~cm$^{-1}$ (i.e. 11\%)~\cite{Boumaza2009,Demich2019}. 
We thus infer that the peak position ($\sim$320~cm$^{-1}$) related to \textit{out-of-plane} O motions, as found in our calculations (see Figs.~\ref{Fig_ImEps_Models} and \ref{Fig_e1_e2_rel}), may suffer from a redshift of $\sim$40~cm$^{-1}$  with respect to the experimental frequency. This further supports the assignment of the 370~cm$^{-1}$ band in $\epsilon_2(\omega)$ to \textit{out-of-plane} motions of $^{[3]}$O atoms.

%
%
A comparison between the various IR spectra of the models discussed here does not allow one to estimate the concentration of fourfold coordinated O atoms. However, based on Fig.~\ref{Fig_ImEps_Models}, several considerations can be made. 
First, the \ce{Al-O} bond-length distribution of model~I is shifted towards longer bonds compared to other models. Since the distribution reflects the variety of structural motifs present in model~I, the shift can be explained by an excessively frequent occurrence of fivefold and sixfold coordinated Al atoms, as well as of fourfold coordinated O atoms, which leads to an increase in the average \ce{Al-O} bond-length (see Sec.~\ref{Sec_StructAnal}). 
Second, the trend (see Fig.~\ref{Fig_AlO_Stretch}) that correlates the \ce{Al-O} bond lengths and  stretching frequencies, together with the distributions of \ce{Al-O} bond-lengths (see Fig.~\ref{Fig_bond_voids}), implies that, in the $\epsilon_2(\omega)$ spectrum of model~I, the bond stretching bands will be redshifted compared with those of models~II and III, thus explaining the lower frequency of the main peak (570~cm$^{-1}$) and rapid decay of $\epsilon_2(\omega)$ spectrum above $\sim$650~cm$^{-1}$. 

%
%
On the other hand, the \textit{out-of-plane} oxygen band peaking at $\sim$320~cm$^{-1}$ in model~II is too sharp [see Fig.~\ref{Fig_ImEps_Models}(b)], which probably indicates an excess of \ce{OAl3} units in the model. 
Among the three models, the model that gives the values of $\epsilon_2(\omega)$ closest to the experiment in terms of the positions of the bands and the overall shape of the spectrum is model~III.
In terms of the global shape of the spectrum, however, we achieve the best description of $\epsilon_1(\omega)$, $\epsilon_2(\omega)$ and $-\mathrm{Im}[1/\epsilon(\omega)]$ by averaging over all our models. This indicates that the set of models used in the present work is sufficiently representative of am-\ce{Al2O3}, although none of the adopted models provide such a good agreement individually. 
On the whole, as regards the intensities of the bands and the positions of the peaks, we have a better agreement between the calculated average IR spectra and those measured in this work compared to Ref.~\cite{Erik1981}.

The averaged $\epsilon_1(\omega)$ spectrum shows a very strained and stretched S-like shape, reminiscent of the one in silica~\cite{luigi2009}. From the high-frequency limit of the calculated $\epsilon_1(\omega)$ we derive a refractive index $n=\sqrt{\epsilon_\infty}$ of $\sim$1.7, which is consistent with the upper values pertaining to very compact am-\ce{Al2O3} films~\cite{Lee1995}. In the low-frequency limit, we obtain for $\epsilon_0$ a value of $\sim$10--11, which belongs to the upper side of the typical experimental range $\sim$8--11 (see Table~\ref{Tab_DielConst}). 

While the formation of am-\ce{Al2O3} is thermodynamically quite unfavourable, it can be stabilized by applying interface, surface, or bulk stress. Indeed, producing thin films and nanoparticles was shown to be a successful approach to stabilize am-\ce{Al2O3}~\cite{Mavric19}. In the present study, the electron-beam deposition technique at extremely low deposition rate is used to prepare homogeneous void-free am-\ce{Al2O3} as indicated by Masakazu Tane~\etal~\cite{Tane2011}. We note that other kinds of am-\ce{Al2O3} thin films, e.g. prepared by RF sputtering, are known to have a grainy structure, which consequently introduces a source of scattering that affects the recorded IR spectra, while ALD prepared films contain carbon and chlorine impurities, which are absent in high-quality electron-beam deposited alumina films.
Furthermore, a background subtraction processing is crucial to obtain a correct dielectric function. In our study, a transmission spectrum of a blank Si substrate is subtracted, allowing one to calculate the dielectric function taking advantage of the absence of IR absorption in the substrate. The ability to avoid measuring the reflectance eliminates the need to correctly indicate the intensity of the reflectance, which is usually related to Au (see Ref.~\cite{Erik1981}), and in fact differs from the reflectance at the substrate interface, thereby introducing a systematic error.
However, in the case of thin films, it is particularly challenging to extract the dielectric function through the Kramers-Kronig relations, because of the multiple reflections and inferences arising from light interaction at the film-substrate interfaces and grain boundaries. Additionally, systematic errors in reflectivity measurements can occur by choosing, storing and measuring a suitable reference for the reflectivity measurements~\cite{Bennett65}. In this work we demonstrate that using the iterative procedure developed in Ref.~\cite{Gera2020} is an efficient way for obtaining the dielectric function from a \textit{single} transmission spectrum.
The value of the dielectric function in the mid-IR range nicely follows the experimental data previously reported by Eriksson~\etal~\cite{Erik1981}, while in the far-IR range Eriksson's data shows lower values. The origin of the discrepancy might arise from the processing method used to prepare the thin films~\cite{Lee2014}. 
The intensity in the $\epsilon_2(\omega)$ spectrum around 580~cm$^{-1}$ measured in this work is larger compared with Eriksson's data. In the light of our theoretical analysis, this may be attributed to a larger content of \ce{AlO6} octahedra in the films used in the present work. Also, the films obtained by evaporation are not necessarily stoichiometric~\cite{Erik1981}, which leads to some variability in the proportions of \ce{AlO4}, \ce{AlO5}, and \ce{AlO6} polyhedra.

\section{Conclusions}
\label{Sec_Concl}
In this paper, we investigate the vibrational spectra, IR and VDOS spectra, in am-\ce{Al2O3} by means of first-principles calculations and experimental measurements. We validate our numerical results by carrying out a critical comparison of our calculated vibrational spectra with two sets of experimental data for the dielectric function of am-\ce{Al2O3}, taken from the literature~\cite{Erik1981} and measured in this work. In particular, the comparison between the latter data set and the dielectric function obtained through \textit{ab initio} calculations results in an improved agreement with respect to the data reported by 
Eriksson~\etal~\cite{Erik1981}.

%
We show that the main features of the $\epsilon_2(\omega)$ spectrum peaking at $\sim$360--380~cm~$^{-1}$ and $\sim$600--630~cm$^{-1}$ arise from the motions of threefold coordinated O atoms, which form the vast majority of O atoms in am-\ce{Al2O3}. 
By using an atomic decomposition analysis, we show that the oxygen motions at $\sim$370~cm$^{-1}$ occur along a direction normal to the plane defined by the three Al nearest neighbors, i.e. \textit{out-of-plane} O motions. 
At frequencies above $\sim$500~cm$^{-1}$, Al-O stretching vibrations, which for threefold coordinated O atoms correspond to \textit{in-plane} motions, occur and underlie the broad band peaking at $\sim$600--630~cm$^{-1}$. 
Al atoms and fourfold coordinated O atoms give a fairly uniform contribution to the VDOS and $\epsilon_2(\omega)$ spectra in the frequency range of 300--700~cm$^{-1}$
without manifesting specific features.
Essentially, by conducting this comprehensive analysis 
we disclaim earlier hypotheses that relate the observed vibrational modes from $\sim$200 to $\sim$400 cm$^{-1}$ in am-\ce{Al2O3} to stretching vibrational modes in \ce{AlO}$_n$ ($n=4$, 5 and 6)
polyhedra.

Finally, we demonstrate that the IR spectrum can be successfully modeled by assuming isotropic Born charges for Al atoms and fourfold coordinated O atoms, while in the case of threefold coordinated O atoms, the anisotropy of the Born charges can well be captured through the use of only two extra parameters in the local reference frame. 
The scope of the proposed Born charges model is quite wide, which allows modeling the spectra of much larger models compared to \emph{ab initio} ones, since it allows for a  direct calculation of IR spectra requiring only the atomic configuration, vibrational modes, and frequencies, e.g. obtained by using \emph{classical} MD simulations.
\newline

\begin{acknowledgements}
L.G. and A.M. acknowledge financing of the Slovenian Research Agency (ARRS) through the Research Core Funding No. P2-0412. A.M. also acknowledges ARRS support 
through Project No. J2-2498, and L.G. acknowledges support through the CEA-ARRS Project No. 0018.
Y.L. acknowledges the financial support from the National Natural Science Foundation of China (22279013, 21872019).
N.S.S. acknowledges support from the French government (ANR) through the national program Investments for the Future, Grant No. ANR-11-LABX022-01 (LabEx MMCD project). We acknowledge Prof. H. Momida for providing us with models of Ref.~\onlinecite{Momi2006}. 
We acknowledge EPFL for providing access to hpc facilities at EPFL (SCITAS) and I. Jerman from National Institute of Chemistry (Slovenia) for recording IR spectra. L.G. acknowledges Dr. Nicolas Salles (CNR-IOM) for providing a $\theta$-alumina supercell, J. Kurtovi\'{c} for test calculations and Prof. S. de Gironcoli (SISSA) for computational resources at CINECA and useful discussions.
\end{acknowledgements}

\vspace{1.5cm}
\bibliography{ms_alumina}

\begin{thebibliography}{83}%
\makeatletter
\providecommand \@ifxundefined [1]{%
 \@ifx{#1\undefined}
}%
\providecommand \@ifnum [1]{%
 \ifnum #1\expandafter \@firstoftwo
 \else \expandafter \@secondoftwo
 \fi
}%
\providecommand \@ifx [1]{%
 \ifx #1\expandafter \@firstoftwo
 \else \expandafter \@secondoftwo
 \fi
}%
\providecommand \natexlab [1]{#1}%
\providecommand \enquote  [1]{``#1''}%
\providecommand \bibnamefont  [1]{#1}%
\providecommand \bibfnamefont [1]{#1}%
\providecommand \citenamefont [1]{#1}%
\providecommand \href@noop [0]{\@secondoftwo}%
\providecommand \href [0]{\begingroup \@sanitize@url \@href}%
\providecommand \@href[1]{\@@startlink{#1}\@@href}%
\providecommand \@@href[1]{\endgroup#1\@@endlink}%
\providecommand \@sanitize@url [0]{\catcode `\\12\catcode `\$12\catcode
  `\&12\catcode `\#12\catcode `\^12\catcode `\_12\catcode `\%12\relax}%
\providecommand \@@startlink[1]{}%
\providecommand \@@endlink[0]{}%
\providecommand \url  [0]{\begingroup\@sanitize@url \@url }%
\providecommand \@url [1]{\endgroup\@href {#1}{\urlprefix }}%
\providecommand \urlprefix  [0]{URL }%
\providecommand \Eprint [0]{\href }%
\providecommand \doibase [0]{https://doi.org/}%
\providecommand \selectlanguage [0]{\@gobble}%
\providecommand \bibinfo  [0]{\@secondoftwo}%
\providecommand \bibfield  [0]{\@secondoftwo}%
\providecommand \translation [1]{[#1]}%
\providecommand \BibitemOpen [0]{}%
\providecommand \bibitemStop [0]{}%
\providecommand \bibitemNoStop [0]{.\EOS\space}%
\providecommand \EOS [0]{\spacefactor3000\relax}%
\providecommand \BibitemShut  [1]{\csname bibitem#1\endcsname}%
\let\auto@bib@innerbib\@empty
\bibitem [{\citenamefont {Hung}\ \emph {et~al.}(2022)\citenamefont {Hung},
  \citenamefont {Yu}, \citenamefont {Foroozani}, \citenamefont {Fritz},
  \citenamefont {Gerthsen},\ and\ \citenamefont {Osborn}}]{Hung2022}%
  \BibitemOpen
  \bibfield  {author} {\bibinfo {author} {\bibfnamefont {C.-C.}\ \bibnamefont
  {Hung}}, \bibinfo {author} {\bibfnamefont {L.}~\bibnamefont {Yu}}, \bibinfo
  {author} {\bibfnamefont {N.}~\bibnamefont {Foroozani}}, \bibinfo {author}
  {\bibfnamefont {S.}~\bibnamefont {Fritz}}, \bibinfo {author} {\bibfnamefont
  {D.}~\bibnamefont {Gerthsen}},\ and\ \bibinfo {author} {\bibfnamefont
  {K.~D.}\ \bibnamefont {Osborn}},\ }\bibfield  {title} {\bibinfo {title}
  {Probing hundreds of individual quantum defects in polycrystalline and
  amorphous alumina},\ }\href
  {https://doi.org/10.1103/PhysRevApplied.17.034025} {\bibfield  {journal}
  {\bibinfo  {journal} {Phys. Rev. Applied}\ }\textbf {\bibinfo {volume}
  {17}},\ \bibinfo {pages} {034025} (\bibinfo {year} {2022})}\BibitemShut
  {NoStop}%
\bibitem [{\citenamefont {Scott}\ \emph {et~al.}(2018)\citenamefont {Scott},
  \citenamefont {Gaskins}, \citenamefont {King},\ and\ \citenamefont
  {Hopkins}}]{Scott2018}%
  \BibitemOpen
  \bibfield  {author} {\bibinfo {author} {\bibfnamefont {E.~A.}\ \bibnamefont
  {Scott}}, \bibinfo {author} {\bibfnamefont {J.~T.}\ \bibnamefont {Gaskins}},
  \bibinfo {author} {\bibfnamefont {S.~W.}\ \bibnamefont {King}},\ and\
  \bibinfo {author} {\bibfnamefont {P.~E.}\ \bibnamefont {Hopkins}},\
  }\bibfield  {title} {\bibinfo {title} {Thermal conductivity and thermal
  boundary resistance of atomic layer deposited high-k dielectric aluminum
  oxide, hafnium oxide, and titanium oxide thin films on silicon},\ }\href
  {https://doi.org/10.1063/1.5021044} {\bibfield  {journal} {\bibinfo
  {journal} {APL Materials}\ }\textbf {\bibinfo {volume} {6}},\ \bibinfo
  {pages} {058302} (\bibinfo {year} {2018})}\BibitemShut {NoStop}%
\bibitem [{\citenamefont {Reuna}\ \emph {et~al.}(2021)\citenamefont {Reuna},
  \citenamefont {Aho}, \citenamefont {Isoaho}, \citenamefont {Raappana},
  \citenamefont {Aho}, \citenamefont {Anttola}, \citenamefont {Hietalahti},
  \citenamefont {Tukiainen},\ and\ \citenamefont {Guina}}]{Reuna2021}%
  \BibitemOpen
  \bibfield  {author} {\bibinfo {author} {\bibfnamefont {J.}~\bibnamefont
  {Reuna}}, \bibinfo {author} {\bibfnamefont {A.}~\bibnamefont {Aho}}, \bibinfo
  {author} {\bibfnamefont {R.}~\bibnamefont {Isoaho}}, \bibinfo {author}
  {\bibfnamefont {M.}~\bibnamefont {Raappana}}, \bibinfo {author}
  {\bibfnamefont {T.}~\bibnamefont {Aho}}, \bibinfo {author} {\bibfnamefont
  {E.}~\bibnamefont {Anttola}}, \bibinfo {author} {\bibfnamefont
  {A.}~\bibnamefont {Hietalahti}}, \bibinfo {author} {\bibfnamefont
  {A.}~\bibnamefont {Tukiainen}},\ and\ \bibinfo {author} {\bibfnamefont
  {M.}~\bibnamefont {Guina}},\ }\bibfield  {title} {\bibinfo {title} {Use of
  nanostructured alumina thin films in multilayer anti-reflective coatings},\
  }\href {https://doi.org/10.1088/1361-6528/abe747} {\bibfield  {journal}
  {\bibinfo  {journal} {Nanotechnology}\ }\textbf {\bibinfo {volume} {32}},\
  \bibinfo {pages} {215602} (\bibinfo {year} {2021})}\BibitemShut {NoStop}%
\bibitem [{\citenamefont {Garc{\'\i}a~Ferr{\'e}}\ \emph
  {et~al.}(2016)\citenamefont {Garc{\'\i}a~Ferr{\'e}}, \citenamefont {Mairov},
  \citenamefont {Ceseracciu}, \citenamefont {Serruys}, \citenamefont
  {Trocellier}, \citenamefont {Baumier}, \citenamefont {Ka{\"\i}tasov},
  \citenamefont {Brescia}, \citenamefont {Gastaldi}, \citenamefont {Vena},
  \citenamefont {Beghi}, \citenamefont {Beck}, \citenamefont {Sridharan},\ and\
  \citenamefont {{Di Fonzo}}}]{Garcia2016}%
  \BibitemOpen
  \bibfield  {author} {\bibinfo {author} {\bibfnamefont {F.}~\bibnamefont
  {Garc{\'\i}a~Ferr{\'e}}}, \bibinfo {author} {\bibfnamefont {A.}~\bibnamefont
  {Mairov}}, \bibinfo {author} {\bibfnamefont {L.}~\bibnamefont {Ceseracciu}},
  \bibinfo {author} {\bibfnamefont {Y.}~\bibnamefont {Serruys}}, \bibinfo
  {author} {\bibfnamefont {P.}~\bibnamefont {Trocellier}}, \bibinfo {author}
  {\bibfnamefont {C.}~\bibnamefont {Baumier}}, \bibinfo {author} {\bibfnamefont
  {O.}~\bibnamefont {Ka{\"\i}tasov}}, \bibinfo {author} {\bibfnamefont
  {R.}~\bibnamefont {Brescia}}, \bibinfo {author} {\bibfnamefont
  {D.}~\bibnamefont {Gastaldi}}, \bibinfo {author} {\bibfnamefont
  {P.}~\bibnamefont {Vena}}, \bibinfo {author} {\bibfnamefont {M.~G.}\
  \bibnamefont {Beghi}}, \bibinfo {author} {\bibfnamefont {L.}~\bibnamefont
  {Beck}}, \bibinfo {author} {\bibfnamefont {K.}~\bibnamefont {Sridharan}},\
  and\ \bibinfo {author} {\bibfnamefont {F.}~\bibnamefont {{Di Fonzo}}},\
  }\bibfield  {title} {\bibinfo {title} {Radiation endurance in {Al$_2$O$_3$}
  nanoceramics},\ }\href {https://doi.org/10.1038/srep33478} {\bibfield
  {journal} {\bibinfo  {journal} {Scientific Reports}\ }\textbf {\bibinfo
  {volume} {6}},\ \bibinfo {pages} {1} (\bibinfo {year} {2016})}\BibitemShut
  {NoStop}%
\bibitem [{\citenamefont {Robertson}\ and\ \citenamefont
  {Falabretti}(2006)}]{Robertson2006}%
  \BibitemOpen
  \bibfield  {author} {\bibinfo {author} {\bibfnamefont {J.}~\bibnamefont
  {Robertson}}\ and\ \bibinfo {author} {\bibfnamefont {B.}~\bibnamefont
  {Falabretti}},\ }\bibfield  {title} {\bibinfo {title} {Band offsets of high
  {$K$} gate oxides on {III-V} semiconductors},\ }\href
  {https://doi.org/doi.org/10.1063/1.2213170} {\bibfield  {journal} {\bibinfo
  {journal} {Journal of Applied Physics}\ }\textbf {\bibinfo {volume} {100}},\
  \bibinfo {pages} {014111} (\bibinfo {year} {2006})}\BibitemShut {NoStop}%
\bibitem [{\citenamefont {Zhao}\ \emph {et~al.}(2022)\citenamefont {Zhao},
  \citenamefont {Xiao}, \citenamefont {Chen}, \citenamefont {Wang},
  \citenamefont {Liang}, \citenamefont {Liu}, \citenamefont {Hung},
  \citenamefont {Gan},\ and\ \citenamefont {Hou}}]{Zhao2022}%
  \BibitemOpen
  \bibfield  {author} {\bibinfo {author} {\bibfnamefont {Z.}~\bibnamefont
  {Zhao}}, \bibinfo {author} {\bibfnamefont {D.}~\bibnamefont {Xiao}}, \bibinfo
  {author} {\bibfnamefont {K.}~\bibnamefont {Chen}}, \bibinfo {author}
  {\bibfnamefont {R.}~\bibnamefont {Wang}}, \bibinfo {author} {\bibfnamefont
  {L.}~\bibnamefont {Liang}}, \bibinfo {author} {\bibfnamefont
  {Z.}~\bibnamefont {Liu}}, \bibinfo {author} {\bibfnamefont {I.}~\bibnamefont
  {Hung}}, \bibinfo {author} {\bibfnamefont {Z.}~\bibnamefont {Gan}},\ and\
  \bibinfo {author} {\bibfnamefont {G.}~\bibnamefont {Hou}},\ }\bibfield
  {title} {\bibinfo {title} {Nature of five-coordinated {Al} in
  $\gamma$-\ce{Al2O3} revealed by ultra-high-field solid-state {NMR}},\ }\href
  {https://doi.org/10.1021/acscentsci.1c01497} {\bibfield  {journal} {\bibinfo
  {journal} {ACS Central Science}\ }\textbf {\bibinfo {volume} {8}},\ \bibinfo
  {pages} {795} (\bibinfo {year} {2022})}\BibitemShut {NoStop}%
\bibitem [{\citenamefont {Mavri{\v{c}}}\ \emph {et~al.}(2019)\citenamefont
  {Mavri{\v{c}}}, \citenamefont {Valant}, \citenamefont {Cui},\ and\
  \citenamefont {Wang}}]{Mavric19}%
  \BibitemOpen
  \bibfield  {author} {\bibinfo {author} {\bibfnamefont {A.}~\bibnamefont
  {Mavri{\v{c}}}}, \bibinfo {author} {\bibfnamefont {M.}~\bibnamefont
  {Valant}}, \bibinfo {author} {\bibfnamefont {C.}~\bibnamefont {Cui}},\ and\
  \bibinfo {author} {\bibfnamefont {Z.~M.}\ \bibnamefont {Wang}},\ }\bibfield
  {title} {\bibinfo {title} {Advanced applications of amorphous alumina: From
  nano to bulk},\ }\href {https://doi.org/10.1016/j.jnoncrysol.2019.119493}
  {\bibfield  {journal} {\bibinfo  {journal} {Journal of Non-Crystalline
  Solids}\ }\textbf {\bibinfo {volume} {521}},\ \bibinfo {pages} {119493}
  (\bibinfo {year} {2019})}\BibitemShut {NoStop}%
\bibitem [{\citenamefont {Fukuhara}\ \emph {et~al.}(2018)\citenamefont
  {Fukuhara}, \citenamefont {Kuroda}, \citenamefont {Hasegawa}, \citenamefont
  {Hashida}, \citenamefont {Kwon},\ and\ \citenamefont {Konno}}]{Fuku2018}%
  \BibitemOpen
  \bibfield  {author} {\bibinfo {author} {\bibfnamefont {M.}~\bibnamefont
  {Fukuhara}}, \bibinfo {author} {\bibfnamefont {T.}~\bibnamefont {Kuroda}},
  \bibinfo {author} {\bibfnamefont {F.}~\bibnamefont {Hasegawa}}, \bibinfo
  {author} {\bibfnamefont {T.}~\bibnamefont {Hashida}}, \bibinfo {author}
  {\bibfnamefont {E.}~\bibnamefont {Kwon}},\ and\ \bibinfo {author}
  {\bibfnamefont {K.}~\bibnamefont {Konno}},\ }\bibfield  {title} {\bibinfo
  {title} {Amorphous aluminum-oxide supercapacitors},\ }\href
  {https://doi.org/10.1209/0295-5075/123/58004} {\bibfield  {journal} {\bibinfo
   {journal} {EPL (Europhysics Letters)}\ }\textbf {\bibinfo {volume} {123}},\
  \bibinfo {pages} {58004} (\bibinfo {year} {2018})}\BibitemShut {NoStop}%
\bibitem [{\citenamefont {Zaborowska}\ \emph {et~al.}(2021)\citenamefont
  {Zaborowska}, \citenamefont {Clozel}, \citenamefont {Olivier}, \citenamefont
  {O'Connell}, \citenamefont {Vanazzi}, \citenamefont {Di~Fonzo}, \citenamefont
  {Azarov}, \citenamefont {J{\'o}{\'z}wik}, \citenamefont {Frelek-Kozak},
  \citenamefont {Diduszko}, \citenamefont {Neethling},\ and\ \citenamefont
  {Jagielski}}]{Zabo2021}%
  \BibitemOpen
  \bibfield  {author} {\bibinfo {author} {\bibfnamefont {A.}~\bibnamefont
  {Zaborowska}}, \bibinfo {author} {\bibfnamefont {M.}~\bibnamefont {Clozel}},
  \bibinfo {author} {\bibfnamefont {E.}~\bibnamefont {Olivier}}, \bibinfo
  {author} {\bibfnamefont {J.}~\bibnamefont {O'Connell}}, \bibinfo {author}
  {\bibfnamefont {M.}~\bibnamefont {Vanazzi}}, \bibinfo {author} {\bibfnamefont
  {F.}~\bibnamefont {Di~Fonzo}}, \bibinfo {author} {\bibfnamefont
  {A.}~\bibnamefont {Azarov}}, \bibinfo {author} {\bibfnamefont
  {I.}~\bibnamefont {J{\'o}{\'z}wik}}, \bibinfo {author} {\bibfnamefont
  {M.}~\bibnamefont {Frelek-Kozak}}, \bibinfo {author} {\bibfnamefont
  {R.}~\bibnamefont {Diduszko}}, \bibinfo {author} {\bibfnamefont {J.~H.}\
  \bibnamefont {Neethling}},\ and\ \bibinfo {author} {\bibfnamefont
  {J.}~\bibnamefont {Jagielski}},\ }\bibfield  {title} {\bibinfo {title}
  {Absolute radiation tolerance of amorphous alumina coatings at room
  temperature},\ }\href {https://doi.org/0.1016/j.ceramint.2021.09.013}
  {\bibfield  {journal} {\bibinfo  {journal} {Ceramics International}\ }\textbf
  {\bibinfo {volume} {47}},\ \bibinfo {pages} {34740} (\bibinfo {year}
  {2021})}\BibitemShut {NoStop}%
\bibitem [{\citenamefont {Valant}\ \emph {et~al.}(2016)\citenamefont {Valant},
  \citenamefont {Luin}, \citenamefont {Fanetti}, \citenamefont {Mavri{\v{c}}},
  \citenamefont {Vyshniakova}, \citenamefont {Siketi{\'c}},\ and\ \citenamefont
  {Kalin}}]{valant2016}%
  \BibitemOpen
  \bibfield  {author} {\bibinfo {author} {\bibfnamefont {M.}~\bibnamefont
  {Valant}}, \bibinfo {author} {\bibfnamefont {U.}~\bibnamefont {Luin}},
  \bibinfo {author} {\bibfnamefont {M.}~\bibnamefont {Fanetti}}, \bibinfo
  {author} {\bibfnamefont {A.}~\bibnamefont {Mavri{\v{c}}}}, \bibinfo {author}
  {\bibfnamefont {K.}~\bibnamefont {Vyshniakova}}, \bibinfo {author}
  {\bibfnamefont {Z.}~\bibnamefont {Siketi{\'c}}},\ and\ \bibinfo {author}
  {\bibfnamefont {M.}~\bibnamefont {Kalin}},\ }\bibfield  {title} {\bibinfo
  {title} {Fully transparent nanocomposite coating with an amorphous alumina
  matrix and exceptional wear and scratch resistance},\ }\href
  {https://doi.org/10.1002/adfm.201600213} {\bibfield  {journal} {\bibinfo
  {journal} {Advanced Functional Materials}\ }\textbf {\bibinfo {volume}
  {26}},\ \bibinfo {pages} {4362} (\bibinfo {year} {2016})}\BibitemShut
  {NoStop}%
\bibitem [{\citenamefont {Harper}\ \emph {et~al.}(2023)\citenamefont {Harper},
  \citenamefont {Iwanowski}, \citenamefont {Payne},\ and\ \citenamefont
  {Simoncelli}}]{harper2023}%
  \BibitemOpen
  \bibfield  {author} {\bibinfo {author} {\bibfnamefont {A.~F.}\ \bibnamefont
  {Harper}}, \bibinfo {author} {\bibfnamefont {K.}~\bibnamefont {Iwanowski}},
  \bibinfo {author} {\bibfnamefont {M.~C.}\ \bibnamefont {Payne}},\ and\
  \bibinfo {author} {\bibfnamefont {M.}~\bibnamefont {Simoncelli}},\
  }\href@noop {} {\bibinfo {title} {Vibrational and thermal properties of
  amorphous alumina from first principles}} (\bibinfo {year} {2023}),\ \Eprint
  {https://arxiv.org/abs/2303.08637} {arXiv:2303.08637 [cond-mat.mtrl-sci]}
  \BibitemShut {NoStop}%
\bibitem [{\citenamefont {Lee}\ \emph {et~al.}(2010)\citenamefont {Lee},
  \citenamefont {Park}, \citenamefont {Yi},\ and\ \citenamefont
  {Moon}}]{Lee2010}%
  \BibitemOpen
  \bibfield  {author} {\bibinfo {author} {\bibfnamefont {S.~K.}\ \bibnamefont
  {Lee}}, \bibinfo {author} {\bibfnamefont {S.~Y.}\ \bibnamefont {Park}},
  \bibinfo {author} {\bibfnamefont {Y.~S.}\ \bibnamefont {Yi}},\ and\ \bibinfo
  {author} {\bibfnamefont {J.}~\bibnamefont {Moon}},\ }\bibfield  {title}
  {\bibinfo {title} {Structure and disorder in amorphous alumina thin films:
  {I}nsights from high-resolution solid-state {NMR}},\ }\href
  {https://doi.org/10.1021/jp105306r} {\bibfield  {journal} {\bibinfo
  {journal} {The Journal of Physical Chemistry C}\ }\textbf {\bibinfo {volume}
  {114}},\ \bibinfo {pages} {13890} (\bibinfo {year} {2010})}\BibitemShut
  {NoStop}%
\bibitem [{\citenamefont {Lee}\ \emph {et~al.}(2009)\citenamefont {Lee},
  \citenamefont {Lee}, \citenamefont {Park}, \citenamefont {Yi},\ and\
  \citenamefont {Ahn}}]{Lee2009}%
  \BibitemOpen
  \bibfield  {author} {\bibinfo {author} {\bibfnamefont {S.~K.}\ \bibnamefont
  {Lee}}, \bibinfo {author} {\bibfnamefont {S.~B.}\ \bibnamefont {Lee}},
  \bibinfo {author} {\bibfnamefont {S.~Y.}\ \bibnamefont {Park}}, \bibinfo
  {author} {\bibfnamefont {Y.~S.}\ \bibnamefont {Yi}},\ and\ \bibinfo {author}
  {\bibfnamefont {C.~W.}\ \bibnamefont {Ahn}},\ }\bibfield  {title} {\bibinfo
  {title} {Structure of amorphous aluminum oxide},\ }\href
  {https://doi.org/10.1103/PhysRevLett.103.095501} {\bibfield  {journal}
  {\bibinfo  {journal} {Physical Review Letters}\ }\textbf {\bibinfo {volume}
  {103}},\ \bibinfo {pages} {095501} (\bibinfo {year} {2009})}\BibitemShut
  {NoStop}%
\bibitem [{\citenamefont {Kim}\ \emph {et~al.}(2014)\citenamefont {Kim},
  \citenamefont {Bassiri}, \citenamefont {Fejer},\ and\ \citenamefont
  {Stebbins}}]{Kim2014}%
  \BibitemOpen
  \bibfield  {author} {\bibinfo {author} {\bibfnamefont {N.}~\bibnamefont
  {Kim}}, \bibinfo {author} {\bibfnamefont {R.}~\bibnamefont {Bassiri}},
  \bibinfo {author} {\bibfnamefont {M.~M.}\ \bibnamefont {Fejer}},\ and\
  \bibinfo {author} {\bibfnamefont {J.~F.}\ \bibnamefont {Stebbins}},\
  }\bibfield  {title} {\bibinfo {title} {The structure of ion beam sputtered
  amorphous alumina films and effects of zn doping: High-resolution 27al nmr},\
  }\href {https://doi.org/https://doi.org/10.1016/j.jnoncrysol.2014.08.022}
  {\bibfield  {journal} {\bibinfo  {journal} {Journal of Non-Crystalline
  Solids}\ }\textbf {\bibinfo {volume} {405}},\ \bibinfo {pages} {1} (\bibinfo
  {year} {2014})}\BibitemShut {NoStop}%
\bibitem [{\citenamefont {Liz{\'a}rraga}\ \emph {et~al.}(2011)\citenamefont
  {Liz{\'a}rraga}, \citenamefont {Holmstr{\"o}m}, \citenamefont {Parker},\ and\
  \citenamefont {Arrouvel}}]{Liz2011}%
  \BibitemOpen
  \bibfield  {author} {\bibinfo {author} {\bibfnamefont {R.}~\bibnamefont
  {Liz{\'a}rraga}}, \bibinfo {author} {\bibfnamefont {E.}~\bibnamefont
  {Holmstr{\"o}m}}, \bibinfo {author} {\bibfnamefont {S.~C.}\ \bibnamefont
  {Parker}},\ and\ \bibinfo {author} {\bibfnamefont {C.}~\bibnamefont
  {Arrouvel}},\ }\bibfield  {title} {\bibinfo {title} {Structural
  characterization of amorphous alumina and its polymorphs from
  first-principles {XPS} and {NMR} calculations},\ }\href
  {https://doi.org/10.1103/PhysRevB.83.094201} {\bibfield  {journal} {\bibinfo
  {journal} {Physical Review B}\ }\textbf {\bibinfo {volume} {83}},\ \bibinfo
  {pages} {094201} (\bibinfo {year} {2011})}\BibitemShut {NoStop}%
\bibitem [{\citenamefont {Nakamura}\ \emph {et~al.}(2013)\citenamefont
  {Nakamura}, \citenamefont {Ishimaru}, \citenamefont {Yasuda},\ and\
  \citenamefont {Nakajima}}]{Nakamura2013}%
  \BibitemOpen
  \bibfield  {author} {\bibinfo {author} {\bibfnamefont {R.}~\bibnamefont
  {Nakamura}}, \bibinfo {author} {\bibfnamefont {M.}~\bibnamefont {Ishimaru}},
  \bibinfo {author} {\bibfnamefont {H.}~\bibnamefont {Yasuda}},\ and\ \bibinfo
  {author} {\bibfnamefont {H.}~\bibnamefont {Nakajima}},\ }\bibfield  {title}
  {\bibinfo {title} {Atomic rearrangements in amorphous \ce{Al2O3} under
  electron-beam irradiation},\ }\href {https://doi.org/10.1063/1.4790705}
  {\bibfield  {journal} {\bibinfo  {journal} {Journal of Applied Physics}\
  }\textbf {\bibinfo {volume} {113}},\ \bibinfo {pages} {064312} (\bibinfo
  {year} {2013})}\BibitemShut {NoStop}%
\bibitem [{\citenamefont {Lee}\ and\ \citenamefont {Ryu}(2018)}]{Lee2018}%
  \BibitemOpen
  \bibfield  {author} {\bibinfo {author} {\bibfnamefont {S.~K.}\ \bibnamefont
  {Lee}}\ and\ \bibinfo {author} {\bibfnamefont {S.}~\bibnamefont {Ryu}},\
  }\bibfield  {title} {\bibinfo {title} {Probing of triply coordinated oxygen
  in amorphous \ce{Al2O3}},\ }\href
  {https://doi.org/10.1021/acs.jpclett.7b03027} {\bibfield  {journal} {\bibinfo
   {journal} {The Journal of Physical Chemistry Letters}\ }\textbf {\bibinfo
  {volume} {9}},\ \bibinfo {pages} {150} (\bibinfo {year} {2018})}\BibitemShut
  {NoStop}%
\bibitem [{\citenamefont {Shi}\ \emph {et~al.}(2019)\citenamefont {Shi},
  \citenamefont {Alderman}, \citenamefont {Berman}, \citenamefont {Du},
  \citenamefont {Neuefeind}, \citenamefont {Tamalonis}, \citenamefont {Weber},
  \citenamefont {You},\ and\ \citenamefont {Benmore}}]{Shi2019}%
  \BibitemOpen
  \bibfield  {author} {\bibinfo {author} {\bibfnamefont {C.}~\bibnamefont
  {Shi}}, \bibinfo {author} {\bibfnamefont {O.~L.}\ \bibnamefont {Alderman}},
  \bibinfo {author} {\bibfnamefont {D.}~\bibnamefont {Berman}}, \bibinfo
  {author} {\bibfnamefont {J.}~\bibnamefont {Du}}, \bibinfo {author}
  {\bibfnamefont {J.}~\bibnamefont {Neuefeind}}, \bibinfo {author}
  {\bibfnamefont {A.}~\bibnamefont {Tamalonis}}, \bibinfo {author}
  {\bibfnamefont {J.~R.}\ \bibnamefont {Weber}}, \bibinfo {author}
  {\bibfnamefont {J.}~\bibnamefont {You}},\ and\ \bibinfo {author}
  {\bibfnamefont {C.~J.}\ \bibnamefont {Benmore}},\ }\bibfield  {title}
  {\bibinfo {title} {The structure of amorphous and deeply supercooled liquid
  alumina},\ }\href {https://doi.org/10.3389/fmats.2019.00038} {\bibfield
  {journal} {\bibinfo  {journal} {Frontiers in Materials}\ }\textbf {\bibinfo
  {volume} {6}},\ \bibinfo {pages} {38} (\bibinfo {year} {2019})}\BibitemShut
  {NoStop}%
\bibitem [{\citenamefont {Hashimoto}\ \emph {et~al.}(2022)\citenamefont
  {Hashimoto}, \citenamefont {Onodera}, \citenamefont {Tahara}, \citenamefont
  {Kohara}, \citenamefont {Yazawa}, \citenamefont {Segawa}, \citenamefont
  {Murakami},\ and\ \citenamefont {Ohara}}]{Hash2022}%
  \BibitemOpen
  \bibfield  {author} {\bibinfo {author} {\bibfnamefont {H.}~\bibnamefont
  {Hashimoto}}, \bibinfo {author} {\bibfnamefont {Y.}~\bibnamefont {Onodera}},
  \bibinfo {author} {\bibfnamefont {S.}~\bibnamefont {Tahara}}, \bibinfo
  {author} {\bibfnamefont {S.}~\bibnamefont {Kohara}}, \bibinfo {author}
  {\bibfnamefont {K.}~\bibnamefont {Yazawa}}, \bibinfo {author} {\bibfnamefont
  {H.}~\bibnamefont {Segawa}}, \bibinfo {author} {\bibfnamefont
  {M.}~\bibnamefont {Murakami}},\ and\ \bibinfo {author} {\bibfnamefont
  {K.}~\bibnamefont {Ohara}},\ }\bibfield  {title} {\bibinfo {title} {Structure
  of alumina glass},\ }\href {https://doi.org/10.1038/s41598-021-04455-6}
  {\bibfield  {journal} {\bibinfo  {journal} {Scientific Reports}\ }\textbf
  {\bibinfo {volume} {12}},\ \bibinfo {pages} {1} (\bibinfo {year}
  {2022})}\BibitemShut {NoStop}%
\bibitem [{\citenamefont {Cortie}\ \emph {et~al.}(2020)\citenamefont {Cortie},
  \citenamefont {Cyster}, \citenamefont {Ablott}, \citenamefont {Richardson},
  \citenamefont {Smith}, \citenamefont {Iles}, \citenamefont {Wang},
  \citenamefont {Mitchell}, \citenamefont {Mole}, \citenamefont {de~Souza},
  \citenamefont {Yu},\ and\ \citenamefont {Cole}}]{Cortie2020}%
  \BibitemOpen
  \bibfield  {author} {\bibinfo {author} {\bibfnamefont {D.~L.}\ \bibnamefont
  {Cortie}}, \bibinfo {author} {\bibfnamefont {M.~J.}\ \bibnamefont {Cyster}},
  \bibinfo {author} {\bibfnamefont {T.~A.}\ \bibnamefont {Ablott}}, \bibinfo
  {author} {\bibfnamefont {C.}~\bibnamefont {Richardson}}, \bibinfo {author}
  {\bibfnamefont {J.~S.}\ \bibnamefont {Smith}}, \bibinfo {author}
  {\bibfnamefont {G.~N.}\ \bibnamefont {Iles}}, \bibinfo {author}
  {\bibfnamefont {X.~L.}\ \bibnamefont {Wang}}, \bibinfo {author}
  {\bibfnamefont {D.~R.~G.}\ \bibnamefont {Mitchell}}, \bibinfo {author}
  {\bibfnamefont {R.~A.}\ \bibnamefont {Mole}}, \bibinfo {author}
  {\bibfnamefont {N.~R.}\ \bibnamefont {de~Souza}}, \bibinfo {author}
  {\bibfnamefont {D.~H.}\ \bibnamefont {Yu}},\ and\ \bibinfo {author}
  {\bibfnamefont {J.~H.}\ \bibnamefont {Cole}},\ }\bibfield  {title} {\bibinfo
  {title} {Boson peak in ultrathin alumina layers investigated with neutron
  spectroscopy},\ }\href {https://doi.org/10.1103/PhysRevResearch.2.023320}
  {\bibfield  {journal} {\bibinfo  {journal} {Phys. Rev. Research}\ }\textbf
  {\bibinfo {volume} {2}},\ \bibinfo {pages} {023320} (\bibinfo {year}
  {2020})}\BibitemShut {NoStop}%
\bibitem [{\citenamefont {Eriksson}\ \emph {et~al.}(1981)\citenamefont
  {Eriksson}, \citenamefont {Hjortsberg}, \citenamefont {Niklasson},\ and\
  \citenamefont {Granqvist}}]{Erik1981}%
  \BibitemOpen
  \bibfield  {author} {\bibinfo {author} {\bibfnamefont {T.}~\bibnamefont
  {Eriksson}}, \bibinfo {author} {\bibfnamefont {A.}~\bibnamefont
  {Hjortsberg}}, \bibinfo {author} {\bibfnamefont {G.}~\bibnamefont
  {Niklasson}},\ and\ \bibinfo {author} {\bibfnamefont {C.-G.}\ \bibnamefont
  {Granqvist}},\ }\bibfield  {title} {\bibinfo {title} {Infrared optical
  properties of evaporated alumina films},\ }\href
  {https://doi.org/10.1364/AO.20.002742} {\bibfield  {journal} {\bibinfo
  {journal} {Applied Optics}\ }\textbf {\bibinfo {volume} {20}},\ \bibinfo
  {pages} {2742} (\bibinfo {year} {1981})}\BibitemShut {NoStop}%
\bibitem [{\citenamefont {Ohwaki}\ and\ \citenamefont
  {Onishi}(1999)}]{Ohw1999}%
  \BibitemOpen
  \bibfield  {author} {\bibinfo {author} {\bibfnamefont {T.}~\bibnamefont
  {Ohwaki}}\ and\ \bibinfo {author} {\bibfnamefont {T.}~\bibnamefont
  {Onishi}},\ }\bibfield  {title} {\bibinfo {title} {Infrared absorption at
  longitudinal optical frequency in amorphous oxides and their mixtures},\
  }\href {https://doi.org/10.1143/JJAP.38.L1191} {\bibfield  {journal}
  {\bibinfo  {journal} {Japanese Journal of Applied Physics}\ }\textbf
  {\bibinfo {volume} {38}},\ \bibinfo {pages} {L1191} (\bibinfo {year}
  {1999})}\BibitemShut {NoStop}%
\bibitem [{\citenamefont {Maruyama}\ and\ \citenamefont
  {Arai}(1992)}]{Arai1991}%
  \BibitemOpen
  \bibfield  {author} {\bibinfo {author} {\bibfnamefont {T.}~\bibnamefont
  {Maruyama}}\ and\ \bibinfo {author} {\bibfnamefont {S.}~\bibnamefont
  {Arai}},\ }\bibfield  {title} {\bibinfo {title} {Aluminum oxide thin films
  prepared by chemical vapor deposition from aluminum acetylacetonate},\ }\href
  {https://doi.org/10.1063/1.106699} {\bibfield  {journal} {\bibinfo  {journal}
  {Applied Physics Letters}\ }\textbf {\bibinfo {volume} {60}},\ \bibinfo
  {pages} {322} (\bibinfo {year} {1992})}\BibitemShut {NoStop}%
\bibitem [{\citenamefont {Chu}\ \emph {et~al.}(1988)\citenamefont {Chu},
  \citenamefont {Bates}, \citenamefont {White},\ and\ \citenamefont
  {Farlow}}]{Chu1988}%
  \BibitemOpen
  \bibfield  {author} {\bibinfo {author} {\bibfnamefont {Y.}~\bibnamefont
  {Chu}}, \bibinfo {author} {\bibfnamefont {J.}~\bibnamefont {Bates}}, \bibinfo
  {author} {\bibfnamefont {C.}~\bibnamefont {White}},\ and\ \bibinfo {author}
  {\bibfnamefont {G.}~\bibnamefont {Farlow}},\ }\bibfield  {title} {\bibinfo
  {title} {Optical dielectric functions for amorphous {Al$_2$O$_3$} and
  $\gamma$-{Al$_2$O$_3$}},\ }\href {https://doi.org/10.1063/1.341367}
  {\bibfield  {journal} {\bibinfo  {journal} {Journal of Applied Physics}\
  }\textbf {\bibinfo {volume} {64}},\ \bibinfo {pages} {3727} (\bibinfo {year}
  {1988})}\BibitemShut {NoStop}%
\bibitem [{\citenamefont {Orosco}\ and\ \citenamefont
  {Coimbra}(2018)}]{Orosco2018}%
  \BibitemOpen
  \bibfield  {author} {\bibinfo {author} {\bibfnamefont {J.}~\bibnamefont
  {Orosco}}\ and\ \bibinfo {author} {\bibfnamefont {C.}~\bibnamefont
  {Coimbra}},\ }\bibfield  {title} {\bibinfo {title} {Optical response of thin
  amorphous films to infrared radiation},\ }\href
  {https://doi.org/10.1103/PhysRevB.97.094301} {\bibfield  {journal} {\bibinfo
  {journal} {Physical Review B}\ }\textbf {\bibinfo {volume} {97}},\ \bibinfo
  {pages} {094301} (\bibinfo {year} {2018})}\BibitemShut {NoStop}%
\bibitem [{\citenamefont {Li}\ \emph {et~al.}(2020{\natexlab{a}})\citenamefont
  {Li}, \citenamefont {Wray}, \citenamefont {Su}, \citenamefont {Tu},
  \citenamefont {Andaraarachchi}, \citenamefont {Jeong}, \citenamefont
  {Atwater},\ and\ \citenamefont {Kortshagen}}]{Li2020ACS}%
  \BibitemOpen
  \bibfield  {author} {\bibinfo {author} {\bibfnamefont {Z.}~\bibnamefont
  {Li}}, \bibinfo {author} {\bibfnamefont {P.~R.}\ \bibnamefont {Wray}},
  \bibinfo {author} {\bibfnamefont {M.~P.}\ \bibnamefont {Su}}, \bibinfo
  {author} {\bibfnamefont {Q.}~\bibnamefont {Tu}}, \bibinfo {author}
  {\bibfnamefont {H.~P.}\ \bibnamefont {Andaraarachchi}}, \bibinfo {author}
  {\bibfnamefont {Y.~J.}\ \bibnamefont {Jeong}}, \bibinfo {author}
  {\bibfnamefont {H.~A.}\ \bibnamefont {Atwater}},\ and\ \bibinfo {author}
  {\bibfnamefont {U.~R.}\ \bibnamefont {Kortshagen}},\ }\bibfield  {title}
  {\bibinfo {title} {Aluminum oxide nanoparticle films deposited from a
  nonthermal plasma: {S}ynthesis, characterization, and crystallization},\
  }\href {https://doi.org/10.1021/acsomega.0c03353} {\bibfield  {journal}
  {\bibinfo  {journal} {ACS Omega}\ }\textbf {\bibinfo {volume} {5}},\ \bibinfo
  {pages} {24754} (\bibinfo {year} {2020}{\natexlab{a}})}\BibitemShut {NoStop}%
\bibitem [{\citenamefont {Momida}\ \emph {et~al.}(2006)\citenamefont {Momida},
  \citenamefont {Hamada}, \citenamefont {Takagi}, \citenamefont {Yamamoto},
  \citenamefont {Uda},\ and\ \citenamefont {Ohno}}]{Momi2006}%
  \BibitemOpen
  \bibfield  {author} {\bibinfo {author} {\bibfnamefont {H.}~\bibnamefont
  {Momida}}, \bibinfo {author} {\bibfnamefont {T.}~\bibnamefont {Hamada}},
  \bibinfo {author} {\bibfnamefont {Y.}~\bibnamefont {Takagi}}, \bibinfo
  {author} {\bibfnamefont {T.}~\bibnamefont {Yamamoto}}, \bibinfo {author}
  {\bibfnamefont {T.}~\bibnamefont {Uda}},\ and\ \bibinfo {author}
  {\bibfnamefont {T.}~\bibnamefont {Ohno}},\ }\bibfield  {title} {\bibinfo
  {title} {Theoretical study on dielectric response of amorphous alumina},\
  }\href {https://doi.org/10.1103/PhysRevB.73.054108} {\bibfield  {journal}
  {\bibinfo  {journal} {Physical Review B}\ }\textbf {\bibinfo {volume} {73}},\
  \bibinfo {pages} {054108} (\bibinfo {year} {2006})}\BibitemShut {NoStop}%
\bibitem [{\citenamefont {Vashishta}\ \emph {et~al.}(2008)\citenamefont
  {Vashishta}, \citenamefont {Kalia}, \citenamefont {Nakano},\ and\
  \citenamefont {Rino}}]{Vash2008}%
  \BibitemOpen
  \bibfield  {author} {\bibinfo {author} {\bibfnamefont {P.}~\bibnamefont
  {Vashishta}}, \bibinfo {author} {\bibfnamefont {R.~K.}\ \bibnamefont
  {Kalia}}, \bibinfo {author} {\bibfnamefont {A.}~\bibnamefont {Nakano}},\ and\
  \bibinfo {author} {\bibfnamefont {J.~P.}\ \bibnamefont {Rino}},\ }\bibfield
  {title} {\bibinfo {title} {Interaction potentials for alumina and molecular
  dynamics simulations of amorphous and liquid alumina},\ }\href
  {https://doi.org/10.1063/1.2901171} {\bibfield  {journal} {\bibinfo
  {journal} {Journal of Applied Physics}\ }\textbf {\bibinfo {volume} {103}},\
  \bibinfo {pages} {083504} (\bibinfo {year} {2008})}\BibitemShut {NoStop}%
\bibitem [{\citenamefont {Guti{\'e}rrez}\ \emph {et~al.}(2010)\citenamefont
  {Guti{\'e}rrez}, \citenamefont {Men{\'e}ndez-Proupin}, \citenamefont
  {Loyola}, \citenamefont {Peralta},\ and\ \citenamefont {Davis}}]{Gut2010}%
  \BibitemOpen
  \bibfield  {author} {\bibinfo {author} {\bibfnamefont {G.}~\bibnamefont
  {Guti{\'e}rrez}}, \bibinfo {author} {\bibfnamefont {E.}~\bibnamefont
  {Men{\'e}ndez-Proupin}}, \bibinfo {author} {\bibfnamefont {C.}~\bibnamefont
  {Loyola}}, \bibinfo {author} {\bibfnamefont {J.}~\bibnamefont {Peralta}},\
  and\ \bibinfo {author} {\bibfnamefont {S.}~\bibnamefont {Davis}},\ }\bibfield
   {title} {\bibinfo {title} {Computer simulation study of amorphous compounds:
  {S}tructural and vibrational properties},\ }\href
  {https://doi.org/10.1007/s10853-010-4579-0} {\bibfield  {journal} {\bibinfo
  {journal} {Journal of Materials Science}\ }\textbf {\bibinfo {volume} {45}},\
  \bibinfo {pages} {5124} (\bibinfo {year} {2010})}\BibitemShut {NoStop}%
\bibitem [{\citenamefont {Li}\ \emph {et~al.}(2020{\natexlab{b}})\citenamefont
  {Li}, \citenamefont {Ando},\ and\ \citenamefont {Watanabe}}]{Li2020}%
  \BibitemOpen
  \bibfield  {author} {\bibinfo {author} {\bibfnamefont {W.}~\bibnamefont
  {Li}}, \bibinfo {author} {\bibfnamefont {Y.}~\bibnamefont {Ando}},\ and\
  \bibinfo {author} {\bibfnamefont {S.}~\bibnamefont {Watanabe}},\ }\bibfield
  {title} {\bibinfo {title} {Effects of density and composition on the
  properties of amorphous alumina: {A} high-dimensional neural network
  potential study},\ }\href {https://doi.org/10.1063/5.0026289} {\bibfield
  {journal} {\bibinfo  {journal} {The Journal of Chemical Physics}\ }\textbf
  {\bibinfo {volume} {153}},\ \bibinfo {pages} {164119} (\bibinfo {year}
  {2020}{\natexlab{b}})}\BibitemShut {NoStop}%
\bibitem [{\citenamefont {Begemann}\ \emph {et~al.}(1997)\citenamefont
  {Begemann}, \citenamefont {Dorschner}, \citenamefont {Henning}, \citenamefont
  {Mutschke}, \citenamefont {G\"urtler}, \citenamefont {K\"ompe},\ and\
  \citenamefont {Nass}}]{Bege1997}%
  \BibitemOpen
  \bibfield  {author} {\bibinfo {author} {\bibfnamefont {B.}~\bibnamefont
  {Begemann}}, \bibinfo {author} {\bibfnamefont {J.}~\bibnamefont {Dorschner}},
  \bibinfo {author} {\bibfnamefont {T.}~\bibnamefont {Henning}}, \bibinfo
  {author} {\bibfnamefont {H.}~\bibnamefont {Mutschke}}, \bibinfo {author}
  {\bibfnamefont {J.}~\bibnamefont {G\"urtler}}, \bibinfo {author}
  {\bibfnamefont {C.}~\bibnamefont {K\"ompe}},\ and\ \bibinfo {author}
  {\bibfnamefont {R.}~\bibnamefont {Nass}},\ }\bibfield  {title} {\bibinfo
  {title} {Aluminum oxide and the opacity of oxygen-rich circumstellar dust in
  the 12-17 micron range},\ }\href {https://doi.org/10.1086/303597} {\bibfield
  {journal} {\bibinfo  {journal} {The Astrophysical Journal}\ }\textbf
  {\bibinfo {volume} {476}},\ \bibinfo {pages} {199} (\bibinfo {year}
  {1997})}\BibitemShut {NoStop}%
\bibitem [{\citenamefont {K\"ubler}(1991)}]{Kubler1991}%
  \BibitemOpen
  \bibfield  {author} {\bibinfo {author} {\bibfnamefont {W.}~\bibnamefont
  {K\"ubler}},\ }\bibfield  {title} {\bibinfo {title} {Properties of \ce{Al2O3}
  thin films prepared by ion-assisted evaporation},\ }\href
  {https://doi.org/10.1016/0040-6090(91)90007-K} {\bibfield  {journal}
  {\bibinfo  {journal} {Thin Solid Films}\ }\textbf {\bibinfo {volume} {199}},\
  \bibinfo {pages} {247} (\bibinfo {year} {1991})}\BibitemShut {NoStop}%
\bibitem [{\citenamefont {Eisele}(1975)}]{Eis1975}%
  \BibitemOpen
  \bibfield  {author} {\bibinfo {author} {\bibfnamefont {K.~M.}\ \bibnamefont
  {Eisele}},\ }\bibfield  {title} {\bibinfo {title} {Charge storage and
  stoichiometry in electron beam evaporated alumina},\ }\href
  {https://doi.org/10.1149/1.2134144} {\bibfield  {journal} {\bibinfo
  {journal} {Journal of The Electrochemical Society}\ }\textbf {\bibinfo
  {volume} {122}},\ \bibinfo {pages} {148} (\bibinfo {year}
  {1975})}\BibitemShut {NoStop}%
\bibitem [{\citenamefont {Shamala}\ \emph {et~al.}(2004)\citenamefont
  {Shamala}, \citenamefont {Murthy},\ and\ \citenamefont {Rao}}]{Sham2004}%
  \BibitemOpen
  \bibfield  {author} {\bibinfo {author} {\bibfnamefont {K.}~\bibnamefont
  {Shamala}}, \bibinfo {author} {\bibfnamefont {L.}~\bibnamefont {Murthy}},\
  and\ \bibinfo {author} {\bibfnamefont {K.~N.}\ \bibnamefont {Rao}},\
  }\bibfield  {title} {\bibinfo {title} {Studies on optical and dielectric
  properties of \ce{Al2O3} thin films prepared by electron beam evaporation and
  spray pyrolysis method},\ }\href {https://doi.org/10.1016/j.mseb.2003.09.036}
  {\bibfield  {journal} {\bibinfo  {journal} {Materials Science and
  Engineering: B}\ }\textbf {\bibinfo {volume} {106}},\ \bibinfo {pages} {269}
  (\bibinfo {year} {2004})}\BibitemShut {NoStop}%
\bibitem [{\citenamefont {Mikhaelashvili}\ \emph {et~al.}(1998)\citenamefont
  {Mikhaelashvili}, \citenamefont {Betzer}, \citenamefont {Prudnikov},
  \citenamefont {Orenstein}, \citenamefont {Ritter},\ and\ \citenamefont
  {Eisenstein}}]{Mikhael1998}%
  \BibitemOpen
  \bibfield  {author} {\bibinfo {author} {\bibfnamefont {V.}~\bibnamefont
  {Mikhaelashvili}}, \bibinfo {author} {\bibfnamefont {Y.}~\bibnamefont
  {Betzer}}, \bibinfo {author} {\bibfnamefont {I.}~\bibnamefont {Prudnikov}},
  \bibinfo {author} {\bibfnamefont {M.}~\bibnamefont {Orenstein}}, \bibinfo
  {author} {\bibfnamefont {D.}~\bibnamefont {Ritter}},\ and\ \bibinfo {author}
  {\bibfnamefont {G.}~\bibnamefont {Eisenstein}},\ }\bibfield  {title}
  {\bibinfo {title} {Electrical characteristics of metal-dielectric-metal and
  metal-dielectric-semiconductor structures based on electron beam evaporated
  \ce{Y2O3}, \ce{Ta2O5} and \ce{A2O3} thin film},\ }\href
  {https://doi.org/10.1063/1.369002} {\bibfield  {journal} {\bibinfo  {journal}
  {Journal of Applied Physics}\ }\textbf {\bibinfo {volume} {84}},\ \bibinfo
  {pages} {6747} (\bibinfo {year} {1998})}\BibitemShut {NoStop}%
\bibitem [{\citenamefont {Perdew}\ \emph {et~al.}(1996)\citenamefont {Perdew},
  \citenamefont {Burke},\ and\ \citenamefont {Ernzerhof}}]{PBE_funct}%
  \BibitemOpen
  \bibfield  {author} {\bibinfo {author} {\bibfnamefont {J.~P.}\ \bibnamefont
  {Perdew}}, \bibinfo {author} {\bibfnamefont {K.}~\bibnamefont {Burke}},\ and\
  \bibinfo {author} {\bibfnamefont {M.}~\bibnamefont {Ernzerhof}},\ }\bibfield
  {title} {\bibinfo {title} {Generalized gradient approximation made simple},\
  }\href {https://doi.org/10.1103/PhysRevLett.77.3865} {\bibfield  {journal}
  {\bibinfo  {journal} {Physical Review Letters}\ }\textbf {\bibinfo {volume}
  {77}},\ \bibinfo {pages} {3865} (\bibinfo {year} {1996})}\BibitemShut
  {NoStop}%
\bibitem [{Pse()}]{PseudiNota}%
  \BibitemOpen
  \href@noop {} {\bibinfo {title} {We used the pseudopotential files {\tt
  o.pbe-mt.upf, al.pbe-rrkj.upf} as available from the qe website
  \href{http://www.quantum-espresso.org}{http://www.quantum-espresso.org}.}}\BibitemShut
  {Stop}%
\bibitem [{EXC()}]{EXC_Nota}%
  \BibitemOpen
  \href@noop {} {\bibinfo {title} {For the purposes of the present work, which
  are the description of the local mode symmetry and clarification the origin
  of the main ir bands, frequency shifts ($\sim$10~cm$^{-1}$) related to the
  choice of the exchange-correlation functional~\cite{Demich2019} are minor
  issues not affecting the conclusion of the present paper.}}\BibitemShut
  {Stop}%
\bibitem [{\citenamefont {Colleoni}\ \emph {et~al.}(2015)\citenamefont
  {Colleoni}, \citenamefont {Miceli},\ and\ \citenamefont
  {Pasquarello}}]{Coll2015}%
  \BibitemOpen
  \bibfield  {author} {\bibinfo {author} {\bibfnamefont {D.}~\bibnamefont
  {Colleoni}}, \bibinfo {author} {\bibfnamefont {G.}~\bibnamefont {Miceli}},\
  and\ \bibinfo {author} {\bibfnamefont {A.}~\bibnamefont {Pasquarello}},\
  }\bibfield  {title} {\bibinfo {title} {Band alignment and chemical bonding at
  the \ce{GaAs/Al2O3} interface: {A} hybrid functional study},\ }\href
  {https://doi.org/10.1063/1.4936240} {\bibfield  {journal} {\bibinfo
  {journal} {Applied Physics Letters}\ }\textbf {\bibinfo {volume} {107}},\
  \bibinfo {pages} {211601} (\bibinfo {year} {2015})}\BibitemShut {NoStop}%
\bibitem [{\citenamefont {Giannozzi}\ \emph {et~al.}(2009)\citenamefont
  {Giannozzi}, \citenamefont {Baroni}, \citenamefont {Bonini}, \citenamefont
  {Calandra}, \citenamefont {Car}, \citenamefont {Cavazzoni}, \citenamefont
  {Ceresoli}, \citenamefont {Chiarotti}, \citenamefont {Cococcioni},
  \citenamefont {Dabo}, \citenamefont {{Dal Corso}}, \citenamefont
  {de~Gironcoli}, \citenamefont {Fabris}, \citenamefont {Fratesi},
  \citenamefont {Gebauer}, \citenamefont {Gerstmann}, \citenamefont
  {Gougoussis}, \citenamefont {Kokalj}, \citenamefont {Lazzeri}, \citenamefont
  {Martin-Samos}, \citenamefont {Marzari}, \citenamefont {Mauri}, \citenamefont
  {Mazzarello}, \citenamefont {Paolini}, \citenamefont {Pasquarello},
  \citenamefont {Paulatto}, \citenamefont {Sbraccia}, \citenamefont {Scandolo},
  \citenamefont {Sclauzero}, \citenamefont {Seitsonen}, \citenamefont
  {Smogunov}, \citenamefont {Umari},\ and\ \citenamefont {Wentzcovitch}}]{QE}%
  \BibitemOpen
  \bibfield  {author} {\bibinfo {author} {\bibfnamefont {P.}~\bibnamefont
  {Giannozzi}}, \bibinfo {author} {\bibfnamefont {S.}~\bibnamefont {Baroni}},
  \bibinfo {author} {\bibfnamefont {N.}~\bibnamefont {Bonini}}, \bibinfo
  {author} {\bibfnamefont {M.}~\bibnamefont {Calandra}}, \bibinfo {author}
  {\bibfnamefont {R.}~\bibnamefont {Car}}, \bibinfo {author} {\bibfnamefont
  {C.}~\bibnamefont {Cavazzoni}}, \bibinfo {author} {\bibfnamefont
  {D.}~\bibnamefont {Ceresoli}}, \bibinfo {author} {\bibfnamefont {G.~L.}\
  \bibnamefont {Chiarotti}}, \bibinfo {author} {\bibfnamefont {M.}~\bibnamefont
  {Cococcioni}}, \bibinfo {author} {\bibfnamefont {I.}~\bibnamefont {Dabo}},
  \bibinfo {author} {\bibfnamefont {A.}~\bibnamefont {{Dal Corso}}}, \bibinfo
  {author} {\bibfnamefont {S.}~\bibnamefont {de~Gironcoli}}, \bibinfo {author}
  {\bibfnamefont {S.}~\bibnamefont {Fabris}}, \bibinfo {author} {\bibfnamefont
  {G.}~\bibnamefont {Fratesi}}, \bibinfo {author} {\bibfnamefont
  {R.}~\bibnamefont {Gebauer}}, \bibinfo {author} {\bibfnamefont
  {U.}~\bibnamefont {Gerstmann}}, \bibinfo {author} {\bibfnamefont
  {C.}~\bibnamefont {Gougoussis}}, \bibinfo {author} {\bibfnamefont
  {A.}~\bibnamefont {Kokalj}}, \bibinfo {author} {\bibfnamefont
  {M.}~\bibnamefont {Lazzeri}}, \bibinfo {author} {\bibfnamefont
  {L.}~\bibnamefont {Martin-Samos}}, \bibinfo {author} {\bibfnamefont
  {N.}~\bibnamefont {Marzari}}, \bibinfo {author} {\bibfnamefont
  {F.}~\bibnamefont {Mauri}}, \bibinfo {author} {\bibfnamefont
  {R.}~\bibnamefont {Mazzarello}}, \bibinfo {author} {\bibfnamefont
  {S.}~\bibnamefont {Paolini}}, \bibinfo {author} {\bibfnamefont
  {A.}~\bibnamefont {Pasquarello}}, \bibinfo {author} {\bibfnamefont
  {L.}~\bibnamefont {Paulatto}}, \bibinfo {author} {\bibfnamefont
  {C.}~\bibnamefont {Sbraccia}}, \bibinfo {author} {\bibfnamefont
  {S.}~\bibnamefont {Scandolo}}, \bibinfo {author} {\bibfnamefont
  {G.}~\bibnamefont {Sclauzero}}, \bibinfo {author} {\bibfnamefont {A.~P.}\
  \bibnamefont {Seitsonen}}, \bibinfo {author} {\bibfnamefont {A.}~\bibnamefont
  {Smogunov}}, \bibinfo {author} {\bibfnamefont {P.}~\bibnamefont {Umari}},\
  and\ \bibinfo {author} {\bibfnamefont {R.~M.}\ \bibnamefont {Wentzcovitch}},\
  }\bibfield  {title} {\bibinfo {title} {{QUANTUM ESPRESSO}: a modular and
  open-source software project for quantum simulations of materials},\ }\href
  {https://doi.org/10.1088/0953-8984/21/39/395502} {\bibfield  {journal}
  {\bibinfo  {journal} {Journal of Physics: Condensed Matter}\ }\textbf
  {\bibinfo {volume} {21}},\ \bibinfo {pages} {395502} (\bibinfo {year}
  {2009})}\BibitemShut {NoStop}%
\bibitem [{\citenamefont {Lamparter}\ and\ \citenamefont
  {Kniep}(1997)}]{Lamp97}%
  \BibitemOpen
  \bibfield  {author} {\bibinfo {author} {\bibfnamefont {P.}~\bibnamefont
  {Lamparter}}\ and\ \bibinfo {author} {\bibfnamefont {R.}~\bibnamefont
  {Kniep}},\ }\bibfield  {title} {\bibinfo {title} {Structure of amorphous
  \ce{Al2O3}},\ }\href {https://doi.org/10.1016/S0921-4526(96)01044-7}
  {\bibfield  {journal} {\bibinfo  {journal} {Physica B: Condensed Matter}\
  }\textbf {\bibinfo {volume} {234}},\ \bibinfo {pages} {405} (\bibinfo {year}
  {1997})}\BibitemShut {NoStop}%
\bibitem [{\citenamefont {Guo}\ \emph {et~al.}()\citenamefont {Guo},
  \citenamefont {Ambrosio},\ and\ \citenamefont {Pasquarello}}]{Guo2019}%
  \BibitemOpen
  \bibfield  {author} {\bibinfo {author} {\bibfnamefont {Z.}~\bibnamefont
  {Guo}}, \bibinfo {author} {\bibfnamefont {F.}~\bibnamefont {Ambrosio}},\ and\
  \bibinfo {author} {\bibfnamefont {A.}~\bibnamefont {Pasquarello}},\ }\href
  {https://doi.org/10.24435/materialscloud:2019.0026/v1} {\bibinfo {title}
  {Oxygen defects in amorphous \ce{Al2O3}, {M}aterials {C}loud {A}rchive
  2019.0026/v1 (2019)}}\BibitemShut {NoStop}%
\bibitem [{\citenamefont {Matsui}(1996)}]{Matsui96}%
  \BibitemOpen
  \bibfield  {author} {\bibinfo {author} {\bibfnamefont {M.}~\bibnamefont
  {Matsui}},\ }\bibfield  {title} {\bibinfo {title} {Molecular dynamics study
  of the structures and bulk moduli of crystals in the system
  \ce{CaO-MgO-Al2O3-SiO2}},\ }\href {https://doi.org/10.1007/BF00199500}
  {\bibfield  {journal} {\bibinfo  {journal} {Physics and Chemistry of
  Minerals}\ }\textbf {\bibinfo {volume} {23}},\ \bibinfo {pages} {345}
  (\bibinfo {year} {1996})}\BibitemShut {NoStop}%
\bibitem [{\citenamefont {Momida}\ \emph {et~al.}(2007)\citenamefont {Momida},
  \citenamefont {Hamada},\ and\ \citenamefont {Ohno}}]{Momi2007b}%
  \BibitemOpen
  \bibfield  {author} {\bibinfo {author} {\bibfnamefont {H.}~\bibnamefont
  {Momida}}, \bibinfo {author} {\bibfnamefont {T.}~\bibnamefont {Hamada}},\
  and\ \bibinfo {author} {\bibfnamefont {T.}~\bibnamefont {Ohno}},\ }\bibfield
  {title} {\bibinfo {title} {First-principles study of dielectric properties of
  amorphous high-k materials},\ }\href {https://doi.org/10.1143/JJAP.46.3255}
  {\bibfield  {journal} {\bibinfo  {journal} {Japanese Journal of Applied
  Physics}\ }\textbf {\bibinfo {volume} {46}},\ \bibinfo {pages} {3255}
  (\bibinfo {year} {2007})}\BibitemShut {NoStop}%
\bibitem [{\citenamefont {Willems}\ \emph {et~al.}(2012)\citenamefont
  {Willems}, \citenamefont {Rycroft}, \citenamefont {Kazi}, \citenamefont
  {Meza},\ and\ \citenamefont {Haranczyk}}]{zeopp}%
  \BibitemOpen
  \bibfield  {author} {\bibinfo {author} {\bibfnamefont {T.~F.}\ \bibnamefont
  {Willems}}, \bibinfo {author} {\bibfnamefont {C.~H.}\ \bibnamefont
  {Rycroft}}, \bibinfo {author} {\bibfnamefont {M.}~\bibnamefont {Kazi}},
  \bibinfo {author} {\bibfnamefont {J.~C.}\ \bibnamefont {Meza}},\ and\
  \bibinfo {author} {\bibfnamefont {M.}~\bibnamefont {Haranczyk}},\ }\bibfield
  {title} {\bibinfo {title} {Algorithms and tools for high-throughput
  geometry-based analysis of crystalline porous materials},\ }\href
  {https://doi.org/10.1016/j.micromeso.2011.08.020} {\bibfield  {journal}
  {\bibinfo  {journal} {Microporous and Mesoporous Materials}\ }\textbf
  {\bibinfo {volume} {149}},\ \bibinfo {pages} {134} (\bibinfo {year}
  {2012})}\BibitemShut {NoStop}%
\bibitem [{\citenamefont {Hung}\ and\ \citenamefont {Vinh}(2006)}]{hung2006}%
  \BibitemOpen
  \bibfield  {author} {\bibinfo {author} {\bibfnamefont {P.}~\bibnamefont
  {Hung}}\ and\ \bibinfo {author} {\bibfnamefont {L.}~\bibnamefont {Vinh}},\
  }\bibfield  {title} {\bibinfo {title} {Local microstructure of liquid and
  amorphous \ce{Al2O3}},\ }\href
  {https://doi.org/10.1016/j.jnoncrysol.2006.09.016} {\bibfield  {journal}
  {\bibinfo  {journal} {Journal of Non-Crystalline Solids}\ }\textbf {\bibinfo
  {volume} {352}},\ \bibinfo {pages} {5531} (\bibinfo {year}
  {2006})}\BibitemShut {NoStop}%
\bibitem [{Sup()}]{SupplInfo}%
  \BibitemOpen
  \href@noop {} {}\bibinfo {note} {See Supplemental Material at
  \url{http://link.aps.org/supplemental/10.1103/PhysRevMaterials.7.045604} for
  further details on thin film characterization, and on the calculation of IR
  active modes in $\theta$-alumina and of participation ratio in am-\ce{Al2O3}.
  For convenience of the reader the stand-alone pdf file has been included at
  the end of the present manuscript.}\BibitemShut {Stop}%
\bibitem [{\citenamefont {Al-Abadleh}\ and\ \citenamefont
  {Grassian}(2003)}]{Abadleh2003}%
  \BibitemOpen
  \bibfield  {author} {\bibinfo {author} {\bibfnamefont {H.~A.}\ \bibnamefont
  {Al-Abadleh}}\ and\ \bibinfo {author} {\bibfnamefont {V.}~\bibnamefont
  {Grassian}},\ }\bibfield  {title} {\bibinfo {title} {{FT-IR} study of water
  adsorption on aluminum oxide surfaces},\ }\href
  {https://doi.org/10.1021/la026208a} {\bibfield  {journal} {\bibinfo
  {journal} {Langmuir}\ }\textbf {\bibinfo {volume} {19}},\ \bibinfo {pages}
  {341} (\bibinfo {year} {2003})}\BibitemShut {NoStop}%
\bibitem [{\citenamefont {Gerakines}\ and\ \citenamefont
  {Hudson}(2020)}]{Gera2020}%
  \BibitemOpen
  \bibfield  {author} {\bibinfo {author} {\bibfnamefont {P.~A.}\ \bibnamefont
  {Gerakines}}\ and\ \bibinfo {author} {\bibfnamefont {R.~L.}\ \bibnamefont
  {Hudson}},\ }\bibfield  {title} {\bibinfo {title} {A modified algorithm and
  open-source computational package for the determination of infrared optical
  constants relevant to astrophysics},\ }\href
  {https://doi.org/10.3847/1538-4357/abad39} {\bibfield  {journal} {\bibinfo
  {journal} {The Astrophysical Journal}\ }\textbf {\bibinfo {volume} {901}},\
  \bibinfo {pages} {52} (\bibinfo {year} {2020})}\BibitemShut {NoStop}%
\bibitem [{Rel()}]{RelaxNota}%
  \BibitemOpen
  \href@noop {} {\bibinfo {title} {{T}he first-principles relaxation of models
  ii and iii with {PBE}, as compared to the original {LDA} configurations from
  {M}omida~\cite{Momi2006} (and named there as model g and model h,
  respectively), shows a tendency to slightly increase the number of fivefold
  and sixfold coordinated {Al} atoms and of the number of fourfold coordinated
  {O} atoms by decreasing the number of fourfold coordinated {Al} atoms and
  twofold {O} atoms.}}\BibitemShut {Stop}%
\bibitem [{\citenamefont {Cui}\ \emph {et~al.}(2018)\citenamefont {Cui},
  \citenamefont {Kast}, \citenamefont {Hammann}, \citenamefont {Afriyie},
  \citenamefont {Woods}, \citenamefont {Plassmeyer}, \citenamefont {Perkins},
  \citenamefont {Ma}, \citenamefont {Keszler}, \citenamefont {Page},
  \citenamefont {Boettcher},\ and\ \citenamefont {Hayes}}]{Cui2018}%
  \BibitemOpen
  \bibfield  {author} {\bibinfo {author} {\bibfnamefont {J.}~\bibnamefont
  {Cui}}, \bibinfo {author} {\bibfnamefont {M.~G.}\ \bibnamefont {Kast}},
  \bibinfo {author} {\bibfnamefont {B.~A.}\ \bibnamefont {Hammann}}, \bibinfo
  {author} {\bibfnamefont {Y.}~\bibnamefont {Afriyie}}, \bibinfo {author}
  {\bibfnamefont {K.~N.}\ \bibnamefont {Woods}}, \bibinfo {author}
  {\bibfnamefont {P.~N.}\ \bibnamefont {Plassmeyer}}, \bibinfo {author}
  {\bibfnamefont {C.~K.}\ \bibnamefont {Perkins}}, \bibinfo {author}
  {\bibfnamefont {Z.~L.}\ \bibnamefont {Ma}}, \bibinfo {author} {\bibfnamefont
  {D.~A.}\ \bibnamefont {Keszler}}, \bibinfo {author} {\bibfnamefont {C.~J.}\
  \bibnamefont {Page}}, \bibinfo {author} {\bibfnamefont {S.~W.}\ \bibnamefont
  {Boettcher}},\ and\ \bibinfo {author} {\bibfnamefont {S.~E.}\ \bibnamefont
  {Hayes}},\ }\bibfield  {title} {\bibinfo {title} {Aluminum oxide thin films
  from aqueous solutions: Insights from solid-state nmr and dielectric
  response},\ }\href {https://doi.org/10.1021/acs.chemmater.7b05078} {\bibfield
   {journal} {\bibinfo  {journal} {Chemistry of Materials}\ }\textbf {\bibinfo
  {volume} {30}},\ \bibinfo {pages} {7456} (\bibinfo {year}
  {2018})}\BibitemShut {NoStop}%
\bibitem [{\citenamefont {Lee}\ and\ \citenamefont {Ahn}(2014)}]{Lee2014}%
  \BibitemOpen
  \bibfield  {author} {\bibinfo {author} {\bibfnamefont {S.~K.}\ \bibnamefont
  {Lee}}\ and\ \bibinfo {author} {\bibfnamefont {C.~W.}\ \bibnamefont {Ahn}},\
  }\bibfield  {title} {\bibinfo {title} {Probing of 2 dimensional
  confinement-induced structural transitions in amorphous oxide thin film},\
  }\href {https://doi.org/https://doi.org/10.1038/srep04200} {\bibfield
  {journal} {\bibinfo  {journal} {Scientific reports}\ }\textbf {\bibinfo
  {volume} {4}},\ \bibinfo {pages} {1} (\bibinfo {year} {2014})}\BibitemShut
  {NoStop}%
\bibitem [{\citenamefont {Kokalj}(1999)}]{Kokal99}%
  \BibitemOpen
  \bibfield  {author} {\bibinfo {author} {\bibfnamefont {A.}~\bibnamefont
  {Kokalj}},\ }\bibfield  {title} {\bibinfo {title} {{XCrySDen}—a new program
  for displaying crystalline structures and electron densities},\ }\href
  {https://doi.org/https://doi.org/10.1016/S1093-3263(99)00028-5} {\bibfield
  {journal} {\bibinfo  {journal} {Journal of Molecular Graphics and Modelling}\
  }\textbf {\bibinfo {volume} {17}},\ \bibinfo {pages} {176} (\bibinfo {year}
  {1999})}\BibitemShut {NoStop}%
\bibitem [{\citenamefont {Bell}\ and\ \citenamefont {Dean}(1972)}]{bell72}%
  \BibitemOpen
  \bibfield  {author} {\bibinfo {author} {\bibfnamefont {R.~J.}\ \bibnamefont
  {Bell}}\ and\ \bibinfo {author} {\bibfnamefont {P.}~\bibnamefont {Dean}},\
  }\bibfield  {title} {\bibinfo {title} {The structure of vitreous silica:
  Validity of the random network theory},\ }\href
  {https://doi.org/10.1080/14786437208223861} {\bibfield  {journal} {\bibinfo
  {journal} {Philosophical Magazine}\ }\textbf {\bibinfo {volume} {25}},\
  \bibinfo {pages} {1381} (\bibinfo {year} {1972})}\BibitemShut {NoStop}%
\bibitem [{\citenamefont {Shcheblanov}\ \emph {et~al.}(2019)\citenamefont
  {Shcheblanov}, \citenamefont {Giacomazzi}, \citenamefont {Povarnitsyn},
  \citenamefont {Kohara}, \citenamefont {Martin-Samos}, \citenamefont
  {Mountjoy}, \citenamefont {Newport}, \citenamefont {Haworth}, \citenamefont
  {Richard},\ and\ \citenamefont {Ollier}}]{Nik2019}%
  \BibitemOpen
  \bibfield  {author} {\bibinfo {author} {\bibfnamefont {N.}~\bibnamefont
  {Shcheblanov}}, \bibinfo {author} {\bibfnamefont {L.}~\bibnamefont
  {Giacomazzi}}, \bibinfo {author} {\bibfnamefont {M.}~\bibnamefont
  {Povarnitsyn}}, \bibinfo {author} {\bibfnamefont {S.}~\bibnamefont {Kohara}},
  \bibinfo {author} {\bibfnamefont {L.}~\bibnamefont {Martin-Samos}}, \bibinfo
  {author} {\bibfnamefont {G.}~\bibnamefont {Mountjoy}}, \bibinfo {author}
  {\bibfnamefont {R.}~\bibnamefont {Newport}}, \bibinfo {author} {\bibfnamefont
  {R.}~\bibnamefont {Haworth}}, \bibinfo {author} {\bibfnamefont
  {N.}~\bibnamefont {Richard}},\ and\ \bibinfo {author} {\bibfnamefont
  {N.}~\bibnamefont {Ollier}},\ }\bibfield  {title} {\bibinfo {title}
  {Vibrational and structural properties of \ce{P2O5} glass: {A}dvances from a
  combined modeling approach},\ }\href
  {https://doi.org/10.1103/PhysRevB.100.134309} {\bibfield  {journal} {\bibinfo
   {journal} {Physical Review B}\ }\textbf {\bibinfo {volume} {100}},\ \bibinfo
  {pages} {134309} (\bibinfo {year} {2019})}\BibitemShut {NoStop}%
\bibitem [{\citenamefont {Thorpe}\ and\ \citenamefont
  {De~Leeuw}(1986)}]{ThorpeLeeuw}%
  \BibitemOpen
  \bibfield  {author} {\bibinfo {author} {\bibfnamefont {M.~F.}\ \bibnamefont
  {Thorpe}}\ and\ \bibinfo {author} {\bibfnamefont {S.~W.}\ \bibnamefont
  {De~Leeuw}},\ }\bibfield  {title} {\bibinfo {title} {Coulomb effects in
  disordered solids},\ }\href {https://doi.org/10.1103/PhysRevB.33.8490}
  {\bibfield  {journal} {\bibinfo  {journal} {Physical Review B}\ }\textbf
  {\bibinfo {volume} {33}},\ \bibinfo {pages} {8490} (\bibinfo {year}
  {1986})}\BibitemShut {NoStop}%
\bibitem [{\citenamefont {Pasquarello}\ and\ \citenamefont {Car}(1997)}]{PC97}%
  \BibitemOpen
  \bibfield  {author} {\bibinfo {author} {\bibfnamefont {A.}~\bibnamefont
  {Pasquarello}}\ and\ \bibinfo {author} {\bibfnamefont {R.}~\bibnamefont
  {Car}},\ }\bibfield  {title} {\bibinfo {title} {Dynamical charge tensors and
  infrared spectrum of amorphous \ce{SiO2}},\ }\href
  {https://doi.org/10.1103/PhysRevLett.79.1766} {\bibfield  {journal} {\bibinfo
   {journal} {Physical Review Letters}\ }\textbf {\bibinfo {volume} {79}},\
  \bibinfo {pages} {1766} (\bibinfo {year} {1997})}\BibitemShut {NoStop}%
\bibitem [{\citenamefont {Povarnitsyn}\ \emph {et~al.}(2020)\citenamefont
  {Povarnitsyn}, \citenamefont {Shcheblanov}, \citenamefont {Ivanov},
  \citenamefont {Timoshenko},\ and\ \citenamefont {Klimentov}}]{misha2020}%
  \BibitemOpen
  \bibfield  {author} {\bibinfo {author} {\bibfnamefont {M.}~\bibnamefont
  {Povarnitsyn}}, \bibinfo {author} {\bibfnamefont {N.}~\bibnamefont
  {Shcheblanov}}, \bibinfo {author} {\bibfnamefont {D.}~\bibnamefont {Ivanov}},
  \bibinfo {author} {\bibfnamefont {V.~Y.}\ \bibnamefont {Timoshenko}},\ and\
  \bibinfo {author} {\bibfnamefont {S.}~\bibnamefont {Klimentov}},\ }\bibfield
  {title} {\bibinfo {title} {Vibrational analysis of silicon nanoparticles
  using simulation and decomposition of raman spectra},\ }\href
  {https://doi.org/10.1103/PhysRevApplied.14.014067} {\bibfield  {journal}
  {\bibinfo  {journal} {Phys. Rev. Applied}\ }\textbf {\bibinfo {volume}
  {14}},\ \bibinfo {pages} {014067} (\bibinfo {year} {2020})}\BibitemShut
  {NoStop}%
\bibitem [{\citenamefont {Umari}\ and\ \citenamefont
  {Pasquarello}(2005)}]{PumaDRM}%
  \BibitemOpen
  \bibfield  {author} {\bibinfo {author} {\bibfnamefont {P.}~\bibnamefont
  {Umari}}\ and\ \bibinfo {author} {\bibfnamefont {A.}~\bibnamefont
  {Pasquarello}},\ }\bibfield  {title} {\bibinfo {title} {Infrared and raman
  spectra of disordered materials from first principles},\ }\href
  {https://doi.org/10.1016/j.diamond.2004.12.007} {\bibfield  {journal}
  {\bibinfo  {journal} {Diamond and Related Materials}\ }\textbf {\bibinfo
  {volume} {14}},\ \bibinfo {pages} {1255} (\bibinfo {year}
  {2005})}\BibitemShut {NoStop}%
\bibitem [{\citenamefont {Lee}\ \emph {et~al.}(1995)\citenamefont {Lee},
  \citenamefont {Cahill},\ and\ \citenamefont {Allen}}]{Lee1995}%
  \BibitemOpen
  \bibfield  {author} {\bibinfo {author} {\bibfnamefont {S.-M.}\ \bibnamefont
  {Lee}}, \bibinfo {author} {\bibfnamefont {D.~G.}\ \bibnamefont {Cahill}},\
  and\ \bibinfo {author} {\bibfnamefont {T.~H.}\ \bibnamefont {Allen}},\
  }\bibfield  {title} {\bibinfo {title} {Thermal conductivity of sputtered
  oxide films},\ }\href {https://doi.org/10.1103/PhysRevB.52.253} {\bibfield
  {journal} {\bibinfo  {journal} {Physical Review B}\ }\textbf {\bibinfo
  {volume} {52}},\ \bibinfo {pages} {253} (\bibinfo {year} {1995})}\BibitemShut
  {NoStop}%
\bibitem [{\citenamefont {Segda}\ \emph {et~al.}(2001)\citenamefont {Segda},
  \citenamefont {Jacquet},\ and\ \citenamefont {Besse}}]{Segda2001}%
  \BibitemOpen
  \bibfield  {author} {\bibinfo {author} {\bibfnamefont {B.}~\bibnamefont
  {Segda}}, \bibinfo {author} {\bibfnamefont {M.}~\bibnamefont {Jacquet}},\
  and\ \bibinfo {author} {\bibfnamefont {J.}~\bibnamefont {Besse}},\ }\bibfield
   {title} {\bibinfo {title} {Elaboration, characterization and dielectric
  properties study of amorphous alumina thin films deposited by r.f. magnetron
  sputtering},\ }\href {https://doi.org/10.1016/S0042-207X(01)00114-2}
  {\bibfield  {journal} {\bibinfo  {journal} {Vacuum}\ }\textbf {\bibinfo
  {volume} {62}},\ \bibinfo {pages} {27} (\bibinfo {year} {2001})}\BibitemShut
  {NoStop}%
\bibitem [{\citenamefont {Gusev}\ \emph {et~al.}(2001)\citenamefont {Gusev},
  \citenamefont {Cartier}, \citenamefont {Buchanan}, \citenamefont {Gribelyuk},
  \citenamefont {Copel}, \citenamefont {Okorn-Schmidt},\ and\ \citenamefont
  {D’emic}}]{Gus2001}%
  \BibitemOpen
  \bibfield  {author} {\bibinfo {author} {\bibfnamefont {E.}~\bibnamefont
  {Gusev}}, \bibinfo {author} {\bibfnamefont {E.}~\bibnamefont {Cartier}},
  \bibinfo {author} {\bibfnamefont {D.}~\bibnamefont {Buchanan}}, \bibinfo
  {author} {\bibfnamefont {M.}~\bibnamefont {Gribelyuk}}, \bibinfo {author}
  {\bibfnamefont {M.}~\bibnamefont {Copel}}, \bibinfo {author} {\bibfnamefont
  {H.}~\bibnamefont {Okorn-Schmidt}},\ and\ \bibinfo {author} {\bibfnamefont
  {C.}~\bibnamefont {D’emic}},\ }\bibfield  {title} {\bibinfo {title}
  {Ultrathin high-k metal oxides on silicon: processing, characterization and
  integration issues},\ }\href {https://doi.org/10.1016/S0167-9317(01)00667-0}
  {\bibfield  {journal} {\bibinfo  {journal} {Microelectronic Engineering}\
  }\textbf {\bibinfo {volume} {59}},\ \bibinfo {pages} {341} (\bibinfo {year}
  {2001})}\BibitemShut {NoStop}%
\bibitem [{\citenamefont {Aguilar-Gama}\ \emph {et~al.}(2015)\citenamefont
  {Aguilar-Gama}, \citenamefont {Ram{\'\i}rez-Morales}, \citenamefont
  {Montiel-Gonz{\'a}lez}, \citenamefont {Mendoza-Galv{\'a}n}, \citenamefont
  {Sotelo-Lerma}, \citenamefont {Nair},\ and\ \citenamefont {Hu}}]{Gama2015}%
  \BibitemOpen
  \bibfield  {author} {\bibinfo {author} {\bibfnamefont {M.~T.}\ \bibnamefont
  {Aguilar-Gama}}, \bibinfo {author} {\bibfnamefont {E.}~\bibnamefont
  {Ram{\'\i}rez-Morales}}, \bibinfo {author} {\bibfnamefont {Z.}~\bibnamefont
  {Montiel-Gonz{\'a}lez}}, \bibinfo {author} {\bibfnamefont {A.}~\bibnamefont
  {Mendoza-Galv{\'a}n}}, \bibinfo {author} {\bibfnamefont {M.}~\bibnamefont
  {Sotelo-Lerma}}, \bibinfo {author} {\bibfnamefont {P.}~\bibnamefont {Nair}},\
  and\ \bibinfo {author} {\bibfnamefont {H.}~\bibnamefont {Hu}},\ }\bibfield
  {title} {\bibinfo {title} {Structure and refractive index of thin alumina
  films grown by atomic layer deposition},\ }\href
  {https://doi.org/10.1007/s10854-014-2111-z} {\bibfield  {journal} {\bibinfo
  {journal} {Journal of Materials Science: Materials in Electronics}\ }\textbf
  {\bibinfo {volume} {26}},\ \bibinfo {pages} {5546} (\bibinfo {year}
  {2015})}\BibitemShut {NoStop}%
\bibitem [{\citenamefont {Evangelisti}\ \emph {et~al.}(2017)\citenamefont
  {Evangelisti}, \citenamefont {Stiefel}, \citenamefont {Guseva}, \citenamefont
  {Nia}, \citenamefont {Hauert}, \citenamefont {Hack}, \citenamefont
  {Jeurgens}, \citenamefont {Ambrosio}, \citenamefont {Pasquarello},
  \citenamefont {Schmutz},\ and\ \citenamefont {Cancellieri}}]{Evan2017}%
  \BibitemOpen
  \bibfield  {author} {\bibinfo {author} {\bibfnamefont {F.}~\bibnamefont
  {Evangelisti}}, \bibinfo {author} {\bibfnamefont {M.}~\bibnamefont
  {Stiefel}}, \bibinfo {author} {\bibfnamefont {O.}~\bibnamefont {Guseva}},
  \bibinfo {author} {\bibfnamefont {R.~P.}\ \bibnamefont {Nia}}, \bibinfo
  {author} {\bibfnamefont {R.}~\bibnamefont {Hauert}}, \bibinfo {author}
  {\bibfnamefont {E.}~\bibnamefont {Hack}}, \bibinfo {author} {\bibfnamefont
  {L.~P.}\ \bibnamefont {Jeurgens}}, \bibinfo {author} {\bibfnamefont
  {F.}~\bibnamefont {Ambrosio}}, \bibinfo {author} {\bibfnamefont
  {A.}~\bibnamefont {Pasquarello}}, \bibinfo {author} {\bibfnamefont
  {P.}~\bibnamefont {Schmutz}},\ and\ \bibinfo {author} {\bibfnamefont
  {C.}~\bibnamefont {Cancellieri}},\ }\bibfield  {title} {\bibinfo {title}
  {Electronic and structural characterization of barrier-type amorphous
  aluminium oxide},\ }\href {https://doi.org/10.1016/j.electacta.2016.12.090}
  {\bibfield  {journal} {\bibinfo  {journal} {Electrochimica Acta}\ }\textbf
  {\bibinfo {volume} {224}},\ \bibinfo {pages} {503} (\bibinfo {year}
  {2017})}\BibitemShut {NoStop}%
\bibitem [{\citenamefont {Parfitt}\ \emph {et~al.}(1995)\citenamefont
  {Parfitt}, \citenamefont {Goldiner}, \citenamefont {Jones},\ and\
  \citenamefont {Was}}]{Parf1995}%
  \BibitemOpen
  \bibfield  {author} {\bibinfo {author} {\bibfnamefont {L.}~\bibnamefont
  {Parfitt}}, \bibinfo {author} {\bibfnamefont {M.}~\bibnamefont {Goldiner}},
  \bibinfo {author} {\bibfnamefont {J.~W.}\ \bibnamefont {Jones}},\ and\
  \bibinfo {author} {\bibfnamefont {G.~S.}\ \bibnamefont {Was}},\ }\bibfield
  {title} {\bibinfo {title} {Residual stresses in amorphous alumina films
  synthesized by ion beam assisted deposition},\ }\href
  {https://doi.org/10.1063/1.358652} {\bibfield  {journal} {\bibinfo  {journal}
  {Journal of Applied Physics}\ }\textbf {\bibinfo {volume} {77}},\ \bibinfo
  {pages} {3029} (\bibinfo {year} {1995})}\BibitemShut {NoStop}%
\bibitem [{\citenamefont {Umari}\ and\ \citenamefont
  {Pasquarello}(2003)}]{PUMA2003}%
  \BibitemOpen
  \bibfield  {author} {\bibinfo {author} {\bibfnamefont {P.}~\bibnamefont
  {Umari}}\ and\ \bibinfo {author} {\bibfnamefont {A.}~\bibnamefont
  {Pasquarello}},\ }\bibfield  {title} {\bibinfo {title} {Polarizability and
  dielectric constant in density-functional supercell calculations with
  discrete k-point samplings},\ }\href
  {https://doi.org/10.1103/PhysRevB.68.085114} {\bibfield  {journal} {\bibinfo
  {journal} {Physical Review B}\ }\textbf {\bibinfo {volume} {68}},\ \bibinfo
  {pages} {085114} (\bibinfo {year} {2003})}\BibitemShut {NoStop}%
\bibitem [{\citenamefont {Eriksson}\ \emph {et~al.}(1982)\citenamefont
  {Eriksson}, \citenamefont {Hjortsberg},\ and\ \citenamefont
  {Granqvist}}]{Erik1982}%
  \BibitemOpen
  \bibfield  {author} {\bibinfo {author} {\bibfnamefont {T.}~\bibnamefont
  {Eriksson}}, \bibinfo {author} {\bibfnamefont {A.}~\bibnamefont
  {Hjortsberg}},\ and\ \bibinfo {author} {\bibfnamefont {C.}~\bibnamefont
  {Granqvist}},\ }\bibfield  {title} {\bibinfo {title} {Solar absorptance and
  thermal emittance of \ce{Al2O3} films on {A}l: {A} theoretical assessment},\
  }\href {https://doi.org/10.1016/0165-1633(82)90020-X} {\bibfield  {journal}
  {\bibinfo  {journal} {Solar Energy Materials}\ }\textbf {\bibinfo {volume}
  {6}},\ \bibinfo {pages} {191} (\bibinfo {year} {1982})}\BibitemShut {NoStop}%
\bibitem [{Not()}]{NotaDigit}%
  \BibitemOpen
  \href@noop {} {\bibinfo {title} {{T}he experimental data of
  refs.~\cite{Erik1981,Erik1982} have been digitized using the software
  \href{https://markummitchell.github.io/engauge-digitizer/}{Engauge Digitizer
  v.10.10.} for the $\epsilon_2(\omega)$ function, we find peaks positions at
  378 and 627~cm$^{-1}$ with an estimated position error of
  $\sim$2.5~cm$^{-1}$.}}\BibitemShut {Stop}%
\bibitem [{\citenamefont {Giacomazzi}\ and\ \citenamefont
  {Umari}(2009)}]{LG2009b}%
  \BibitemOpen
  \bibfield  {author} {\bibinfo {author} {\bibfnamefont {L.}~\bibnamefont
  {Giacomazzi}}\ and\ \bibinfo {author} {\bibfnamefont {P.}~\bibnamefont
  {Umari}},\ }\bibfield  {title} {\bibinfo {title} {First-principles
  investigation of electronic, structural, and vibrational properties of
  a-\ce{Si3N4}},\ }\href {https://doi.org/10.1103/PhysRevB.80.144201}
  {\bibfield  {journal} {\bibinfo  {journal} {Physical Review B}\ }\textbf
  {\bibinfo {volume} {80}},\ \bibinfo {pages} {144201} (\bibinfo {year}
  {2009})}\BibitemShut {NoStop}%
\bibitem [{\citenamefont {Galeener}\ and\ \citenamefont
  {Lucovsky}(1976)}]{Gal76}%
  \BibitemOpen
  \bibfield  {author} {\bibinfo {author} {\bibfnamefont {F.}~\bibnamefont
  {Galeener}}\ and\ \bibinfo {author} {\bibfnamefont {G.}~\bibnamefont
  {Lucovsky}},\ }\bibfield  {title} {\bibinfo {title} {Longitudinal optical
  vibrations in glasses: \ce{GeO2} and \ce{SiO2}},\ }\href
  {https://doi.org/10.1103/PhysRevLett.37.1474} {\bibfield  {journal} {\bibinfo
   {journal} {Physical Review Letters}\ }\textbf {\bibinfo {volume} {37}},\
  \bibinfo {pages} {1474} (\bibinfo {year} {1976})}\BibitemShut {NoStop}%
\bibitem [{\citenamefont {Giacomazzi}\ \emph {et~al.}(2006)\citenamefont
  {Giacomazzi}, \citenamefont {Umari},\ and\ \citenamefont
  {Pasquarello}}]{LG2009}%
  \BibitemOpen
  \bibfield  {author} {\bibinfo {author} {\bibfnamefont {L.}~\bibnamefont
  {Giacomazzi}}, \bibinfo {author} {\bibfnamefont {P.}~\bibnamefont {Umari}},\
  and\ \bibinfo {author} {\bibfnamefont {A.}~\bibnamefont {Pasquarello}},\
  }\bibfield  {title} {\bibinfo {title} {Vibrational spectra of vitreous
  germania from first-principles},\ }\href
  {https://doi.org/10.1103/PhysRevB.74.155208} {\bibfield  {journal} {\bibinfo
  {journal} {Physical Review B}\ }\textbf {\bibinfo {volume} {74}},\ \bibinfo
  {pages} {155208} (\bibinfo {year} {2006})}\BibitemShut {NoStop}%
\bibitem [{\citenamefont {Heid}\ \emph {et~al.}(2000)\citenamefont {Heid},
  \citenamefont {Strauch},\ and\ \citenamefont {Bohnen}}]{Heid2000}%
  \BibitemOpen
  \bibfield  {author} {\bibinfo {author} {\bibfnamefont {R.}~\bibnamefont
  {Heid}}, \bibinfo {author} {\bibfnamefont {D.}~\bibnamefont {Strauch}},\ and\
  \bibinfo {author} {\bibfnamefont {K.-P.}\ \bibnamefont {Bohnen}},\ }\bibfield
   {title} {\bibinfo {title} {\emph{Ab initio} lattice dynamics of sapphire},\
  }\href {https://doi.org/10.1103/PhysRevB.61.8625} {\bibfield  {journal}
  {\bibinfo  {journal} {Physical Review B}\ }\textbf {\bibinfo {volume} {61}},\
  \bibinfo {pages} {8625} (\bibinfo {year} {2000})}\BibitemShut {NoStop}%
\bibitem [{\citenamefont {Dicks}\ \emph {et~al.}(2019)\citenamefont {Dicks},
  \citenamefont {Cottom}, \citenamefont {Shluger},\ and\ \citenamefont
  {Afanas’ev}}]{Dicks2019}%
  \BibitemOpen
  \bibfield  {author} {\bibinfo {author} {\bibfnamefont {O.~A.}\ \bibnamefont
  {Dicks}}, \bibinfo {author} {\bibfnamefont {J.}~\bibnamefont {Cottom}},
  \bibinfo {author} {\bibfnamefont {A.~L.}\ \bibnamefont {Shluger}},\ and\
  \bibinfo {author} {\bibfnamefont {V.~V.}\ \bibnamefont {Afanas’ev}},\
  }\bibfield  {title} {\bibinfo {title} {The origin of negative charging in
  amorphous {Al$_2$O$_3$} films: the role of native defects},\ }\href
  {https://doi.org/10.1088/1361-6528/ab0450} {\bibfield  {journal} {\bibinfo
  {journal} {Nanotechnology}\ }\textbf {\bibinfo {volume} {30}},\ \bibinfo
  {pages} {205201} (\bibinfo {year} {2019})}\BibitemShut {NoStop}%
\bibitem [{\citenamefont {Giacomazzi}\ and\ \citenamefont
  {Pasquarello}(2007)}]{Giacomazzi2007}%
  \BibitemOpen
  \bibfield  {author} {\bibinfo {author} {\bibfnamefont {L.}~\bibnamefont
  {Giacomazzi}}\ and\ \bibinfo {author} {\bibfnamefont {A.}~\bibnamefont
  {Pasquarello}},\ }\bibfield  {title} {\bibinfo {title} {Vibrational spectra
  of vitreous {SiO}$_2$ and vitreous {GeO}$_2$ from first principles},\ }\href
  {https://doi.org/10.1088/0953-8984/19/41/415112} {\bibfield  {journal}
  {\bibinfo  {journal} {Journal of Physics: Condensed Matter}\ }\textbf
  {\bibinfo {volume} {19}},\ \bibinfo {pages} {415112} (\bibinfo {year}
  {2007})}\BibitemShut {NoStop}%
\bibitem [{\citenamefont {Schubert}\ \emph {et~al.}(2000)\citenamefont
  {Schubert}, \citenamefont {Tiwald},\ and\ \citenamefont
  {Herzinger}}]{Schubert2000}%
  \BibitemOpen
  \bibfield  {author} {\bibinfo {author} {\bibfnamefont {M.}~\bibnamefont
  {Schubert}}, \bibinfo {author} {\bibfnamefont {T.}~\bibnamefont {Tiwald}},\
  and\ \bibinfo {author} {\bibfnamefont {C.}~\bibnamefont {Herzinger}},\
  }\bibfield  {title} {\bibinfo {title} {Infrared dielectric anisotropy and
  phonon modes of sapphire},\ }\href {https://doi.org/10.1103/PhysRevB.61.8187}
  {\bibfield  {journal} {\bibinfo  {journal} {Physical Review B}\ }\textbf
  {\bibinfo {volume} {61}},\ \bibinfo {pages} {8187} (\bibinfo {year}
  {2000})}\BibitemShut {NoStop}%
\bibitem [{\citenamefont {Boumaza}\ \emph {et~al.}(2009)\citenamefont
  {Boumaza}, \citenamefont {Favaro}, \citenamefont {L{\'e}dion}, \citenamefont
  {Sattonnay}, \citenamefont {Brubach}, \citenamefont {Berthet}, \citenamefont
  {Huntz}, \citenamefont {Roy},\ and\ \citenamefont {T{\'e}tot}}]{Boumaza2009}%
  \BibitemOpen
  \bibfield  {author} {\bibinfo {author} {\bibfnamefont {A.}~\bibnamefont
  {Boumaza}}, \bibinfo {author} {\bibfnamefont {L.}~\bibnamefont {Favaro}},
  \bibinfo {author} {\bibfnamefont {J.}~\bibnamefont {L{\'e}dion}}, \bibinfo
  {author} {\bibfnamefont {G.}~\bibnamefont {Sattonnay}}, \bibinfo {author}
  {\bibfnamefont {J.}~\bibnamefont {Brubach}}, \bibinfo {author} {\bibfnamefont
  {P.}~\bibnamefont {Berthet}}, \bibinfo {author} {\bibfnamefont
  {A.}~\bibnamefont {Huntz}}, \bibinfo {author} {\bibfnamefont
  {P.}~\bibnamefont {Roy}},\ and\ \bibinfo {author} {\bibfnamefont
  {R.}~\bibnamefont {T{\'e}tot}},\ }\bibfield  {title} {\bibinfo {title}
  {Transition alumina phases induced by heat treatment of boehmite: {A}n
  {X}-ray diffraction and infrared spectroscopy study},\ }\href
  {https://doi.org/10.1016/j.jssc.2009.02.006} {\bibfield  {journal} {\bibinfo
  {journal} {Journal of Solid State Chemistry}\ }\textbf {\bibinfo {volume}
  {182}},\ \bibinfo {pages} {1171} (\bibinfo {year} {2009})}\BibitemShut
  {NoStop}%
\bibitem [{\citenamefont {Kao}\ and\ \citenamefont {Wei}(2000)}]{Kao2000}%
  \BibitemOpen
  \bibfield  {author} {\bibinfo {author} {\bibfnamefont {H.-C.}\ \bibnamefont
  {Kao}}\ and\ \bibinfo {author} {\bibfnamefont {W.-C.}\ \bibnamefont {Wei}},\
  }\bibfield  {title} {\bibinfo {title} {Kinetics and microstructural evolution
  of heterogeneous transformation of $\theta$-alumina to $\alpha$-alumina},\
  }\href {https://doi.org/https://doi.org/10.1111/j.1151-2916.2000.tb01198.x}
  {\bibfield  {journal} {\bibinfo  {journal} {Journal of the American Ceramic
  Society}\ }\textbf {\bibinfo {volume} {83}},\ \bibinfo {pages} {362}
  (\bibinfo {year} {2000})}\BibitemShut {NoStop}%
\bibitem [{\citenamefont {Dorsey}(1968)}]{Dorsey1968}%
  \BibitemOpen
  \bibfield  {author} {\bibinfo {author} {\bibfnamefont {G.~A.}\ \bibnamefont
  {Dorsey}},\ }\bibfield  {title} {\bibinfo {title} {Far infrared absorption of
  hydrous and anhydrous aluminas},\ }\href
  {https://doi.org/10.1021/ac60262a036} {\bibfield  {journal} {\bibinfo
  {journal} {Analytical Chemistry}\ }\textbf {\bibinfo {volume} {40}},\
  \bibinfo {pages} {971} (\bibinfo {year} {1968})}\BibitemShut {NoStop}%
\bibitem [{\citenamefont {Tarte}(1967)}]{Tarte1967}%
  \BibitemOpen
  \bibfield  {author} {\bibinfo {author} {\bibfnamefont {P.}~\bibnamefont
  {Tarte}},\ }\bibfield  {title} {\bibinfo {title} {Infra-red spectra of
  inorganic aluminates and characteristic vibrational frequencies of {AlO$_4$}
  tetrahedra and {AlO$_6$} octahedra},\ }\href
  {https://doi.org/10.1016/0584-8539(67)80100-4} {\bibfield  {journal}
  {\bibinfo  {journal} {Spectrochimica Acta Part A: Molecular Spectroscopy}\
  }\textbf {\bibinfo {volume} {23}},\ \bibinfo {pages} {2127} (\bibinfo {year}
  {1967})}\BibitemShut {NoStop}%
\bibitem [{\citenamefont {Demichelis}\ \emph {et~al.}(2010)\citenamefont
  {Demichelis}, \citenamefont {Civalleri}, \citenamefont {Ferrabone},\ and\
  \citenamefont {Dovesi}}]{Demich2019}%
  \BibitemOpen
  \bibfield  {author} {\bibinfo {author} {\bibfnamefont {R.}~\bibnamefont
  {Demichelis}}, \bibinfo {author} {\bibfnamefont {B.}~\bibnamefont
  {Civalleri}}, \bibinfo {author} {\bibfnamefont {M.}~\bibnamefont
  {Ferrabone}},\ and\ \bibinfo {author} {\bibfnamefont {R.}~\bibnamefont
  {Dovesi}},\ }\bibfield  {title} {\bibinfo {title} {On the performance of
  eleven {DFT} functionals in the description of the vibrational properties of
  aluminosilicates},\ }\href {https://doi.org/10.1002/qua.22301} {\bibfield
  {journal} {\bibinfo  {journal} {International Journal of Quantum Chemistry}\
  }\textbf {\bibinfo {volume} {110}},\ \bibinfo {pages} {406} (\bibinfo {year}
  {2010})}\BibitemShut {NoStop}%
\bibitem [{\citenamefont {Giacomazzi}\ \emph {et~al.}(2009)\citenamefont
  {Giacomazzi}, \citenamefont {Umari},\ and\ \citenamefont
  {Pasquarello}}]{luigi2009}%
  \BibitemOpen
  \bibfield  {author} {\bibinfo {author} {\bibfnamefont {L.}~\bibnamefont
  {Giacomazzi}}, \bibinfo {author} {\bibfnamefont {P.}~\bibnamefont {Umari}},\
  and\ \bibinfo {author} {\bibfnamefont {A.}~\bibnamefont {Pasquarello}},\
  }\bibfield  {title} {\bibinfo {title} {Medium-range structure of vitreous
  {SiO$_2$} obtained through first-principles investigation of vibrational
  spectra},\ }\href {https://doi.org/10.1103/PhysRevB.79.064202} {\bibfield
  {journal} {\bibinfo  {journal} {Phys. Rev. B}\ }\textbf {\bibinfo {volume}
  {79}},\ \bibinfo {pages} {064202} (\bibinfo {year} {2009})}\BibitemShut
  {NoStop}%
\bibitem [{\citenamefont {Tane}\ \emph {et~al.}(2011)\citenamefont {Tane},
  \citenamefont {Nakano}, \citenamefont {Nakamura}, \citenamefont {Ogi},
  \citenamefont {Ishimaru}, \citenamefont {Kimizuka},\ and\ \citenamefont
  {Nakajima}}]{Tane2011}%
  \BibitemOpen
  \bibfield  {author} {\bibinfo {author} {\bibfnamefont {M.}~\bibnamefont
  {Tane}}, \bibinfo {author} {\bibfnamefont {S.}~\bibnamefont {Nakano}},
  \bibinfo {author} {\bibfnamefont {R.}~\bibnamefont {Nakamura}}, \bibinfo
  {author} {\bibfnamefont {H.}~\bibnamefont {Ogi}}, \bibinfo {author}
  {\bibfnamefont {M.}~\bibnamefont {Ishimaru}}, \bibinfo {author}
  {\bibfnamefont {H.}~\bibnamefont {Kimizuka}},\ and\ \bibinfo {author}
  {\bibfnamefont {H.}~\bibnamefont {Nakajima}},\ }\bibfield  {title} {\bibinfo
  {title} {Nanovoid formation by change in amorphous structure through the
  annealing of amorphous \ce{Al2O3} thin films},\ }\href
  {https://doi.org/https://doi.org/10.1016/j.actamat.2011.04.008} {\bibfield
  {journal} {\bibinfo  {journal} {Acta Materialia}\ }\textbf {\bibinfo {volume}
  {59}},\ \bibinfo {pages} {4631} (\bibinfo {year} {2011})}\BibitemShut
  {NoStop}%
\bibitem [{\citenamefont {Bennett}\ and\ \citenamefont
  {Ashley}(1965)}]{Bennett65}%
  \BibitemOpen
  \bibfield  {author} {\bibinfo {author} {\bibfnamefont {J.~M.}\ \bibnamefont
  {Bennett}}\ and\ \bibinfo {author} {\bibfnamefont {E.~J.}\ \bibnamefont
  {Ashley}},\ }\bibfield  {title} {\bibinfo {title} {Infrared reflectance and
  emittance of silver and gold evaporated in ultrahigh vacuum},\ }\href
  {https://doi.org/10.1364/AO.4.000221} {\bibfield  {journal} {\bibinfo
  {journal} {Appl. Opt.}\ }\textbf {\bibinfo {volume} {4}},\ \bibinfo {pages}
  {221} (\bibinfo {year} {1965})}\BibitemShut {NoStop}%
\end{thebibliography}%

\ifarXiv
    \foreach \x in {1,...,\numbersupplementpages}
    {
        \clearpage
        \includepdf[pages={\x,{}}]{\supplementfilename}
    }
\fi
\end{document}